%% file: paper.tex
\documentclass[nonacm,screen,manuscript,USenglish]{acmart}

\usepackage{listings}
\usepackage[WOprobsubsuperscripts]{auxlib}

\floatname{algoflt}{Algorithm}
\floatstyle{ruled}
\newfloat{algoflt}{tbp}{loa}
\crefname{algoflt}{Algorithm}{Algorithms}
\def\dcmnumberstyle{}
\def\dcmbasicstyle{}
\makeatletter
\lst@AddToHook{OnEmptyLine}{\addtocounter{lstnumber}{-1}\vspace{-0.5\baselineskip}}
\lstnewenvironment{lstalgo}[2][htb]{%
\lstset{%
	mathescape,%
	tabsize=4,%
	numbers=left,%
	numberstyle=\footnotesize\dcmnumberstyle,%
	numberblanklines=false,%
	basicstyle=\rmfamily\dcmbasicstyle,%
	columns=fullflexible,%
	emph={%
		[1]if, else, then, and, or, for, do, while, return, exit, not, output, initialize %
		},
	emphstyle={%
		[1]\bfseries%
		},%
	escapeinside={/*}{*/},
	escapebegin={\ifmmode\else\hspace{0.5em}\hfill\color{gray}\em\small$\triangleright$ \fi},
	escapeend={\ifmmode\else\fi}
	}%
\setlength{\topsep}{0pt}%
\setlength{\parindent}{0pt}%
\@skiphyperreftrue%
\let\@vspace\@vspace@orig%
\global\def\dcmnumberstyle{\addtocounter{lstnumber}{-1}\color{white}\global\def\dcmnumberstyle{}}%
\def\dcmnolineno{\global\def\dcmnumberstyle{\addtocounter{lstnumber}{-1}\color{white}\global\def\dcmnumberstyle{}}}%
\algoflt[#1]\caption{#2}\hspace{2em}\minipage{\textwidth-2em}%
}{\endminipage\@skiphyperreffalse\endalgoflt}
\makeatother

\let\corollary\relax
\let\lemma\relax
\let\proposition\relax

\declaretheorem[style=plain, sibling=theorem]{corollary}
\declaretheorem[style=plain, sibling=theorem]{lemma}
\declaretheorem[style=plain, sibling=theorem]{proposition}
\declaretheorem[style=plain, sibling=theorem]{observation}
\theoremstyle{plain}
\newtheorem{property}{Property}
\crefname{property}{Property}{Properties}

\def\vardef#1{\expandafter\def\csname #1\endcsname##1{\ensuremath{##1\mathtt{.#1}}}}
\def\col#1#2{\ensuremath{#1\mathtt{.col}[#2]}}
\def\colarr#1{\ensuremath{#1\mathtt{.col}}}
\def\kwdef#1{\expandafter\def\csname #1\endcsname{\ensuremath{\mathtt{#1}}}}

\def\clustercol#1#2{\ensuremath{#1\mathtt{.cluster\_col}[#2]}}
\def\clustercolarr#1{\ensuremath{#1\mathtt{.cluster\_col}}}
\def\clustergen#1{\ensuremath{#1\mathtt{.cluster\_gen}}}

\vardef{gen}
\vardef{mode}
\kwdef{propagate}
\kwdef{twoChoices}
\kwdef{TC}
\kwdef{NIL}
\vardef{ticks}
\vardef{gensize}
\vardef{size}
\vardef{flag}
\vardef{share}
\vardef{wakeup}

\def\Bin{\operatorname{Bin}}
\def\Beta{\operatorname{Beta}}
\def\PE{\operatorname{PE}}
\def\Exp{\operatorname{Exp}}

\makeatletter
\def\?#1{}
\def\whp{w.h.p\@ifnextchar.{.\whpFootnote/\?}{\@ifnextchar,{.\whpFootnote/}{\@ifnextchar){.\whpFootnote/}{.\whpFootnote/\ }}}}
\def\Whp{W.h.p\@ifnextchar.{.\whpFootnote/\?}{\@ifnextchar,{\whpFootnote/.}{.\whpFootnote/\ }}}

\def\whpFootnote/{\global\def\whpFootnote/{}\footnote{%
    The expression \emph{with high probability (w.h.p.)} refers to a
    probability of at least \makebox{$1 - n^{-\Omega(1)}$}.
}}
\makeatother

\title{Positive Aging Admits Fast Asynchronous Plurality Consensus}

\author{Gregor Bankhamer}
\email{gbank@cs.sbg.ac.at}
\author{Robert Elsässer}
\email{elsa@cs.sbg.ac.at}
\affiliation{%
  \institution{University of Salzburg}
  \country{Austria}
}

\author{Dominik Kaaser}
\email{dominik.kaaser@uni-hamburg.de}
\affiliation{%
  \institution{Universität Hamburg}
  \country{Germany}}

\author{Matjaž Krnc}
\email{matjaz.krnc@upr.si}
\affiliation{%
  \institution{University of Primorska}
  \country{Slovenia}
}

\begin{document}

\begin{abstract}
\input{01-abstract}
\end{abstract}
\maketitle

\vspace{\fill}
© 2020 Copyright is held by the owner/author(s). This is the author's version of the work. It is posted here for personal use, not for redistribution. An extended abstract was published by  ACM in the Proceedings of the ACM Symposium on Principles of Distributed Computing (PODC'20), August 3--7, 2020, Virtual Event \url{https://doi.org/10.1145/3382734.3406506}

\newpage
\tableofcontents
\newpage

\input{10-introduction}
\input{11-related-work-long.tex}
\input{12-model}

\input{13-results}

\input{30-asynchronous}
\input{40-multi-leader}
\input{50-breaking-lower-bounds}

\input{60-conclusion}
\begin{acks}
The authors would like to thank Felix Biermeier and Janko Gravner for helpful discussions and important hints.

All authors were partially supported by the Austrian Science Fund (FWF) under grant no. P 27613 ("Distributed Voting in Large Networks").
The first and the second author received funding from the European Union's Horizon 2020 research and innovation programme under Grant Agreement no. 824115 (HiDALGO).
The fourth author acknowledges partial support of the Slovenian Research Agency (research programs P1-0383, P1-0297 and research projects J1-1692, J1-9187) 
and the European Commission for funding the InnoRenew CoE project (Grant Agreement no. 739574) under the Horizon 2020 Widespread-Teaming program and the Republic of Slovenia.
\end{acks}

\bibliographystyle{ACM-Reference-Format}
\bibliography{paper,DBLPelizer}
\clearpage
\appendix
\section*{Appendix}
\input{93-analysis-asynchronous}
\input{94-analysis-decentralized}

\input{95-analysis-accelerated}

\input{96-analysis-extensions}

\input{97-polya-eggenberger}
\end{document}

%% file: 01-abstract.tex
We study distributed plurality consensus among $n$ nodes, each
 of which initially holds one of $k$ opinions. The goal is to eventually agree
on the initially dominant opinion. We consider an asynchronous communication model in which each node is equipped with a random clock. Whenever the clock of a node ticks, it may open communication channels to a constant number of other nodes, chosen uniformly at random or from a list of constantly many addresses acquired in previous steps. The tick rates and the delays for establishing communication channels (channel delays) follow some probability distribution. Once a channel is established, communication between nodes can be performed instantaneously.

We consider distributions for the waiting times between ticks and channel delays that have constant mean and the so-called positive aging property. In this setting, asynchronous plurality consensus is fast: if the initial bias between the largest and second largest opinion is at least $\sqrt{n} \log n$, then after $O( \log \log_\alpha k \cdot \log k + \log \log n)$ time all but a $1/ \polylog n$ fraction of nodes have the initial plurality opinion. Here $\alpha$ denotes the initial ratio between the largest and second largest opinion. After additional $O(\log n)$ steps all nodes have the same opinion w.h.p., and this result is tight.

If additionally the distributions satisfy a certain density property, which is common in many 
well-known distributions,
we show that consensus is reached in 
$O(\log \log_\alpha k + \log \log n)$ time for all but $n/\polylog n$ nodes, w.h.p. This implies that for a large range of initial configurations
partial consensus can be reached significantly faster in this asynchronous communication model
than in the synchronous setting. 

To obtain these results, we first assume the existence of a designated base station and later present fully distributed algorithms. Additionally, we derive tail bounds on the Pólya-Eggenberger distribution, which might be of independent interest.

%% file: 10-introduction.tex
\section{Introduction}
\label{sec:introduction}

\emph{Plurality Consensus} is a fundamental problem in distributed computing.

We are given a set of $n$ nodes, each of which starts with its own initial
opinion (or \emph{color}) from a set of size $k$.
The goal is to design an efficient distributed protocol which ensures
that all nodes agree on the opinion, which is initially supported by the most nodes, 
provided a sufficiently large initial bias is given.

In failure-rate distributions, the concept of \emph{aging} describes how a component or a 
system improves or deteriorates with age. ``No aging'' means that the age of a component has no effect on the distribution of residual
lifetime of the component. This unique case describes a Poisson-clock based 
survival distribution, which is widely used to describe asynchronous models.
The family of \emph{positive aging} distributions 
describes the more general situation where the residual lifetime 
decreases or remains the same with increasing age of a component~\cite{LX06}.
Such situations are common 
in reliability engineering where components tend to become worn out with time
due to increased wear and tear, as well as 
in real-life waiting time scenarios. 
Prominent members of this family of distributions include the exponential, Rayleigh, Weibull (with shape parameter at least 1), and Gamma (with parameter at least 1) distributions.

In this paper we consider an asynchronous communication model, where nodes are equipped with a random clock.
If the clock of a node advances, then the node is activated, and we say that this node \emph{ticks}. 
Upon a tick, nodes may start establishing communication channels to constantly many other nodes. The opening of communication channels is subject to random delays, and communication partners may be chosen uniformly at random or from a list containing constantly many node addresses acquired in previous communication steps.
As long as both -- the ticking time and the channel delay -- satisfy
the positive aging property, our protocols guarantee fast
convergence to the initial plurality opinion. Moreover, if these distributions 
also satisfy what we call the \emph{$q$-density} property (see \cref{prop:tick-in-log}) --  fulfilled by a number of well-known
distributions (e.g.~exponential, Rayleigh or Weibull with shape parameter at least $1$) -- then all but $n/\polylog n$ nodes agree \whp on the initially dominant opinion
significantly faster than in the corresponding synchronous setting for a large range of initial configurations. 
In that sense, our algorithms break the lower bound for plurality consensus in the synchronous model, see \cref{sec:lowerbound-overview}.

%% file: 11-related-work-long.tex
\subsection{Related Work}

\paragraph{Synchronous Protocols}

Plurality consensus in the synchronous model is closely related to randomized rumor spreading. Two early papers \cite{DBLP:journals/iandc/HassinP01,DBLP:conf/cocoon/NakataIY99} focused on 
\emph{pull voting} in networks modeled as a graph. This process is executed in synchronous rounds during
which each node contacts a neighbor uniformly at random and takes its
opinion. If each node is initially assigned
one of two possible opinions, the probability for one opinion to win is
proportional to the number of edges incident at nodes supporting this opinion.
Bounds on the convergence time -- the number of rounds until one opinion prevails --
have been derived in \cite{DBLP:journals/iandc/HassinP01,DBLP:journals/siamdm/CooperEOR13,DBLP:conf/icalp/BerenbrinkGKM16,DBLP:conf/soda/KanadeMS19}.

While pull voting requires convergence time $\Omega(n)$, multiple variants have been introduced
to significently improve the performance. In \cite{DBLP:conf/icalp/CooperER14} the two-choices voting process is introduced,
which has convergence time $O(\log n)$ in case the initial bias is large enough.
In this process, each node contacts two random enighbors, and if the two opinions coincide, then the opinion is adopted.
In addition, further variants of pull voting have been studied. See, e.g., the work by \cite{DBLP:journals/dam/AbdullahD15} on five-sample voting, 
or the more general analysis of multi-sample voting \cite{DBLP:journals/questa/CruiseG14} on the complete graph.

Making the step from pull voting with two opinions to plurality consensus, the authors of \cite{DBLP:journals/dc/BecchettiCNPST17} analyzed the $3$-majority dynamics for $k$ opinions.
In this protocol, each node samples three neighbors and adopts the
majority opinion among the sample, breaking ties uniformly at random. 
The authors prove a tight running time of
$\Theta(k\cdot\log n)$ for this protocol, given a sufficiently large
bias. In \cite{DBLP:conf/soda/BecchettiCNPS15}, the three-state population protocol from \cite{DBLP:journals/dc/AngluinAE08} 
is adopted and generalized to $k$ opinions. The resulting
bound on the running time depends on the $2$-norm of the initial opinion
configuration.
More recently, a detailed study and comparison of
the $3$-majority dynamics and the related two-choices process has been performed by
\cite{DBLP:conf/podc/BerenbrinkCEKMN17}. Subsequently, a tight analysis of these processes was
presented in \cite{DBLP:conf/podc/GhaffariL18}. Together, \cite{DBLP:conf/podc/GhaffariL18} and
\cite{DBLP:conf/podc/BerenbrinkCEKMN17} cover a large range of parameters $k$.

In \cite{DBLP:conf/icalp/BerenbrinkFGK16}, two plurality consensus protocols are proposed. Both
assume a complete graph and realize communication via the random phone call
model. The first protocol is very simple and, \whp, achieves plurality
consensus within $O(\log(k)\cdot\log\log_{\alpha} n + \log\log n)$ rounds \whp
using $\Theta(\log\log k)$ bits of additional memory. The second, more
sophisticated protocol achieves plurality consensus within
$O(\log(n)\cdot\log\log_{\alpha} n)$ rounds \whp using only $4$ overhead bits.
Here, $\alpha$ denotes the initial ratio between the largest and
second-largest opinion. They require an initial absolute bias of
$\omega(\sqrt{n}\log^2 n)$. 
In \cite{DBLP:conf/podc/ElsasserFKMT17} and \cite{DBLP:conf/podc/GhaffariP16a}, two similar protocols were presented which achieve (almost) the same running time bounds.

\paragraph{Asynchronous Protocols}
Population protocols \cite{DBLP:journals/dc/AngluinAER07} are a model for asynchronous distributed computation. In
the basic variant, nodes are modeled as finite state machines. The protocols
run in discrete time steps, where in each step a pair of nodes is chosen
uniformly at random to interact. The interacting nodes update their
states according to a simple deterministic rule. 

In \cite{DBLP:journals/dc/AngluinAE08}, a three-state population protocol for majority (consensus with two opinions) was proposed
that converges after $O(n \log{n})$ interactions ($O(\log{n})$ parallel time) \whp.
If there is a bias of at least $\omega(\sqrt n\log n)$, the protocol converges
to the majority \whp.

Two similar four-state protocols that solve exact majority were presented in \cite{DBLP:journals/siamco/DraiefV12,DBLP:journals/dc/MertziosNRS17}. The protocols are guaranteed to converge to the initial majority opinion regardless of the initial bias, but they require $\Omega(n^2)$ interactions in expectation.
Recently, a large number of papers has considered the stabilization time for exact majority, see~\cite{DBLP:conf/podc/AlistarhGV15,DBLP:conf/soda/AlistarhAEGR17, DBLP:conf/soda/AlistarhAG18,DBLP:conf/wdag/BerenbrinkEFKKR18,DBLP:journals/corr/abs-1805-04586,BKKP20}. 
The currently best known protocol from \cite{BKKP20} requires $O(\log{n})$ states and $O(\log^{3/2}{n})$ parallel time.

Plurality consensus and the related dual problem of coalescing random walks \cite{AF14}
have also been considered in certain asynchronous models.
For an arbitrary number of initial random walks which evolve according to some reversible Markov chain generator, the expected coalescence time is bounded by 
the largest hitting time of an element in the state space \cite{Oli10}. This time corresponds to the expected 
time needed for the corresponding pull voting process to converge.
In \cite{DBLP:conf/icalp/CooperR16}, the so-called linear voting model has been introduced, which covers a number of synchronous and asynchronous 
voting protocols. They show that the expected time of
asynchronous pull voting on a graph with minimum degree $d_{\min}$ and conductance $\Phi$ 
is bounded by $O(nm/(d_{\min} \Phi))$. Here, asynchronicity means that at each step one single node 
is selected u.a.r., and this node chooses a random neighbor for communication.
So-called discordant
voting processes have been considered in \cite{DBLP:journals/siamdm/CooperDFR18}, where in every time step a pair of nodes with different opinions is selected for an interaction. In \cite{DBLP:conf/esa/BerenbrinkFKMW16}, plurality consensus in general graphs and for general bias is solved
using load balancing in different communication models.
In \cite{DBLP:conf/podc/ElsasserFKMT17}, plurality consensus in a synchronous and an asynchronous model is considered.
In the asynchronous case, they assume that each node has a Poisson clock ticking with rate $1$. Whenever the clock of a node ticks, it may choose up to a 
constant number of random neighbors, and revise its opinion based on the set of received opinions. They show that if initially the size of the largest opinion 
exceeds the size of the second largest one by some factor $(1+\epsilon)$, $\epsilon>1$ constant, and the number of opinions is 
$O(\exp(\log n/\log \log n))$, then (partial) consensus is achieved in time $O(\log n)$ \whp. Note that there are no communication delays and once a communication partner is chosen, communication happens instantaneously.

%% file: 12-model.tex
\subsection{Model} \label{sec:model}

 Our model comes with two different forms of asynchronicity, the waiting time between local operations (\emph{ticking time}) and the delay required to engage in communication (\emph{channel delay}). For the ticking time, every node is equipped with a random clock following a distribution with the \emph{positive aging} property.
This property (also known as \emph{decreasing conditional survival} or \emph{increasing failure rate})
is defined as follows.
\begin{property}[Positive Aging] \label{prop:positive-aging}
Let $\mathcal{T}$ be a non-negative distribution and $X \sim \mathcal{T}$. Then $\mathcal{T}$ 
has the \emph{positive aging} property if and only if $P(X > s) \geq P(X > t + s\, |\, X > t)$ for all $s, t>0$.%
\footnote{Our results (except \cref{thm:further-acceleration}) still hold if we require this to only hold for $s > C$ for some constant $C$. For the sake of readability of our analysis we assume that $C=0$.}
\end{property}

When a node ticks, it may start establishing communication channels to a constant 
number of nodes, chosen either uniformly at random or from a list of constantly many addresses acquired in some previous communication steps.
In contrast to the synchronous case, we assume that after initiating a
communication channel, some time is required to build up a connection
to the sampled node. This time -- the channel delay -- is also assumed
to follow a distribution with the positive aging property.
Once the channels to all requested nodes are established,
messages can be exchanged.
For such an exchange of messages no additional time is required.
This reflects the fact that in various scenarios  
(e.g.\ three-way handshake, DNS lookup, or key-exchange for encryption)
the time required for
opening a communication channel may dominate the time required for the entire
communication. 
For both the ticking time and the channel delay we assume that their distributions take values from a
non-negative domain with constant mean.

\paragraph{Remembering Node Addresses}
Many of the results in synchronous and asynchronous plurality consensus assume that each node may only contact random neighbors \cite{DBLP:conf/podc/ElsasserFKMT17,DBLP:conf/podc/GhaffariP16a,DBLP:conf/esa/BerenbrinkFKMW16}.
In our work we assume that nodes may remember the addresses of constantly many nodes, which may be reused for communication in future steps. This allows nodes to communicate with a designated base station or set of leader nodes. We note that such a modification of the random phone call model in rumor spreading leads
 to improvements of the running time \cite{DBLP:journals/endm/DoerrFF11, DBLP:journals/dc/AvinE18, DBLP:conf/podc/HaeuplerM14} or computational complexity \cite{DBLP:conf/soda/ElsasserS08} of standard push-pull protocols. Also in plurality consensus remembering node IDs has lead to extended results in certain cases, see, e.g., \cite{DBLP:conf/wdag/CooperERRS15}.

%% file: 13-results.tex
\subsection{Our Results} \label{own-results} \label{sec:our-results}

We are given $n$ nodes, each of which holds initially one of $k$ different opinions. We
assume that $2\leq k \leq n^{\varepsilon}$ for any constant $0<\varepsilon < 1/2$. Let $a_0$ and $b_0$
be the (relative) size of the initially largest and second largest opinion, respectively.
We assume that the initial (absolute) bias $n \cdot (a_0  - b_0)$ is at least 
$\sqrt{n} \log n$ and we use $\alpha$ to denote the corresponding relative
bias, defined as $\alpha = a_0 / b_0$.

\paragraph{Algorithmic Approach} \phantomsection \label{sec:approach}

Similar to the protocols mentioned above, our plurality consensus algorithms employ 
well-known population dynamics. In particular, we use pull voting
and the $2$-majority dynamics (also called the \emph{two-choices} process).
The nodes pass through
a sequence of numbered stages, which
we call \emph{generations}. 
The intuition is that a certain generation implies a certain
chance for the nodes to have the initially dominant opinion.
This latter property makes the concept of generations a crucial part of our algorithms.

The essential idea of our approach is the following. Every time a node $v$ becomes active, it may 
sample two nodes. Depending on the sample, it may perform one of the following two actions. A so-called \emph{two-choices step} is executed if 
\begin{enumerate}
  \item[(i)] the two sampled nodes are in the same, $i$-th generation,
  \item[(ii)] this generation is at least as high as $v$'s generation,
  \item[(iii)] they have the same opinion, and 
  \item[(iv)] the total number of nodes of that generation is large enough.
\end{enumerate}
In this case $v$ adopts the sampled opinion and proceeds to generation $i+1$.
Otherwise, the node $v$ performs a so-called \emph{propagation step}, where it adopts the
generation and opinion of the node with the highest generation among the sample, provided this generation is higher than its own (breaking ties arbitrarily).
In the analysis we will
 show that the ratio between the largest and second largest opinions grows
rapidly as the generations become higher. As a consequence, any node has the initial plurality opinion once it reaches a certain generation.

\paragraph{Positive Aging in Plurality Consensus}
Many important distributions we consider for clock ticks and channel delays do not allow consensus among all nodes in time less than $\Omega(\log{n})$. However, as we show later, \emph{partial consensus} can be achieved much faster.
Here, partial consensus means that all but at most $n/\polylog n$ nodes agree on the initial majority opinion.
In particular,
we show that in our setting 
partial consensus is reached in $O(\log \log_{\alpha} k \cdot \log k + \log \log n)$ time \whp.
Afterwards, $O(\log n)$ further steps suffice for all nodes to agree on the initial majority opinion, \whp.

We apply aforementioned algorithmic approach and use the concept of generations as well as the method of alternating between two-choices and propagation steps.
In order to determine the time when a 
two-choices step may be performed (see requirement (iv) above), we introduce a leader-based mechanism, which allows the system to be aware of the 
moments in time when the number of nodes in the highest generation is large enough (which, in turn, results in the creation of a new generation).

We first present an algorithm where we assume that there is one predefined base station in the system. This base station has
a restricted amount of memory ($O(\log n)$ bits) and if a node sends a request to this node, then it answers with
the values stored in this memory. More precisely, the base station has a value for the highest generation
allowed to be created in the system (initially set to $1$), and it stores a bit which indicates whether the
nodes should perform
two-choices or propagation steps.

When a node $v$ is activated by a tick,
it contacts the base station and two randomly chosen nodes.
If the base station's bit
allows two-choices and the generation stored in it's memory is higher than the generation of
$v$, then $v$ performs a two-choices step -- if conditions (i)-(iii) are fulfilled as described above (see \nameref{sec:approach}).
Once the base station allows the
creation of a new generation, that is, its bit is set so that two-choices steps are allowed, it starts counting the number of so-called incoming
\emph{signals} sent out by the nodes. After a linear number of signals have been
received, it flips its bit to allow propagation. This ensures that for a constant time frame the nodes promote
themselves to a new generation using only the two-choices dynamics  and thus a new generation of a certain size is created by the two-choices mechanism only.

If a node receives a bit from the base station which allows propagation, it performs a propagation step
as described in the algorithmic approach above.
When a node contacts the base station, it sends its generation number
to it so that the base station can maintain the number of nodes in the highest generation created so far.
Once the majority of all nodes are in the highest generation,
the base station allows the nodes to promote themselves to a higher generation by setting
the corresponding bit accordingly and allowing two-choices steps. These
alternating two-choices/propagation stages are
repeated until the last generation created is monochromatic \whp.
A formal description of this protocol is given in \cref{sec:Asynchronous-1leader}.

Finally, we extend the algorithm described above to a distributed system without a predefined base station in \cref{sec:Decentralized-system}. First,
we partition almost all nodes into clusters of size ${\polylog n}$. During this procedure,
leaders emerge in all these clusters. Then, these leaders act in a distributed manner to coordinate
the actions of the nodes, and we derive an algorithm that mimics the procedure designed for the case with a base station.
This allows us to show a similar result as in the previous case, however, without
assuming the existence of a designated base station.

\paragraph{Comparison with Related Work}

For initial configurations with $k=\Theta(1)$ our protocols match the optimal $O(\log n)$ convergence time for full consensus. A similar result is achieved by \cite{DBLP:journals/dc/AngluinAE08, DBLP:conf/podc/ElsasserFKMT17} with respect to the Poisson-clock model and population protocols.
If $k=\omega(1)$ then our protocols reach partial consensus faster than related approaches \cite{DBLP:conf/esa/BerenbrinkFKMW16,DBLP:conf/podc/ElsasserFKMT17,DBLP:conf/icalp/CooperR16}
that operate in a comparable asynchronous model (i.e., Poisson-clock model, population protocols and sequential model of \cite{DBLP:conf/esa/BerenbrinkFKMW16} with $O(\log n)$ bits of memory per node).
Some of this improvement is related to the fact that our model allows nodes to remember (and reuse)  addresses of constantly many nodes (see \cref{sec:model}).

Our algorithmic approach can also be implemented in the synchronous round-based model. This algorithm achieves (full) plurality consensus in $O(\log k \cdot \log \log_{\alpha} n + \log \log n)$ rounds \whp. Note that this matches current state-of-the-art results of approaches operating in the synchronous setting (e.g. \cite{DBLP:conf/icalp/BerenbrinkFGK16,DBLP:conf/podc/ElsasserFKMT17,DBLP:conf/podc/GhaffariP16a}).
The basic idea is to define a sequence of rounds $\{t_i\}_{i \geq 1}$ at which each node is allowed to perform a two-choices step. Then, at every $t_i$,  a new generation $i$ is created via two-choices step \whp. This sequence of time steps is chosen in such a way that throughout the steps $t_i, t_i + 1, \dots, t_{i+1}-1$ the generation created at time $t_i$ grows to a constant fraction of nodes. We achieve this by setting $t_{i+1} - t_{i} = C \cdot \log k$ for some sufficiently large constant $C$.

\paragraph{Breaking the Lower Bound for Synchronous Consensus Processes}
Many well-known distributions such as exponential, Rayleigh or Weibull with shape parameter at least $1$ satisfy besides positive aging also the $q$-density property (\cref{prop:tick-in-log}, formally defined in \cref{sec:lowerbound-overview}). This property guarantees that  within any time frame of length $1/\log n$ any node ticks and establishes its communication channels to constantly many nodes with probability at least $1/\polylog n$. 
If the distribution of the waiting time between two ticks as well as of the channel delays satisfy this additional property, then the partial consensus time can significantly be
reduced. We show that under these conditions, in time $O(\log \log_\alpha k + \log \log n)$ all but $n/\polylog n$ nodes agree on the initial majority opinion \whp. For a large range
of initial configurations, this convergence time 
is significantly better than any synchronous algorithm can achieve with the same limitations on the number of communication partners of a node per time step as in the
asynchronous model. 
 Note that a similar phenomenon has been observed in rumor spreading w.r.t.~synchronous vs.~asynchronous algorithms \cite{DBLP:conf/soda/FountoulakisPS12}.
Furthermore we show that, assuming that communication can be performed instantly and nodes are activated according to Poisson clocks, partial consensus can be reached in time as low as $O(\log \log n)$ for an initial bias of at least $2\sqrt{n} \log^4 n$.
This is a significant improvement over the $O(\log n)$ (partial) convergence time of \cite{DBLP:conf/podc/ElsasserFKMT17}. While their model does not allow node addresses to be stored, they otherwise operate in this Poisson clock based model and consider a much higher initial bias of $\alpha >( 1+\varepsilon)$ for constant $\varepsilon >0 $. See \cref{sec:lowerbound-overview} for further discussion.

\paragraph{Tail Bounds on the Pólya-Eggenberger distribution with $s=1$}
We model parts of our analysis with the help of a so-called Pólya-Eggenberger urn process \cite{PE23}. The process starts with $a$ black and $b$ white balls and consists of $n$ steps in total. In each step, a black ball is added with probability corresponding to the fraction of black balls currently in the system. Otherwise, a white ball is added to the urn. The related distribution -- called Pólya-Eggenberger distribution -- models the number of black balls added throughout these $n$ steps, and is denoted by $\text{PE}_1(a,b,n)$ in the following. It is known (e.g. page 181 of \cite{JK77}) that this distribution is equivalent to the binomial distribution $\text{Bin}(n,P)$, where the success probability $P$ is drawn a priori from the beta distribution $\text{Beta}(a,b)$. 
Using this representation together with a recently developed tight bound on the Beta distribution \cite{ZZ19}, we state a result that might be of independent interest. Additional discussion, including a proof of this statement, can be found in \cref{sec:polya-eggenberger} starting on page \pageref{sec:polya-eggenberger}.

\begin{restatable}{theorem}{pethmtotal}
\label{pe-thm-total}
Let $A \sim \PE_1(a,b,n- (a+b))$,  $\mu := (a/(a+b)) n$ and $a+b \geq 1$ as well as $n \geq a+b$. Then, for any $\delta$ with $0 < \delta < \sqrt{a}$ it holds for some universal constant $c_2>0$ that
\[
    P\Big(a + A < \mu - \sqrt{a} \cdot \frac{n}{a+b} \cdot  \delta\Big) < 4 \exp(-c_2 \cdot \delta^2)
\]
\[
    P\Big(a + A > \mu + \sqrt{a} \cdot \frac{n}{a+b} \cdot  \delta \Big) < 4 \exp(-c_2 \cdot \delta^2)
\]
\end{restatable}

%% file: 30-asynchronous.tex

\section{Protocol with a Base Station}
\label{sec:Asynchronous-1leader} \label{sec:centralized}

The main difficulty in analyzing our asynchronous protocols lies in the fact
that we cannot predict (accurately) when a new generation has to be created, since the nodes lack a global notion of time. This
is further complicated by the fact that nodes cannot easily decide based on their local view when to execute two-choices and propagation steps. As a first approach, we therefore resort to a so-called \emph{base station} that is constrained to $O(\log n)$ bits of memory. Later, we present a fully distributed algorithm, which does not require any base station. Our intermediate result is the following.

\begin{theorem}
\label{thm:async-centralized} \label{thm:single-leader}
Assume a designated base station is present. The protocol defined in
\cref{alg:The-basic-procedure-asy-1} reaches  partial consensus in
$$O\left(\log\log_{\alpha}k\cdot\log k+\log\log n\right)$$ time \whp. Within
additional $O(\log{n})$ time, all nodes have the initially dominant opinion
\whp.
\end{theorem}

\begin{lstalgo}[t]{Consensus protocol for node $u$.\label{alg:The-basic-procedure-asy-1} \label{alg:asynchronous}}
initialize $(\gen{u},\, \col{u}{0}) \gets (0, \text{initial color of node }u)$

for each tick of node $u$ do
	send signal $0$ to the base station $\ell$. $\label{alg2ln1}$
	if a previous tick is still being processed then
		skip the remainder of the procedure $\label{ln:blocked-tick}$

	sample nodes $v_1\text{ and }v_2$ u.a.r.
	wait $\text{for}$ communication channels to $\ell$, $v_1,\text{ and }v_2$ to open $\label{ln:est-channels}$
	w.l.o.g. assume $\gen{v_1} \geq \gen{v_2}$

	if $\mode{\ell} = \propagate$ and $\gen{v_1}>\gen{u}$ then /* Propagation \label{ln:prop-asy}*/
		$\left( \gen{u},\, \col{u}{\gen{v_1}}\right) \gets \left( \gen{v_1},\, \col{v_1}{\gen{v_1}} \right)$
		send signal $\gen{u}$ to the base station $\ell$ $\label{ln:sl-ln:notify-1}$
	
	if $\mode{\ell} = \TC$ and $\gen{\ell} > \gen{u}\dcmnolineno$ and /* Two-Choices */
	$\col{v_1}{\gen{\ell}-1} = \col{v_2}{\gen{\ell}-1} \neq \NIL$ then $\label{ln:2ch-asy}$
		$\left( \gen{u},\, \col{u}{\gen{\ell}}\right) \gets \left( \gen{\ell},\, \col{v_1}{\gen{\ell}-1} \right)$
		send signal $\gen{u}$ to the base station $\ell$ $\label{ln:sl-ln:notify-2} \label{alg2ln12}$
\end{lstalgo}

\begin{lstalgo}[t]{Consensus protocol for the base station.\label{alg:leader-increment}}
initialize $\left(\gen{\ell},\, \mode{\ell},\, \gensize{\ell},\, \ticks{\ell}\right) \gets
	\left(1,\, \TC,\, 0,\,0\right)$

for each incoming signal $i$ do
	if $i = 0$ then
		$\ticks{\ell} \gets \ticks{\ell}+1$
		if $\ticks{\ell}= \mathcal{H}(C_1) \cdot n$ /* allow propagation \label{ln:gossip}*/
			$\mode{\ell} \gets \propagate$

	if $i = \gen{\ell}$ then
		$\gensize{\ell} \gets \gensize{\ell} + 1$
		if $\gensize{\ell} \geq n/2$ /* start next generation \label{ln:next-gen} \label{ln:one-leader-ln6}*/
			$\left(\gen{\ell},\, \mode{\ell},\, \gensize{\ell},\, \ticks{\ell}   \right) \gets \left(\gen{\ell}+1,\, \TC,\, 0,\, 0\right)$
\end{lstalgo}

\subsection{Our Protocol} \label{sec:async-description}

We analyze the protocol defined in \cref{alg:The-basic-procedure-asy-1},
where we assume that a base station is present. 
This base station receives signals from
nodes and performs simple counting
operations, which are defined in \cref{alg:leader-increment}. It's purpose is to orchestrate the distributed computation by
providing two variables, $\mathtt{gen}$
and $\mathtt{mode}$. 
The variable $\mathtt{gen}$ represents the currently highest allowed
generation in the system, initially set to $1$. The variable $\mathtt{mode}$, initially set to $\mathtt{TC}$ (meaning two-choices),
indicates whether nodes in generation $\mathtt{gen}$ should perform two-choices steps.

When a node ticks, it requests the state of the base station and uses its variable $\mathtt{mode}$ to decide which operation
to execute (see \cref{ln:prop-asy} and \cref{ln:2ch-asy} of \cref{alg:The-basic-procedure-asy-1}).
If a tick occurs while waiting for the channel(s) in \cref{ln:est-channels} to be established, we only allow $v$ to send out a $0$-signal to the base station. The remaining operations are skipped in such a case.
Note that a $0$-signal may need time to reach the base station (the channel opening delay), but nodes
do not need to wait for the actual channel to be established. 

Besides knowledge of $n$, we require that the base station knows upper and lower bounds on the means of the waiting time and channel delay distributions (hidden in the constant $\mathcal{H}(C_1)$). 

For simplicity of presentation we defined 
\cref{alg:The-basic-procedure-asy-1} 
in such a way that node $u$ stores the opinion of generation $i$ as
$\col{u}{i}$. Note that this is done in the pseudocode for presentation
purposes only. For our analysis, it suffices that nodes store their current
opinion and the opinion of the previous generation, $\col{u}{\gen{u}}$ and
$\col{u}{\gen{u} - 1}$, respectively. If a node $u$ does not hold any opinion for generation $i$,
we say that $\col{u}{i} = \NIL$. This is initially the case for all $i > 0$ and might occur, e.g., if node $u$ jumps two generations in a propagation step.
For the range of initial configurations we consider, $O(\log k + \log \log n)$  bits are required for the transmission and storage of the color and generation values.

\paragraph{Notation and Conventions} 
\label{sec:intro-notation}
We define $\mathrm{g}_i(t)$ to be the fraction of
nodes of generation $i$ at time $t$. 
Furthermore, we denote by $c_{j,i}(t)$ the fraction of these $\mathrm{g}_i(t)\cdot n$ nodes  which  have $\col{v}{i} = j$, and let $p_{i}(t) = \sum_{j} c_{j,i}(t)^2$. Note that $1/k \leq p_{i}(t)$ holds as long as $g_i(t) > 0$. Let $\alpha_i(t)$ denote the relative ratio between the most and second-most dominant color in generation $i$ at time $t$. We denote by $t_i$ the point in time when generation $i$ was first allowed by the base station, and let $t_i(\gamma)$ correspond to the time when generation $i$ globally reaches cardinality $\gamma \cdot n$.
Throughout the analysis we may fix a generation $i$ and time $t$ and let
$a$ and $b$ be the opinions with the largest and the second largest support in generation $i$ at time $t$, respectively.
We then define $a_i(t) = c_{a,i}(t)$ and $b_i(t) = c_{b,i}(t)$ for easier readability.
Furthermore, for variables with generation subscript $i$ we sometimes omit the parameter $t$ to denote time $t_{i+1}$ (e.g., $a_i = a_i(t_{i+1})$ ).
Also, if we say that a node $v$ is of color $j$ at some time $t$, we mean $\col{v}{\gen{v}} = j$. Similarly, we will say $v$ takes (or adopts) color $j$, if $v$ increases its generation to some generation $i$ and sets $\col{v}{i} \gets j$.

\subsection{Core Concepts of our Analysis}
\label{sec:sl-overview}

\paragraph{Time Measures}
At the core of the analysis lies the so-called \emph{time unit}. A \emph{time unit} denotes the number of time steps $C_1$ with the following property: Within any time interval of length $C_1$, each node establishes with probability $0.9$ the channels to three nodes chosen for communication. The crucial point is that this time unit is \emph{independent} of the nodes execution history. If the distributions of the channel delays and the time between ticks have the positive aging property, we show that such a time unit $C_1$ is of constant length. 
Unless explicitly stated otherwise, we measure the time in \emph{time units}. 

Counting $0$-signals in \cref{alg:leader-increment} allows the base station to approximate the time accurately. 
Here, $\mathcal{H}(\cdot)$ is a linear functions, which is specified in detail as part of \cref{lem:signals-per-time} on page \pageref{lem:signals-per-time}. 
Additionally, $\mu_0$ and $\mu_{\ell}$ denote the means of the distributions for waiting time and establishing communication channels.
\begin{restatable}{corollary}{signalcounting}
	\label{lem:counting-time}
	Consider a set of nodes $U$ sending $0$-signals to a designated node $v$ upon each activation, where $|U|  \geq \log^{2 + \varepsilon} n$ for some constant $\varepsilon > 0$.
	Let $T = \Omega(1)$.
	Then, $v$ receives $\mathcal{H}(T) \cdot |U|$ many $0$-signals in 
	\begin{enumerate}
		\item at least $T$ and
		\item at most $\mathcal{S}(T) := (\mathcal{H}(T) + 1)\cdot 16 \cdot \max\{\mu_0, \mu_\ell\} = O(T)$ time steps \whp.
	\end{enumerate} 
\end{restatable}
In this section, the designated node $v$ is the base station, and $U$ contains all other nodes.
\paragraph{Time Between Consecutive Generations}
We now consider a fixed generation $i$. That means, we consider the time frame $[t_i, t_{i+1})$ in which the base station has $\gen{\ell} = i$. We are interested in an upper bound on the time frame $t_{i+1} - t_i$. Starting from time $t_i$, we know by \cref{lem:counting-time} that after $\Theta(1)$ time units the condition in \cref{ln:gossip} of \cref{alg:leader-increment} becomes satisfied \whp. Throughout this time, sufficiently many nodes promote themselves to generation $i$ via two-choices steps. 

\begin{restatable}{proposition}{tcgrowth}
\label{lem:1st-step-1}
Fix some generation $i$ and assume that $g_{i-1} \geq 1/2$.
Let $t_i + t'$ denote the time when the base station allows promotions to generation $i$ via propagation. Then, $g_{i}(t_i +t') \geq p_{i-1} / 5$ \whp.
\end{restatable}

 From time $t_i + t'$ until $t_{i+1}$,  the base station only allows propagation steps. Therefore, one can see the set of nodes of generation $i$ as a set of informed nodes, which grows by pull broadcasting (cf.\ \cite{DBLP:conf/focs/KarpSSV00}). That is, the set of nodes of generation $i$  increases by a constant factor in every time unit \whp.
 
\begin{restatable}{proposition}{propgrowth}
\label{prop:1leader-gossiping-one-step}
Fix some generation $i$ and let $t_i + t'$ denote the time when the two-choices phase of generation $i$ ends. Then,
$t''=\log_{1.4} (3 / p_{i-1})$
time units after the base station starts allowing propagation steps, the cardinality of the $i$-th generation exceeds $n/2$ \whp.
\end{restatable}

Remember that as soon as $t_i(1/2)$ is reached, generation $i+1$ is allowed by the base station (see \cref{ln:next-gen} of \cref{alg:leader-increment}). Therefore, it follows that $t_{i+1} - t_i =O( \log( 1 / p_{i-1}))$. For the proofs of the previous two statements and a more detailed discussion we refer to \cref{sec:time-to-inc-gen}.

\paragraph{Concentration Results}
We again consider some fixed generation $i$.  Let $a$ and $b$ be the largest and second largest opinion in generation $i-1$ at time $t_i$. We show that the color fractions $a_i(t)$ and $b_i(t)$ are well concentrated around their expectation.
 Throughout the analysis we assume that color $b$ still has significant support, i.e.,  $b_{i-1} \gg 1/\sqrt{n}$. \label{def:x1_gg_x2} Here $x_1 \gg x_2$ means that there exists a constant $\varepsilon >0 $ s.t.\ $x_1 \geq x_2 \cdot n^{\varepsilon}$.  Otherwise $a_{i-1} = 1- o(1)$, and within $O(1)$ generations, the first \emph{monochromatic} generation is reached. A monochromatic generation $i^*$ w.r.t.\ color $a$ is a generation 
 where all nodes $v$ either have $\col{v}{i^*}=a$ or $\col{v}{i^*} = \NIL$ at any time $t$.
 
We start by focusing on the time frame $[t_i, t_i + t']$, where $t'$ is defined s.t. at time $t_i + t'$ the two-choices phase of generation $i$ ends. 
Observe that a node $v$ that attempts a two-choices step (see \cref{ln:2ch-asy} in \cref{alg:asynchronous}) at time exactly $t_i$, samples two nodes $v_1, v_2$ with defined color value and $\col{v_1}{i-1} = \col{v_2}{i-1}$ with probability exactly $c_{j,i-1}^2 \cdot g_{i-1}^2$.
As in the time frame $[t_i , t_i + t']$ the base station only allows two-choices steps to generation $i$, no other node $v'$ will modify its $\col{v'}{i-1}$ field. Hence, any node that joins generation $i$ throughout $[t_i, t_i +t']$ takes some fixed color $j$ with probability exactly $c_{j,i}^2 / p_{i-1}$. This allows us to state the following.

\begin{restatable}{lemma}{tcconc}
\label{lem:initial-bias-1}
Let $a$ and $b$ be the largest and second largest opinion in generation $i-1$ at time $t_i$ and assume that $a_{i-1} > b_{i-1} \gg 1/\sqrt{n}$. Let $t_{i}+t'$ be the time
when the propagation phase for the $i$-th generation begins. Then \whp
\begin{align*}
a_i(t_i + t')  &= \frac{(a_{i-1})^2}{p_{i-1}} \left(1  \pm \frac{1}{a_{i-1}} \sqrt{\frac{\log n}{n}} \right), \text{ and} \\ 
	b_i(t_i + t') &= \frac{(b_{i-1})^2}{p_{i-1}}  \left(1 \pm \frac{1}{b_{i-1}} \sqrt{\frac{\log n}{n}} \right).
\end{align*}
\end{restatable}
Note that this implies that $a_{i}(t_i +t') / b_{i}(t_i + t') > (a_{i-1} / b_{i-1})^2  \cdot (1 - o(1))$, i.e., the ratio between the most and second-most dominant color fractions roughly squares throughout the two-choices phase. From $t_i + t'$ until $t_{i+1}$, the base station only allows propagation steps. The idea is to show that throughout the propagation phase, this ratio does not deviate by much.
Each time a node performs a successful propagation step it does so based on randomly sampled neighbors. Hence, if we denote by $t^{(1)}, t^{(2)}, ... , t^{(r)}$ with $r = n \cdot (1/2 - g_i(t_i +t'))$ the points in time at which nodes join generation $i$ throughout $[t_i +t', t_{i+1}]$, then the sequence $[c_{j,i}(t^{(\ell)})]_\ell$ forms a martingale for any color $j$. 
However, standard techniques (namely Azuma-Hoeffding) fail to provide tight enough bounds. Instead, we model the number of $j$-colored nodes that join throughout the remainder of generation $i$ with the help of a Pólya-Eggenberger process. The idea is the following. 
We consider an urn, initially containing $n \cdot g_{i}(t_i + t')$ many balls -- one for each node of generation $i$ at time $t_i + t'$ -- with a $c_{j,i}(t_i + t')$ fraction of these balls being black. Each time a node $v$ joins generation $i$ at time $t^{(h)}$ for some $1 \leq h \leq r$, we draw a randomly selected ball from the urn. In case we draw a black ball, we assign color $j$ to $v$ and add a black ball to the urn. Otherwise, we conclude that $v$ did take some color other than $j$ and add a white ball to the urn. We repeat this approach for every of the $r$ nodes that join throughout the propagation phase. The number of black balls added throughout this process corresponds exactly to the number of nodes that take color $j$ in $[t_i + t', t_{i+1}]$. We discuss this process in the Pólya-Eggenberger section (\cref{sec:polya-eggenberger}) and use the corresponding results to show the following.

\begin{restatable}{lemma}{propconc}
\label{lem:polya-color-bound}
Let $a$ and $b$ be the largest and second largest opinion in generation $i-1$ at time $t_i$ and assume that $a_{i-1} > b_{i-1} \gg 1/ \sqrt{n}$. Let $t_i + t'$ be the time when the propagation phase of generation $i$ begins. Then \whp
\begin{align*}
	a_{i} &=  a_{i}(t_i +t') \left( 1 \pm O \left( \sqrt{ \frac{\log n}{n}}  \frac{1}{a_{i-1}}  \right) \right), \text{ and } \\
		b_{i} &=  b_{i}(t_i + t') \left( 1 \pm O \left( \sqrt{ \frac{\log n}{n}}  \frac{1}{b_{i-1}}  \right)\right).
\end{align*}
\end{restatable}

Combining \cref{lem:polya-color-bound} and \cref{lem:initial-bias-1}, we can describe how color fractions behave throughout generation $i$, and we show that the bias almost squares when generation $i+1$ is arises.

\begin{restatable}{lemma}{phaseconc}
\label{cor:one-phase-bias-1}
Let $a$ and $b$ be the largest and second largest opinion in generation $i-1$ at time $t_i$ and assume that $a_{i-1} > b_{i-1} \gg 1/\sqrt{n}$.
Let $b'$ be the second largest opinion in generation $i$ at time $t_{i+1}$.
If $a_{i-1} - b_{i-1} \geq \log n /  \sqrt{n}$, then \whp
\begin{enumerate}
    \item $a$ is the largest opinion in generation $i$ at time $t_{i+1}$,
    \item $\alpha_{i} > (\alpha_{i-1})^{1.5}$, and
    \item $a_{i} - b'_{i} \geq  \log n /  \sqrt{n}$.
\end{enumerate}
\end{restatable}
A repeated application of the above gives us that the initially most supported color stays dominant, and after $O(\log \log_{\alpha_0} n)$ generations the second-most dominant color is of insignificant size. This implies that after $O(1)$ further generations the first monochromatic generation appears \whp. The proofs for the above statment  can be found in be found in \cref{sec:conc-results-single-leader}.

\paragraph{Putting Everything Together}
Summarizing, we established that 
$t_{i+1} - t_{i} = O( \log (1 /p_{i-1})) = O(\log k)$.
As the relative bias is roughly squared each time a new generation is created, the generation $\log_{1.5} \log_{\alpha_0} n  + O(1)$ will be monochromatic.  
Note that from this point on (i) every further generation will also be monochromatic, and (ii) at least $n/2$ nodes carry the majority opinion. Hence, $O(\log \log n)$ time units suffice to reach partial consensus. This translates into a required time of $O(\log \log_\alpha n \cdot \log k  + \log \log n)$. 
 This bound can be tightened slightly to yield the result stated in \cref{thm:single-leader}
by observing that $\alpha_{i-1} \geq k$ implies $t_{i+1}  - t_{i} = O(1)$.

%% file: 40-multi-leader.tex
\section{Decentralized Protocol}
\label{sec:Decentralized-system}

The centralized approach with a predefined base station from \cref{sec:centralized}
violates the distributed computing paradigm and has several drawbacks. Most
notably, a huge number of requests is induced on the base station in each time step
and thus the base station becomes the bottleneck of the execution of the protocol.
Furthermore, the system becomes highly vulnerable against attacks, since an
adversary can compromise the entire computation by taking over the base station. To
avoid these drawbacks and decentralize the computation, we introduce
some changes to our protocols, which guarantee a maximum congestion of
$O(\polylog n)$ per node.

The execution of the protocol runs in
two parts, \emph{clustering} and \emph{consensus}. In the clustering part we
first use a distributed algorithm to cluster the nodes into groups of roughly
$\polylog{n}$ nodes and each cluster elects its own \emph{leader}. In the consensus part we define the behavior
of the leaders of different clusters and their interactions with non-leader
nodes, such that all of them collaborate in order to emulate the 
protocol described in \cref{sec:centralized}.
For both parts, the required storage per node as well as the size of information exchanged through each
communication channel can be bounded by $O(\log n)$ bits.
Formally, we show the following statement.
\begin{theorem}
\label{thm:multi-leader}
The decentralized protocol reaches partial consensus in $O\left(\log\log_{\alpha}k\cdot\log
k+\log\log n\right)$ time \whp. Within additional $O(\log{n})$ time, all nodes
have the initially dominant opinion \whp.
\end{theorem}

\paragraph{The Clustering Algorithm}
In the first part, all but a fraction of $O(1/\polylog n)$ nodes are
partitioned into clusters of polylogarithmic size, each containing a
distinguished node which is the \emph{leader} of this cluster. 
Our clustering algorithm achieves this \whp in $O(\log \log n)$ time. It also ensures that, \whp, each such cluster has size at least $\log^{c-1} n$, where $c>4$ is an arbitrary constant that is governed by the algorithm.
In that way, we no longer have one designated base station, but
$\Theta (n/\polylog{n} )$ decentralized cluster leaders. Additionally, these
cluster leaders trigger the start of the consensus algorithm. The clustering
algorithm is presented and analyzed in  \cref{sec:clustering,sec:extended-clustering}.

\subsection{Description of the Consensus Protocol}
\label{sec:multileader_proc}
After the above-mentioned clustering algorithm, all nodes have to perform our consensus protocol, however the nodes that emerged as leaders throughout the clustering protocol also have to carry out so called leader tasks. We start by describing the protocol for the follower nodes as it does not differ much from the centralized procedure (see \cref{alg:asynchronous}).

 \paragraph{The Follower Perspective}
 Each time the clock of a node $v$ ticks, it sends a $0$-signal
 to its leader and (unless an execution started by a previous tick is still in progress) executes the following algorithm.
 It 
opens channels to three nodes $v_1$, $v_2$ and $v_3$ chosen uniformly at random,
as well as to its own leader $l$ and to $l_3$, the leader of node $v_3$. As soon as
all connections are established, $v$ requests the current opinion and
generation from $v_1$ and $v_2$. Furthermore, the state of the leader $l_3$ is
pulled. Recall that once the channels are established, this information can be
retrieved instantly and simultaneously. The 
possible actions of $v$ are very similar as in the centralized protocol; however,  they  depend on the generation number and 
propagation bit of the (almost) uniformly sampled $l_3$ instead of its own leader $l$.
If the information provided by $v_1$ and $v_2$, together with the state of $l_3$
satisfies the two-choices conditions, then a two-choices step is performed. More precisely, if 
\begin{itemize}
\item $v_1$ and $v_2$ have non-$\NIL$ color values for generation $i-1$ as well as  $\col{v_1}{i-1} = \col{v_2}{i-1}$, and 
\item the highest generation allowed by $l_3$ is $i$, and $l_3$ allows promotion via two-choices steps 
\end{itemize} 
then, $v$ will 
adopt the opinion of $v_1$ and $v_2$ and set its generation to
$i$.  
If according to $l_3$ a propagation step is to be performed, then $v$
executes a propagation step just as in the centralized procedure (see \cref{ln:prop-asy} of \cref{alg:asynchronous}). That is, $v$ adopts the color and generation of either $v_1$ or $v_2$ in case one of them is of generation higher than $v$. 
Finally, $v$ transfers state information of $l_3$ to its own leader $l$, 
together with $v$'s possibly increased generation value.
 
\paragraph{The Leader's Perspective}

\begin{figure}
\centering
\includegraphics[scale=0.8]{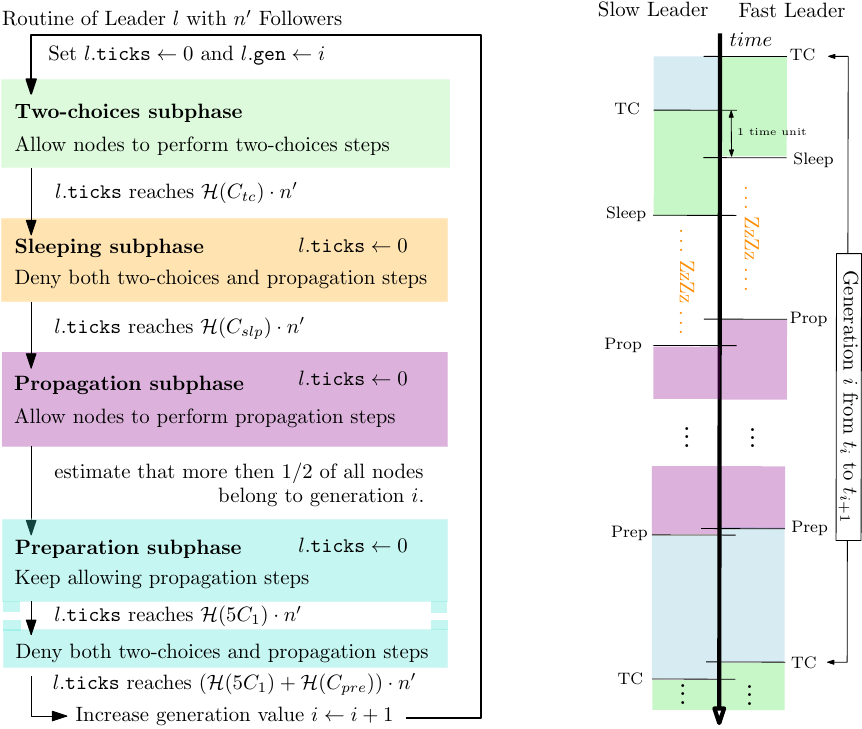}
\caption{Routines executed by leader nodes throughout the consensus mode. On the right side, we describe how the set of leaders progresses through some fixed generation almost synchronously. }
\label{fig:async-leader_simple}
\end{figure}

As opposed to the centralized case, where the base station simply switches between two-choices and propagation mode, leaders now pass through two additional phases.
These two additional phases, called \emph{sleeping} and \emph{preparation} phase, ensure that leaders progress through their generations quite synchronously. For one, achieve that leaders start allowing any fixed generation $i$ at roughly the same time. Additionally,  prevent leaders from allowing two-choices steps while other leaders allow propagation (or vice versa), in order to reuse many parts of the analysis of the centralized case, where the two-choices and propagation phase are properly separated.

With this in mind, the leaders procedure can be described as follows. 
Consider some leader $l$ that just started allowing nodes to promote themselves to a new generation $i$. 
This leader will employ a counter $\ticks{l}$ (just as in the centralized case, see \cref{alg:leader-increment}) in order to count all $0$-signals it receives from its followers.
At the beginning of a generation $i$, the leader starts by allowing two-choices steps 
towards generation $i$, and keeps counting the received $0$-signals of its followers to measure time. 
After receiving sufficient $0$-signals (an amount linear in the number of its followers), the leader enters the so-called \emph{sleeping sub-phase}. 
Note that the $0$-signal counting threshold is set to ensure that \whp there exists a one time-unit frame in which {all} leaders \emph{simultaneously} allow promotions via two-choices before the first leader enters the sleeping phase. 
 
During this sleeping sub-phase, 
which lasts for a constant amount of time, 
the leader again counts incoming $0$-signals to measure time, but neither allows two-choices nor propagation steps.
This forces leaders to wait for some time before entering the propagation phase and allowing promotion via propagation, 
preventing an interleaving of two-choices and propagation phases throughout the system.
Recall that $l$ receives the state information of randomly sampled leaders $l_3$ at each execution of its followers. 
In case $l$ is currently in the sleeping phase and some leader $l_3$ already allows propagation steps,  $l$ will stop sleeping and switches to the propagation phase immediately. 
This way we ensure that no leader is left asleep while some of them may already be finishing   their propagation phase.

After the sleeping sub-phase ends, the leader starts allowing propagation steps and thereby enters the propagation phase. The idea behind this sub-phase is the same as in the centralized case, to quickly spread generation $i$.
However, when it comes to determining when the next generation $i+1$ should be allowed, a more elaborate mechanism than then the one from the centralized algorithm in \cref{sec:Asynchronous-1leader} is needed. In the centralized protocol, the base station simply incremented a counter each time a node promotes to generation $i$. 
As in this decentralized case each leader only has a limited view consisting of its followers, a different approach needs to be employed to estimate the time at which at least $1/2$ of all nodes belong to generation $i$. We interrupt our explanation of the leaders protocol to explain how this can be achieved.

\paragraph{Estimating Global Properties}
Recall that each follower sends the state of the randomly sampled leader $l_3$ to its own leader $l$ upon each execution of the follower procedure. This state information  allows leaders to harvest some information about the global state of the network.
Indeed, if a leader receives $\polylog n$ of such randomly sampled leader-states, it may accurately predict the (global) fraction of leaders satisfying a certain property.
For example, 
let $R$ be such a leader-property which is satisfied whenever the majority of this leaders followers is of generation $i$.
Clearly, a leader $l$ can determine this property by maintaining an $\gensize{l}$ variable. 
Suppose now that $l$ receives $\polylog n$ consecutive  messages regarding random leaders $l_3$ satisfying property $R$. In this case $l$ can be (almost) sure that globally $1/2$ of all nodes are already in generation $i$.
A detailed description of this sampling mechanism together with its analysis can be found in the full version  \cref{sec:sampling}. \\

Using the above approach, the leaders are only allowed to enter the preparation sub-phase after estimating that at least $1/2$ of all nodes belong to generation $i$. 
This guarantees \whp that no leader will start this sub-phase too early.
Upon entering the preparation sub-phase, a leader will still allow propagation steps for some time, but additionally it will again count the incoming $0$-signals. This is done to 
ensure further $\Theta(1)$ waiting time
after which all the leaders are guaranteed to have reached this sub-phase \whp. 
Afterwards, the leader denies both two-choices and propagation steps for $\Theta(1)$ time, which 
prevents propagation steps from occurring during the two-choices phase of the next generation $i+1$.
Finally, the leader resets its counters, increases its highest allowed generation to $i+1$ and starts passing through the 4 sub-phases as part of generation $i+1$.

A visualization of the leaders procedure is given in the left image of \cref{fig:async-leader_simple}. A more detailed explanation of the above algorithm (including the values of the required constants $C_{tc}, C_{br}$ and $C_{pre}$) can be found in \cref{sec:ml-implementing}.

\subsection{Core Concepts of the Analysis}
\label{sec:ml-analysis-concepts}
Roughly, the correctness of our algorithm follows from the analysis results of the centralized approach. To show this we start by the following observations: (i) a follower node $v$ will perform two-choices or propagation steps based on the leader $l_3$ that is chosen independently of the nodes $v_1,v_2$ and $l$, (ii) if at some point all leaders allowed the same generation and sub-phase (e.g. two-choices), then the protocol mimics the behavior of the centralized approach, and (iii) leaders progress through some fixed generation $i$ almost synchronously. To further elaborate on the third point, we now state a selection of the most important invariants which are maintained as the leaders progress through the mentioned sub-phases of arbitrary generation $i$.  We employ the same notation as defined on page \pageref{sec:intro-notation}, with the exception of $t_i$ now denoting the time at which the \emph{fastest}  leader starts allowing generation $i$. The proofs can be found in \cref{sec:ml-analysis} as parts of \cref{prop:ml-timings, prop:ml-generation-change} as well as in \cref{lem:ml-prep-prop}. $C_{br}$ and $C_1$ are constants defined in \cref{sec:ml-variables}.

\begin{lemma}
Fix some generation $i$. Under assumption that all leaders start allowing this generation within time frame of $C_{br} /C_1$ time units, the following statements hold \whp:
\begin{enumerate}
	\item \label{enum:tc-one-tu}All leaders allow two-choices steps towards generation $i$ for at least one simultaneous time unit.
	\item \label{enum:no-overlap}Starting at $t_i$, no leader allows any propagation steps until every leader exits the two-choices sub-phase.
	\item \label{enum:fast-prop}The last leader enters the propagation phase at most $O(1)$ time after the first does so.
	\item \label{enum:tc-parent}No leader enters the preparation phase before $t_i(1/2)$.
	\item \label{enum:enter-gen}Every leader allows generation $i+1$ before time $t_{i+1} + C_{br}/C_1$.
\end{enumerate}
\end{lemma}
Note that \cref{enum:enter-gen} implies that \whp all the above statements hold in the following generations as well. 
To this end, define $t'$ such that at time $t_i + t'$ even the slowest leader has just finished its two-choices phase. In the analysis of the centralized approach, we established that if the base station allows two-choices steps for (at least) one full time unit, then \cref{lem:1st-step-1} follows. Hence, \cref{enum:tc-one-tu} allows us to carry over this result. Furthermore, by \cref{enum:fast-prop} it follows that leaders quickly allow nodes to start spreading generation $i$ via pull propagation, implying the statement of \cref{prop:1leader-gossiping-one-step}. Therefore, the time between $t_{i+1}$ and $t_{i}$ follows the asymptotic bounds as in the centralized case. Also,  \cref{enum:tc-parent} guarantees that majority of all nodes belong to generation $i$ before the two-choices phase of the next generation starts.

When it comes to the concentration of color fractions and evolution of the bias, \cref{enum:no-overlap} is of importance. 
It implies that \whp leaders never allow two-choices and propagation steps at the same time. Hence, each time a node in $[t_i, t_i + t']$ promotes to generation $i$, it is a result of a successful two-choices step. Due to similar reasons as in the centralized case (see paragraph before \cref{lem:initial-bias-1}), and because $l_3$ is selected independently from $v_1$ and $v_2$, such a promotion will cause the node to take color $j$ with probability $c_{j,i-1}^2 / p_{i-1}$. This is the main ingredient of the proof of \cref{lem:initial-bias-1}. Furthermore, starting at time $t_i + t'$, nodes will join generation $i$ via propagation steps only. This allows us to again model the set of nodes that take some color $j$ when promoting to generation $i$ during $[t_i + t', t_{i+1}]$ with a Pólya-Eggenberger process. This leads to the statement of \cref{lem:polya-color-bound} and finally \cref{cor:one-phase-bias-1}. 

Summarizing, we show the same asymptotic guarantees as in the centralized case for both the required number of generations, as well as for the increase of bias with each further generation. A detailed discussion regarding these results can be found in \cref{sec:ml-analysis}.

\subsection{Termination}
This algorithm as well as the centralized algorithm in \cref{sec:Asynchronous-1leader} guarantee that  the nodes eventually reach partial and full consensus, \whp. However, without additional modifications neither of both procedures terminate such that
nodes eventually know that they are in (global) consensus and cease the execution of the protocol.
In \cref{sec:ext-termination} we present an extension to our algorithms that achieves proper termination. 

%% file: 50-breaking-lower-bounds.tex
\section{Breaking the Lower Bound on Synchronized Protocols}
\label{sec:lowerbound-overview}

In this section we first outline the \emph{Accelerated Consensus Protocol}, a modification of the decentralized protocol from \cref{sec:Decentralized-system} and then we argue
that this protocol breaks a lower bound on plurality consensus protocols in the synchronous model.
As before, we assume that all but $n/\polylog n$ nodes are partitioned into clusters of size at least $\polylog n$.
For the accelerated protocol we now assume that in addition to the positive aging property the distributions for the waiting time between ticks and the channel delays are $q$-dense for some constant $q > 0$.

\begin{property} [$q$-dense distribution]
\label{prop:tick-in-log}
Let $\mathcal{T}$ be a non-negative distribution and $X \sim \mathcal{T}$. Then $\mathcal T$
is \emph{$q$-dense} if and only if there exists a constant $t>0$ such that $P(X < s) > s^{q}$ for all $0< s < t$.
\end{property} 

The main difference to the decentralized protocol is the following. All nodes
in a cluster share the same generation and color, which are stored at the
cluster leader. Each time a follower performs a two-choices or propagation
step, the shared variable of its cluster leader is updated (instead of its own as part of the decentralized procedure). So whenever a
two-choices or propagation step updates color or generation, this change is reflected at the leader. Similar, each time a node is queried for its color or generation, it will answer with its leaders shared values instead. That way, followers only act as proxies and help to achieve consensus among the shared color values that are stored at each cluster. 

\cref{prop:tick-in-log} together with positive aging guarantees that in every time frame of length $O(1/\log n)$ a follower of each
cluster ticks and establishes communication channels to all chosen nodes \whp, as long as the clusters are of large enough (polylogarithmic) size.
In case of the decentralized protocol, leaders spend most of their time in the propagation sub-phase (which is the only sub-phase taking $\omega(1)$ time each generation). 
Now, consider the Accelerated Consensus Protocol, and assume that at some point during generation $i$, all leaders allow propagation to generation $i$. As at least one follower of each cluster ticks within every time frame of $O(1/\log n)$, this can be seen as spreading generation $i$ between clusters via pull broadcast at an $\Omega(\log n)$ accelerated rate. 
This way, the time between two consecutive generations, $t_{i+1} - t_{i}$, can be reduced to $O(1)$ \whp.
More details and an analysis can be found in \cref{sec:breaking-the-lower-bound}.

\begin{theorem}
\label{thm:accelerated}
Assume that the initial absolute bias is greater than $2 \sqrt{n} \log^{c'} n$ for any constant  $c' > 4q + 4$. Then the Accelerated Consensus Protocol reaches partial consensus in $O(\log \log_{\alpha} k + \log \log n)$ time
\whp.
\end{theorem}

For a simple lower bound on synchronous protocols, we consider the classical
synchronous model \cite{DBLP:conf/soda/BecchettiCNPS15,DBLP:journals/dc/BecchettiCNPST17}, where we assume that each node may communicate with
$O(\polylog n)$ nodes per round. Additionally, we assume that the nodes do not
know the set of initial opinions (however $k$ may be known to the nodes).  For a node to adopt a certain opinion in this model, it must have interacted at least
once with a node that knows about the existence of this opinion. As each node may communicate
with at most $O(\polylog n)$ other nodes in each round, in order to spread the
initially dominant opinion $a$ (with initial relative support $a_0$) to at least $n/\polylog n$ nodes, one needs $\Omega( \log (1/ a_0) / \log \log n)$ time steps. 

To compare the running time of the asynchronous protocol with this lower bound, consider for example an initial configuration with $\alpha = 2$ and $k = n^{\varepsilon}$ for some constant $0 < \varepsilon < 1/2$. If, initially, all opinions besides the majority opinion have roughly the same support, then our algorithm requires $O(\log \log n)$ time to reach partial
consensus \whp. Any protocol operating in the synchronous round-based model requires $\Omega(\log n / \log \log n)$ time for this task. 

\paragraph{Further Acceleration}
In case all nodes are activated by Poisson clocks with mean $1$, and the exchange of information can be performed instantly, above protocol can be further improved. 
Instead of being constrained to approximate time frames of (at least) constant length via counting of $0$-signals (see \cref{lem:counting-time}), leaders can approximate time frames of length $1/\polylog n$ accurately in this setting -- as long as their cluster is of large enough polylogarithmic size. This is implied by the so-called \emph{memoryless} property of the exponential distribution, as well as the fact that instant communication implies that during some time frame $[t',t'']$, leaders will only receive $0$-signals that were initiated exactly during this time frame. This allows us to speed up not only the propagation phase but also every other phase by a factor of $\Omega(\log^2 n)$ -- in some sense this can be seen as reducing the length of a time unit to $O(1/ \log^2 n)$. This allows full consensus between leaders to be reached after $O(1)$ time. The total running time is then dominated by the clustering procedure and the time followers require to collect the final color values of their leaders. We show the following in \cref{sec:ext-poisson-accelerated}.

\begin{restatable}{theorem}{poissonextension}
	\label{thm:further-acceleration}
	Assume the waiting time between ticks follows $\Exp(1)$ and information between nodes can be exchanged instantly.
	Then, the Accelerated Consensus Protocol can be modified s.t. for an initial bias of at least $2 \sqrt{n} \log^{4} n$, it reaches partial consensus in time $O(\log \log n)$. 
\end{restatable}

%% file: 60-conclusion.tex
\section{Conclusion}
\label{sec:discussion}

In this paper we considered the plurality consensus problem for the setting where we require a certain initial bias between the largest and second largest opinion.
We focused on a particular variant of an asynchronous communication model and showed that asynchronous plurality consensus is fast:  after $O( \log \log_\alpha k \cdot \log k + \log \log n)$ time steps all but a $1/ \polylog n$ fraction of nodes have the initial majority opinion. 
Furthermore, we modify these algorithms such that for a large range of initial configurations and distributions, partial consensus is achieved faster than in \emph{any} algorithm that operates in the corresponding synchronous setting.

In the future we would like to look at several related questions which are still open. 
One possible extension would be to model communication delays on a message basis instead of a channel basis.
However in such a model it seems that one cannot avoid the interleaving of the two-choices sub-phase with the propagation sub-phase within the same generation.
An even more ambitious question would be to try analyze the leaderless variant of the protocol: each time a node ticks it samples two random nodes and executes a propagation step or a two choices step (whichever possible).
In such a setting there are no limitations when, e.g., a higher generation is allowed. While this approach raises many technical difficulties related to the analysis of the running time, our experimental results show that this leaderless algorithm, despite its simplicity, behaves similarly as the ones described in this paper.

%% file: 93-analysis-asynchronous.tex

\section{Preliminaries}
Since our model allows a variety of distributions to be used, and as most of our standard notations are expressed in terms of \emph{time units},
we devote the first subsection to the concept of measuring time, where we cover several
important properties regarding our time measurements.
Additionally, in \cref{subsec:notions} we will define some further notation and conventions we did not cover at the end of \cref{sec:async-description}.

\subsection{Measuring Time}
\label{subsec:Latency-time-statements}

In the context of our asynchronous communication model,
let $\mathcal{T}_0$ denote the distribution of the waiting time between two ticks of a node.
Furthermore let $\mathcal{T}_f$ and $\mathcal{T}_\ell$ correspond to the 
distributions of the time required to open communication channels to the base station (leader in the decentralized case) or to
a follower, respectively. We assume that $\mathcal{T}_0$, $\mathcal{T}_f$, and $\mathcal{T}_\ell$ each fulfill the
positive aging property (\cref{prop:positive-aging}) defined in \cref{sec:model}.

Taking a closer look at \cref{alg:The-basic-procedure-asy-1}, when a node ticks it will start establishing communication channels (to three nodes) unless it is still waiting for channel openings triggered by a previous tick.
Note that as long as a node waits for establishing communication channels after a tick, it is not allowed to start opening further communication channels, if during that time another tick occurs.
However, in any case a $0$-signal (see \cref{alg2ln1} of \cref{alg:The-basic-procedure-asy-1}) is sent to the base station.

Any tick that is not blocked due to ongoing channel establishments initiated by a previous tick, will cause the node to wait for time distributed as $\max\{ \mathcal{T}_f,\mathcal{T}_f, \mathcal{T}_\ell \}$ until all communication channels are opened. Since the signal sent to the base station in \cref{alg2ln1} of \cref{alg:The-basic-procedure-asy-1} does not need a confirmation, it is assumed that it does not cause any waiting time; however, the arrival time still follows the distribution $\mathcal{T}_\ell$.
Now, our model allows information from all partners to be read atomically and instantly as soon as all channels are established. That is, \emph{no time} passes between reading the information and deciding which action to take.

This brings us to the notion of a \emph{time unit}, as described in \cref{sec:sl-overview}. Remember that a \emph{time unit} denotes the number of time steps $C_1$, with the following property: Within any interval of such length $C_1$, each node establishes all three communication channels required for one execution of \cref{alg:asynchronous}, with probablility 0.9. Another important requirement for this {time unit} is to be independent of the nodes actions before the start of this time interval.
In the following we establish that,
within our distribution assumptions, this \emph{time unit} is of constant length.

We simplify the analysis of the time unit by assuming that the communication channels for any fixed node are opened one after another instead of concurrently. That is, the distribution $\mathcal{T}_0 + \mathcal{T}_f + \mathcal{T}_f + \mathcal{T}_\ell$ gives us the next time all channels have opened, as long as we are at the exact point in time where at which the previous execution finished. It is easy to see that upper bounds on the time unit in this modified process carry over to the original scenario. The following lemma is specified in a general manner (such that it can be applied also in the decentralized where additional communication channels need to be established, see \cref{sec:Decentralized-system}). Keep in mind that in the case of \cref{alg:The-basic-procedure-asy-1} we have $h=3$,  $\mathcal{T}_0$ is the same is defined above,  $\mathcal{T}_1, \mathcal{T}_2 = \mathcal{T}_f$, as well as $\mathcal{T}_3 = \mathcal{T}_\ell$.

\begin{lemma}
\label{lem:time-unit-major}
Consider some node $v$.
Then, the time until $v$ finishes its next full execution of \cref{alg:asynchronous} can be majorized by $\sum_{i=0}^h \mathcal{T}_i$, independent of the node's execution history. Here $\mathcal{T}_0$ denotes the distribution for the time between ticks and $\mathcal{T}_i$ for $1 \leq i \leq h$ are the distributions that denotes the required time to open the $i$-th communication channel, respectively.
\end{lemma}
\begin{proof}
W.l.o.g.\ assume that the current time is $0$, and $t$ time steps ago the previous execution of $v$ ended. Furthermore let $X_i \sim \mathcal{T}_i$ for $0 \leq i \leq h$ describe the  waiting times of the current execution. In order to majorize the remaining waiting time, we assume that all channels at time $(-t)$ are opened after another instead of simultaneously.
As we assume $v$ to be currently waiting for the next time all channels are opened, there must exist values $i$ and $t'$  such that (i) $X_0 + X_1 + .. + X_{i-1} = t'$ with $t' < t$, and (ii) $X_{i} > t - t'$. Fixing those values of $i$ and $t'$, we can state the probability that the remaining waiting time exceeds some arbitrary time unit $s$ as
\begin{align}
\nonumber
	&\; P \Big(t' + X_i + X_{i+1} + ... + X_k > s + t ~|~ X_{i} > t -t' \Big)  \\
	=&\; P \Big( X_i > (s- X_{i+1} - ... - X_k) + (t - t') ~|~ X_{i} > t - t' \Big).\label{eq:time-unit-property}
\end{align}
Now we use that \cref{prop:positive-aging} (Positive Aging) holds for the distribution of $X_i$. That is, for all $e \in \mathbb{R}, f\in \mathbb{R}_+$ it holds that $P(X_i > e)  \geq P(X_i > e + f ~|~ X_i +f)$. Hence, setting $e:=(s- X_{i+1} - \dots - X_k)$, and $f=(t-t')$ implies that the probability in (\ref{eq:time-unit-property}) can be upper-bounded by $P(X_i + X_{i+1} + ... + X_k > s ) < P(X_0 + X_1 + ... + X_k > s)$.
\end{proof}

\begin{corollary}
\label{cor:time-unit-bound}
  Let $\mu_0, \mu_f$ and $\mu_\ell$ denote the means of $\mathcal{T}_0, \mathcal{T}_f$ and $\mathcal{T}_\ell$ respectively.
  Then a {time unit} is of constant length $C_1 < 40 \cdot (\mu_0 + 2\mu_f + \mu_\ell)$.
\end{corollary}
\begin{proof}
	By \cref{lem:time-unit-major} we know that the time until some node  $v$ completes the next execution of \cref{alg:asynchronous} can be majorized by $X_0 + X_1 + X_2 + X_3$ where $X_0 \sim \mathcal{T}_0$ and $X_1,X_2 \sim \mathcal{T}_f$ as well as $X_3 \sim \mathcal{T}_\ell$. A Markov inequality application yields for any $i$ that
	$P(X_i > 40 E[X_i]) < 1/40$. Therefore $P(X_0 + X_1 + X_2 + X_3 < 40( E[X_0] + E[X_1] + E[X_2] + E[X_3])) > 9/10$ when applying union bounds.
\end{proof}

One may of course also achieve a sharper bound on the \emph{time unit} $C_1$ when considering the distributions $\mathcal{T}_0$, $\mathcal{T}_f$ and $\mathcal{T}_\ell$ directly.

\begin{example}
\label{ex:exp-delays}
		If $\mathcal{T}_0= \text{Exp}(1)$ and $\mathcal{T}_f = \mathcal{T}_\ell = \text{Exp}(\lambda)$, then $C_1 < 8 / \min\{\lambda,1\}$ time steps. Here $\text{Exp}$, denotes the exponential distribution and $\lambda > 0$ is a constant.
\end{example}

\begin{proof}
		The exponential distribution fulfills \cref{prop:positive-aging}, which is just a weaker version of \emph{memorylessness}. That is, according to  \cref{lem:time-unit-major} we can majorize a time unit by $\mathcal{T}_0 + \mathcal{T}_f + \mathcal{T}_f + \mathcal{T}_\ell$, which can in turn be majorized by the  Erlang distribution $\text{Erl} ( \min\{1,\lambda\},4)$. Using its CDF we get
		\begin{equation*}
		P(\mathcal{T}_0 + \mathcal{T}_f + \mathcal{T}_f + \mathcal{T}_\ell \leq x ) \geq 1 - e^{-\min \{1, \lambda \} \cdot x} \sum_{i=0}^{3} \frac{(\min \{1, \lambda \} \cdot x)^i}{i !}.
\end{equation*}
	Plugging in,  for example, the value $x=\frac{8}{\min\{\lambda,1\}}$ guarantees a probability of more than $90\%$.
\end{proof}

Before heading further we present a list of properties that are implied by the positive aging property.

\begin{lemma}
\label{lem:survival-properties}
Let $X$ be a non-negative random variable whose distribution fulfills \cref{prop:positive-aging} and has constant mean $\mu$. Then
	\begin{enumerate}
		\item $P(X > s) \cdot P(X > t) \geq P(X > s+t)$ for all $s,t >0$,
		\item $P(X > x)^{2^i} \geq P(X > x \cdot 2^i)$ for all $x > 0$ and $i \in \mathbb{N}_0$,
		\item $(1/2)^{2^i} \geq P(X > 2\mu \cdot 2^i)$ for all $i \in \mathbb{N}_0$,
		\item \label{item:survival-sq-constant}$E[X^2]$ is a constant smaller than $72\mu^2$, in particular,  $V[X]$ is constant.
	\end{enumerate}
\end{lemma}
\begin{proof}
	The first statement follows directly from the main property of conditional probabilities. The second follows from the first when setting $s=t=x$ and (inductively) repeating this approach $i$ times. The third statement follows from the second and the Markov inequality as $(1/2) > (X > 2\mu)$.
	
The proof of the fourth statement is more involved. Consider a random variable $X^2$. Additionally we define $X_{e} \sim \text{Exp}( (1/2) \ln (2))$, which implies that $P(X_e > x) = (1/ \sqrt{2})^{x}$. Note that $P(X^2 > x^2) = P(X > x)$ and therefore the third statement implies for all $i \in \mathbb{N}_0$ that
\[
	P(X^2 > 4\mu^2 \cdot 2^{2i}) \leq (1/2)^{2^i} = P(X_e^2 > 2^{2(i+1)}).
\]
Define functions $f(i) = P(X^2 > 4\mu^2 \cdot 2^{2i})$ and $f_e(i) = P(X_e^2 > 2^{2i})$, which are both monotonously decreasing.
Above inequality showed that for any arbitrary $i \geq 0$ it holds that
$f(i) \leq f_e(i+1)$. Monotonicity implies for all $j \in [i,i+1]$  that $f(j) \leq f_e(i+1)$ as well as $f_e(i+1) \leq f_e(j)$ must hold. Hence we conclude that
\[
	\forall j \in [i,i+1]: f(j) < f_e(j),
\]
which implies for \emph{any} $j' \geq 0$ that $f(j') < f_e(j')$.  Setting  $j'= (1/2) \log (x)$ implies that
\begin{equation*}
	P(X^2 > 4 \mu^2 \cdot x) \leq P(X_e^2 > x),
\end{equation*}
for \emph{any} $x \geq 1$. Finally, we consider the second moment of $X$ as
\begin{align*}
	E[X^2] &= \int_{0}^{\infty} P(X^2 > y) dy = \int_{0}^{4\mu^2} P(X^2 > y) dy + \int_{4 \mu^2}^{\infty} P(X^2 > y) dy \\
	&< 4 \mu^2 + 4 \mu^2 \int_{1}^{\infty} P(X^2 > 4 \mu^2 x) dx \\
	&\leq 4 \mu^2 + 4 \mu^2 \int_{1}^{\infty} P(X_e^2 > x) dx ,
\end{align*}
where at the start of the second line we crudely bounded the first integral and substituted $y= 4\mu^2 x$ in the second. Therefore it follows that
\begin{equation*}
	E[X^2] < 4\mu^2 + 4\mu^2 \cdot E[X_e^2].
\end{equation*}
Remember, that $X_e$ follows and exponential distribution with $\lambda = (1/2) \cdot \log_e(2)$. As both variance and mean of $X_e$ are well known we deduce that
\begin{equation*}
	E[X^2] \leq 4\mu^2 + 4\mu^2 \cdot \left(\frac{2 \cdot \sqrt{2}}{\log_e (2)}\right)^2 = O(1). \qedhere
\end{equation*}

\end{proof}
Note that the second and third statement governs information about the distributions right tail. That is, $P(X > 2\mu \cdot x)$ decreases exponentially fast in $x$.

Having covered the important concepts regarding the time measurements in our model, we are now ready to define the basic notions used in throughout the analysis of our algorithms.

\subsection{Description of Notation and Conventions}\label{subsec:notions}

We start by noting that in \cref{sec:async-description} on page \pageref{sec:intro-notation}, we already described  most of the employed notation.
There, we define $t_i ,t_i(\gamma), c_{j,i}(t), \allowbreak g_{i}(t), p_{i}(t), \alpha_{i}(t)$ and explain that we usually fix $a$ and $b$ to the largest and second largest opinion in some generation $i-1$ at time $t_i$. Furthermore, we noted that we sometimes omit the function parameter $t$ for the ease of readability.

In addition to previous definitions, by $\mathcal{T}_0$, $\mathcal{T}_f$ and $\mathcal{T}_\ell$ with means $\mu_0$ and $\mu_f, \mu_\ell$, we denote the distributions of the nodes time between ticks and channel delays to follower and leader nodes (in this case the base station), respectively.
Also, we will denote with $\Bin(n,p)$, $\text{Beta}(a,b)$, $\PE_1(a,b,n)$ and $\text{Exp}(\lambda)$ the binomial, beta, Pólya-Eggenberger (see \cref{sec:polya-eggenberger}) and exponential distribution, respectively. Further conventions include that, unless the base of a logarithm is explicitly given, 
$\log n=\log_{2}n$ while $\ln n=\log_{e}n$. Complementing the definition of $x_1 \gg x_2$ on page \pageref{def:x1_gg_x2}, we say $x_1 \ll x_2$ if
there exists a constant $\varepsilon > 0$ s.t. $x_1 \cdot n^{\varepsilon}\leq x_2$. Additionally, $x_1\sim x_2$ if $x_1\not\ll x_2$ and $x_1\not\gg x_2$.  Also, we will sometimes specify concentration statements in the form of $x_1  = x_2 \cdot (1 \pm \delta)$ for some values $x_1, x_2$ and error term $\delta$. Formally this denotes $x_1 \geq x_2 \cdot  (1 - \delta)$ $\land$ $x_1 \leq x_1 \cdot (1 + \delta)$.

\begin{remark}
	\label{rem:alpha-b}
	Let $a$ and $b$ denote the largest and second largest opinion in generation $i$ at time $t_{i+1}$. Then the following statements hold 
	\begin{enumerate}
		\item $b_i \gg \frac{1}{\sqrt{n}} \qquad \Leftrightarrow  \qquad \alpha_i \ll \sqrt{n}$
		\item $b_i \sim \frac{1}{\sqrt{n}} \qquad  \Leftrightarrow \qquad \alpha_i \sim \sqrt{n}$
		\item $b_i \ll \frac{1}{\sqrt{n}} \qquad \Leftrightarrow \qquad \alpha_i \gg \sqrt{n}$.
	\end{enumerate}
\end{remark}
\begin{proof}
	Let $b_i \ll\frac{1}{\sqrt{n}}$ or $b_i \sim \frac{1}{\sqrt{n}}$. Because $k \ll \sqrt{n}$, it holds that $a_i \geq (1- (k-1) b_i) = 1 - o(1)$ even if all other colors are of equal size $b_i$. In this case $\alpha_i = a_i / b_i$ is determined by $1/b_i$, since $a_i$ is roughly 1.

	Now, if $b_i \gg \frac{1}{\sqrt{n}}$ it follows that $\alpha_i = a_i / b_i \ll a_i  \sqrt{n} \leq \sqrt{n}$, since $a_i$ is bounded by 1.
\end{proof}

\section{Analysis of the Asynchronous Model with a Base Station} \label{sec:analysis-centralized}\label{subsec:The-Analysis-1leader}

In this section we describe the main ingredients of our analysis.
We first consider the $0$-signal counting mechanism that is employed by the base station to measure global time. That is, we will see that at least $1$ and at most $O(1)$ time units after the creation of generation $i$ at $t_i$, the condition in \cref{ln:gossip} of \cref{alg:asynchronous} will be fulfilled. This effectively guarantees that the two-choices lasts for at least $1$ time unit, implying that the set of nodes of generation $i$ after the two-choices phase is large enough. Next, we upper bound the time
needed for the algorithm to increase the amount of nodes of generation to at least $n/2$ by
propagation steps, and thereby bound the required time between the creation of two successive generations. In the following part of the analysis we consider the behavior of the color fractions of nodes of generation $i$ in the time frame $[t_i, t_{i+1}]$.
We will see that the two-choices phase causes the ratio between two colors in the following generation to roughly square. We then establish that until the end of the propagation phase, these fractions remain highly concentrated.
We achieve this result by fitting our process to a so-called Pólya-Eggenberger urn model and using the tail bounds from \cref{sec:polya-eggenberger} on the corresponding distribution. These results
are then used to show that \whp from one generation to the next the ratio 
between the largest and second largest opinion
is squared (up to some small error term). From this we then compute how many generations are needed
in order to guarantee a monochromatic generation \whp and conclude the proof.

\subsection{Measuring Time}
\label{sl-analysis-time}
Consider \cref{ln:gossip} of \cref{alg:leader-increment}. The general idea is to ensure that in any generation the two-choices phase lasts for at least one time unit. To that end we study how accurately the counting mechanism in \cref{ln:gossip} approximates the global time. To state a general result that can later also be used in \cref{sec:Decentralized-system}, where multiple leaders are assumed to be present, we relax the size of $U$. Keep in mind that throughout this section $v$ is the base station, and $U=V$.

\begin{lemma}
\label{lem:signals-per-time}
Consider a set of nodes $U$ sending $0$-signals to a designated node $v$ upon each activation, where $|U| \geq \log^{2+ \varepsilon} n$ for some constant $\varepsilon > 0$. Fix $[t, t+ L]$,  a time interval of length $\Omega(1) \leq L \leq O( \log n)$ and let $W$ be the amount of $0$ signals received by $v$ throughout this interval.
Then, it holds that
	\begin{equation*}
		\mathcal{L}(L)\cdot |U| < W < \mathcal{H}(L) \cdot |U|,
	\end{equation*}
	with $\mathcal{L}(L) :=  \frac{1}{4}\Bigl\lfloor \frac{L}{4 \mu_m}\Bigr\rfloor (1 - o(1))$ and $\mathcal{H}(L) := ( \frac{8\mu_\ell}{\mu_0} + \frac{2L}{\mu_0} + 3C' + 3)(1 + o(1))$ for $\mu_m = \max(\mu_0, \mu_\ell)$ and $C' < 600$.
\end{lemma}
\begin{proof}
We will prove the lower and upper bounds separately. \\
\textbf{Lower bound.}
Let $\mu_m = \max\{\mu_0, \mu_\ell \}$ and consider some time interval of length $4 \mu_m$ time steps.
Assuming the current global time is at the start of this interval, we are interested in the amount of nodes ticking in the first half $2\mu$ of this interval.
Consider some node $v_i$ and let $X_i$ denote a r.v. with $X_i \sim \mathcal{T}_0$. As $\mathcal{T}_0$ follows \cref{prop:positive-aging}, we can lower bound the probability that $v_i$ ticks in the next $2\mu_m$ time steps by $P(X_i < 2\mu_m)$, independent of the nodes previous ticks. Using the results of \cref{lem:survival-properties} it follows that $P(X_i < 2\mu_m) > 1/2$. A Chernoff bound application yields that at least $\frac{|U|}{2} \cdot (1 - o(1))$ nodes will tick throughout the first $2\mu_m$ time steps \whp.

Upon a node ticks and sends a $0$-signal, additional time distributed according to $\mathcal{T}_\ell$ is required for the signal to arrive at the leader. Note that,  a signal sent throughout the first $2\mu_m$ time steps will land inside the $4\mu_m$ sized interval, if its delivery takes  at most $2\mu_m$ time to arrive. We can repeat the above approach, applying Markov and then Chernoff bounds to deduce that $\frac{|U|}{2} \cdot (1- o(1)) \cdot \frac{1}{2} \cdot (1- o(1)) \approx \frac{|U|}{4}$ signals will be received in this interval. As we are interested in an interval of length $L$, we apply this result  $\lfloor L / 4 \mu_m \rfloor$ times and deduce that at least $\frac{|U|}{4} \cdot \lfloor \frac{L}{4 \mu_m} \rfloor (1 - o(1))$ $0$-signals will be received by the leader throughout the interval $[t, t+L]$ \whp.\\
\textbf{Upper bound.}
We start by bounding the number of ticks inside a time interval of length $L$. Consider some node $v$ and assume for now that the previous tick finished just before the interval started.
Denote by $\{ X_1, ... ,X_i \}$ the next $i$ tick waiting times of $v$, where $X_j \sim \mathcal{T}_0$.  Then, if $X:=\sum_{j=1}^i X_j < L$ we can say that $v$ ticked at least $i$ times throughout the interval. Clearly $E[X] = \mu_0 \cdot i$ and together with the inequality in Theorem 3.5 of \cite{DBLP:journals/im/ChungL06} we deduce that
\begin{equation*}
	P(X < E[X] - (i\mu_0 - L)) = P(X < L) \leq \exp \left(- \frac{(i\mu_0-L)^2}{2i E[X_j^2]} \right).
\end{equation*}
where $E[X_j^2]$ is a constant  according to \cref{lem:survival-properties}. Note that this probability decreases exponentially fast for increasing $i$ as long as $\mu_0 \cdot i$ is sufficiently larger than $L$. Therefore, if we let $Y_v$ denote the number of ticks taken by $v$ throughout the time interval of $L$, we can deduce that roughly $P(\sum_{j=1}^i X_j < L ) = P(Y_v \geq i) < e^{-\Omega(i)}$ for large enough $i$. If $L = O(\log n)$ one can immediately see that $P(Y_v \geq C \cdot \log n)< 1 /n^2$ for large enough constant $C$ depending on $\mathcal{T}_0$ and $L$. For the expected value we crudely estimate

\begin{equation*}
	E[Y_v] = \sum_{i=0}^\infty P(Y_v \geq i) < \frac{2L}{\mu_0} + \sum_{i > 2L/ \mu_0}^\infty \exp \left(-\frac{i \mu_0^2}{8E[X_j^2]} \right) = \frac{2L}{ \mu_0} + C' ,
\end{equation*}
where the second sum corresponds to a geometric series and therefore $C'$ a constant again depending on $\mathcal{T}_0$. With the help of \cref{item:survival-sq-constant} of \cref{lem:survival-properties} one may crudely bound $C' \leq 600$.

Consider now the set $U$ and define $Y = \sum_{v \in U} \frac{Y_v}{ C \log n}$. Observe that all $Y_v$ are independent from each other and \whp it holds that $0 < \frac{Y_v}{C\log n}< 1$. This allows us -- considering only the probability space in which all  $Y_v$ are smaller than $C \log n$  -- to apply Chernoff bounds on $Y$. That is, $E[Y] = |U| \cdot  E[Y_v] \cdot \frac{1}{C \log n} = \omega(\log n)$ because of $|U| > \log^{2 + \varepsilon} n$ and this immediately yields $Y < E[Y] (1 + o(1))$ \whp. When undoing the $C \cdot \log n$ normalization we get that during an interval of length $L$  at most $(\frac{2L}{\mu_0} + C') |U| (1 + o(1))$ ticks occur in the interval. Remember that initially we assumed that at the start of the interval no ticks are in progress. To account for this, we add $1 \cdot |U|$ to the expression above.
Summarizing, we now know that during an interval off length $L = \Omega(1)$ at most
\begin{equation}
\label{eq:ticks-per-interval}
	R(L) := \left(\frac{2L}{\mu_0} + C' + 1\right) |U| (1 + o(1)).
\end{equation}
many ticks will occur \whp.

We are, however, interested in bounding the number of received signals. Consider again a time interval $[t, t + L]$ of length $L$. To upper bound the number of received signals in this interval, we assume the algorithm has already been running for $O(\log^2 n)$ many time steps, even though it might have just started.
Consider now the signals originating from ticks inside the interval $[t-2\mu_\ell,t]$. We crudely assume that every tick in this interval  corresponds to a received signal in the interval $[t,t+L]$. That is we have $R(2\mu_\ell)$ of them when using the result of (\ref{eq:ticks-per-interval}). Next take a look at the interval $[t - 4\mu_\ell, t-2\mu_\ell]$ and assume that our target interval is actually $[t,\infty)$. Let $S$ be the set of ticks occurring in the interval $[t - 4\mu_\ell, t-2\mu_\ell]$ and for each $s \in S$ consider the corresponding signal delay $X_s \sim \mathcal{T}_\ell$. A signal started from a tick in this set will arrive in $[t,\infty)$ with probability less than $P[X_s > 2\mu_\ell] < 1/2$ (follows from \cref{lem:survival-properties}). Using Chernoff bounds we get that at most
$R(2\mu_{\ell}) \cdot 1/2 \cdot (1 + o(1))$ such ticks will arrive at a time step in $[t, \infty)$. In general we can apply the results of \cref{lem:survival-properties} to derive that a signal originating from $[t - 2^{i+1} \mu_\ell, t - 2^i \mu_\ell]$ will hit $[t,\infty)$ with probability at most $(1/2)^{2^{i}}$. Therefore when applying Chernoff bounds we deduce that at most
\begin{equation}
\label{lem:eq-h-double-exponential}
R(2^{i} \mu_{\ell}) \cdot \left( \frac{1}{2} \right)^{2^i} (1 + o(1))=
 \left(2^{i+1} \cdot \frac{\mu_\ell}{\mu_0} + C' + 1\right) |U| (1 + o(1)) \cdot \left(\frac{1}{2} \right)^{2^i}
\end{equation}
many signals originate from a tick within such an interval \whp -- as long as  $i  < c \cdot  \log \log |U|$ for some constant $c$.
From any interval of further distance to $t$, i.e. $i > c \cdot \log \log |U|$, at most $O(\log n)$ signals will arrive \whp. As $O( \log^2 n)$ steps suffice for our algorithm to reach consensus, we only need to consider intervals with $i=O(\log \log n)$. Hence, in total, at most $O(\log n \cdot \log \log n) = |U| \cdot o(1)$ signals that originate from intervals with $i > c \cdot \log \log |U|$ will arrive \whp. Combining this with  (\ref{lem:eq-h-double-exponential}), which is dominated by a double-exponentially shrinking term, we get that at most $R(2\mu_\ell) + |U| \cdot o(1)$ many signals started in the interval $[t - O(\log^2 n), t - 2\mu_\ell]$ will arrive in $[t,\infty)$.
Finally, we count the received signals which originate from ticks inside $[t,t + L]$. We crudely assume that every tick inside this interval corresponds to a received signal, leading to further $(\frac{2L}{\mu_0} + C' + 1) |U| (1 + o(1))$ received signals. All the above is summarized in the following table
\begin{center}
\begin{tabular}{|c|c|} \hline
	Interval of origin & Number of signals received \whp\\ \hline
	$[t, t+ L]$ & $ \leq R(L)$ \\
	$[t-2 \mu_\ell, t]$ & $ \leq R(2\mu_\ell)$ \\
	$[t -  O(\log^2 n), t- 2\mu_\ell]$& $ \leq R(2 \mu_\ell) + |U| \cdot o(1)$ \\ \hline
\end{tabular}
\end{center}
When hiding some terms into $|U| \cdot o(1)$ this sums up to
\begin{equation*}
	\mathcal{H}(L) := (8 \frac{\mu_\ell}{\mu_0} + \frac{2L}{\mu_0} + 3C' + 3) |U| (1 + o(1)) \qedhere
\end{equation*}
\end{proof}

The above lemma directly implies the statement that was given in the introduction.

\signalcounting*

In \cref{alg:leader-increment} the base station counts until $\mathcal{H}(C_1)$. As $C_1$ bounds the time unit, above result implies that the base station will allow two-choices steps for at least one time unit per generation.

\begin{corollary}
\label{lem:tc-time}
Let  $t_i$ be the time the base station first allows generation $i$ and let $t'$ denote the
number of time units starting from $t_i$ until the condition in \cref{ln:gossip} of \cref{alg:leader-increment} is satisfied. Then, $t_{i} + 1 < t_i + t' < t_i + O(1)$ \whp.
\end{corollary}

\subsection{Total time to Increase a Generation}
\label{sec:time-to-inc-gen}
In this section we examine the time difference between the time points $t_i$ and $t_{i+1}$, i.e., starting from the time the base station allowed generation $i$ we are interested how long it takes until it allows generation $i+1$ for the first time.
Remember that we want to allow the fraction of nodes in generation $i$ to grow until at least $1/2$ before generation $i+1$ starts. Fixing some generation $i$, we denote by $t_i + t'$ the time at which the two-choices phase ended. At this point, as we will see in \cref{lem:1st-step-1}, at least $n\cdot  p_{i-1}/5$  nodes are of generation $i$ \whp. Throughout the remainder of generation $i$, the base station only allows propagation steps (see \cref{ln:prop-asy} of \cref{alg:The-basic-procedure-asy-1}) until it detects that at least $n/2$ of all nodes belong to generation $i$. Note that this process corresponds to simple pull broadcasting with the goal of spreading generation $i$. That is, after $\log(1 / p_{i-1}) = O(\log k)$ steps the desired amount of nodes of generation $i$ is  reached.

\tcgrowth*

\begin{proof} We want to show that the counting time of our base station
(i.e., $t'$ time units) suffices for the generation $i$ to
grow to contain at least a $\frac{p_{i-1}}{5}$-th fraction of nodes.
Throughout the time frame $[t_i,t_i + t']$ nodes join generation $i$ due to fulfilling the conditions in \cref{ln:2ch-asy} of \cref{alg:The-basic-procedure-asy-1} only.
Assuming a node of generation $i-1$ finishes an execution of \cref{alg:asynchronous} at time exactly $t_{i}$, it would sample two nodes of the same color and generation $i$ with probability exactly $p_{i-1} \cdot g_{i-1}^2 = 1/4 \cdot p_{i-1}$.  By \cref{lem:tc-parent-stable} we get that this indeed corresponds to the probability of  promoting to generation $i$ via \cref{ln:2ch-asy} of \cref{alg:asynchronous}. This is because \cref{item:tc-parent-stable-1, item:tc-parent-stable-2} of \cref{lem:tc-parent-stable} imply that the probability of $v$ promoting and taking some fixed color  $j$ is $c_{j,i}(t_i)^2 \cdot g_{i-1}^2$. Summing this probability over all colors $j$ yields the aforementioned result.

Now, \cref{lem:tc-time} states that, \whp, the time frame $[t_i ,t_i + t']$ is of length at least $1$ time unit. The definition of a time unit implies that each node will perform one full execution in $[t_i,t_i +t']$ with probability at least $0.9$. Combining this with the above allows us to minorize $g_{i}(t_i + t')$ by $\Bin(n, 0.9 \cdot (1/4) \cdot p_{i-1})$. As $p_{i-1} \geq (1/k)$, the result follows from a Chernoff bound application.
\end{proof}

In the time frame $[t_i + t', t_{i+1}]$, the base station only allows its followers to promote via propagation, which corresponds to pull gossiping w.r.t. generation $i$.

\propgrowth*
\begin{proof}
By construction, during the time frame $\left[t_i+t',t_{i}\left(\frac{1}{2}\right)\right]$,
 nodes will only join generation $i$ via propagation steps.
 To examine the growth of the set of nodes of generation $i$ during one time unit, we consider an arbitrary time frame $[t,t+1]$ with $t,(t+1) \in \left[t_i+t',t_{i}\left(\frac{1}{2}\right)\right]$.
We define $x=g_i(t)$ and $x'=g_i(t+1)$, where
by \cref{lem:1st-step-1} we have that $x\geq\frac{p_{i-1}}{5}$.
If during the time interval $\left[t,t+1\right]$, an
arbitrary node $v$
(i) arrived from generation at most $i-1$, (ii) sampled a node from generation $i$,
and (iii) executed a complete operation in the mentioned time-unit, then surely $v$ increased its generation to $i$.
In fact, it is enough to only consider such promotions which may be modeled directly as
\begin{align*}
x' & \geq x+\frac{1}{n} \Bin \left(n \left(1-x\right), 0.9\cdot x\right) \overset{\text{\whp}}{>} 1.4 \cdot x 
\end{align*}
where in the first step we crudely neglect the increase in probability for propagation steps to succeed by assuming that $x$ does not increase throughout the time interval $[t,t+1]$. To prove
that $g_i(t_i + t' +t'') \geq 0.5$, it is enough to iterate the
above process $t''$ times. Indeed,
\begin{align*}
g_i (t_i +t' + t'') & \geq1.4^{t''}\cdot\frac{p_{i-1}}{5}\geq\frac{1}{2}. \qedhere
\end{align*}
\end{proof}

Hence, \cref{prop:1leader-gossiping-one-step} gives us that, starting from $t_{i} + t'$, $O(\log(1/p_{i-1}))= O(\log k)$ time units of propagation suffice to reach $t_{i+1}$. Furthermore, by \cref{lem:tc-time} we know that the counting mechanism on the base stations  end ensures that the two-choices phase lasts for constant time only. This directly leads to the following statement.

\begin{corollary}
\label{cor:1leader-gossiping-one-generation}
The time between the start of two consecutive generations $t_{i+1} - t_{i}$ is less than  $O(\log(1 / p_{i-1}))$ time units \whp.
\end{corollary}

\subsection{Concentration Results}
\label{sec:conc-results-single-leader}

In this section, we examine how the bias behaves throughout some fixed generation $i$. That is, starting with $\alpha_{i-1}$ we will see that the bias evolves and almost squares until the start of the following generation. More precisely, as long as the second largest opinion is still of non-negligible size, we have that $\alpha_{i} > (\alpha_{i-1})^{1.5}$. Similar as in the previous section, we will split the concentration analysis into two parts and start with statements concerning actions in the time frame  $[t_i, t_i +t']$ -- the time at which the base station starts allowing propagation steps.

\paragraph{Concentration during the Two-Choices Phase}
We fix some generation $i$ in the time frame $[t_i,t_i + t']$ at which the base station has  $\mode{\ell} = \TC$ and only allows promotion to generation $i$ via two-choices steps.
Assume that a node samples two neighbors at time exactly $t_i$. Then, with probability $c_{j,i-1}^2 \cdot g_{i-1}^2$, it hits two nodes of generation $i-1$ and color $j$. In order to reflect the idea of a two-choices step as part of our algorithmic approach (see \cref{sec:approach}), we want this to be the probability that the node promotes to generation $i$ and take color $j$. However, this probability may deviate throughout the time-frame $[t_i,t_{i} + t']$, e.g., if some nodes leave generation $i-1$ by promoting to generation $i$.

To circumvent this problem, we carefully specified the two-choices step in \cref{ln:next-gen} of \cref{alg:asynchronous}.
During a $[t_i, t_i + t']$ a node $v$ of generation less than $i$, promotes to generation $i$ whenever it samples two nodes $v_1$ and $v_2$ s.t. $\col{v_1}{i-1} = \col{v_2}{i-1}$ and both these values are defined. However, this implies the desired property we stated above and is formalized as follows.

\begin{lemma}\label{lem:tc-parent-stable}
Consider some fixed generation $i$ throughout $[t_i ,t_i +t']$ and define
\[ S_{i-1,j}(t) = \{v ~|~ v \text{ has } (\col{v}{i-1} = j)  \text{ at time t}\}.
\]
Assume a node $v$ of generation $i-1$ finished establishing all required communication channels at $t \in [t_i, t_i +t']$. Then,
\begin{enumerate}
	\item \label{item:tc-parent-stable-1}$v$ will promote to generation $i$ and take color $j$ if and only if both sampled nodes $v_1$ and $v_2$ lie in $S_{i-1,j}(t)$.
	\item \label{item:tc-parent-stable-2}$\forall t \in [t_i ,t_i +t']: S_{i-1,j}(t) = S_{i-1,j}(t_i)$ and $|S_{i-1,j}(t_i)| / n = c_{j,i-1} \cdot g_{i-1}$.
	\item \label{item:tc-parent-stable-3}$S_{i-1,j} (t) \cap S_{i-1,j'}(t) = \emptyset$ for every pair of colors $j,j'$ with $j \neq j'$.
\end{enumerate}
\end{lemma}
\begin{proof}
	The first point follows directly from \cref{ln:next-gen} of \cref{alg:asynchronous} and the fact that $\mode{\ell} = \TC$ in $[t_i, t_i +t']$.

	Next, the second point. Fix again some color $j$. It is easy to see that nodes are not removed from $S_{i-1,j}(t)$ throughout  $[t_i, t_i +t']$ as nodes only take color values when promoting to higher generations and never overwrite old color values. This implies that $S_{i-1,j}(t_i) \subseteq S_{i-1,j}(t)$ for any $t \in [t_i , t_i  + t']$. As in $[t_i , t_i + t']$ only two-choices steps to generation $i$ are allowed, no node $v$ sets its $\col{v}{i-1}$ field during $[t_i , t_i + t']$. Therefore, $S_{i-1,j}(t) \subseteq S_{i-1,j}(t_i)$ which combined with the above implies that $S_{i-1,j}(t) = S_{i-1,j}(t_i)$. As $S_{i-1,j}(t_i)$ is the set of color $j$ nodes at time $t_i$, it immediately follows that $|S_{i-1,j}(t_i)| / n = c_{j,i-1}(t_i) \cdot g_{i-1}$.

	Regarding the final statement. When following \cref{alg:asynchronous}, nodes only change their color iff they increase their generation. That is, it is impossible for any node  $v$ to overwrite a color value stored in $\col{v}{i-1}$.
\end{proof}

This way, given the set $\mathcal{G}$ of nodes that promoted to generation $i$ via two-choices, we can model the number of nodes of generation $i$ and color $j$ at $t_i + t'$ with the help of a binomial distribution. More formally, we can show that \cref{lem:initial-bias-1} holds, which we restate for convenience.

\tcconc*

\begin{proof}
We start by giving a lower bound on $b_i(t_i + t')$, the number of $b-$colored nodes in generation $i$ at time $t_i + t'$.
 To that end, we define $\mathcal{G}$, the set of nodes of generation $i$ at the end of the two-choices phase with $|\mathcal{G}| = n \cdot g_i(t_i + t')$. During the time frame $[t_i ,t_i + t']$, every node is promoted to generation $i$ due to \cref{ln:2ch-asy} of \cref{alg:asynchronous} only. Consider one such node $v \in \mathcal{G}$. By  \cref{lem:tc-parent-stable} it follows that the execution of \cref{alg:asynchronous} that lead to $v$'s promotion to generation $i$ did so with probability exactly $\sum_j c_{j,i-1}^2 \cdot  g_{i-1}^2 = p_{i-1} \cdot g_{i-1}^2$.

Above observation leads to the following two-step process. First, we determine $\mathcal{G}$ and assume that the color $\col{v}{i}$ of nodes $v$ in $\mathcal{G}$  is still unknown. Second, we uncover the color of each node in $\mathcal{G}$ after another to derive the amount of them taking color $b$. It is important to note, \cref{item:tc-parent-stable-2} of \cref{lem:tc-parent-stable} guarantees that the order in which we uncover the nodes does not matter, i.e., the probability for the next revealed node taking color $b$ will always be $(b_{i-1})^2 / p_{i-1}$. Hence, we can model $b_{i}(t_i + t')$ with the help of a binomial distribution and apply Chernoff bounds as follows:
\[
	\frac{1}{|\mathcal{G}|} \cdot \Bin \left( |\mathcal{G}| , \frac{(b_{i-1})^2}{p_{i-1}}\right) \overset{\whp}{>} \frac{(b_{i-1})^2}{p_{i-1}} \left(1 - O\left( \frac{1}{(b_{i-1})^2} \cdot \sqrt{\frac{\log n}{n}} \right) \right).
\]
The high probability guarantee follows from the fact that, according to \cref{lem:1st-step-1}, $|\mathcal{G}| = \Omega(n \cdot p_{i-1})$ \whp. A repetition of above analysis also yields an upper bound on $b_{i}(t_i +t')$ as well as corresponding bounds on $a_{i-1}(t_i +t')$.
\end{proof}

Assuming that the currently second-most dominant color $b$ has sufficient support in generation $i-1$, i.e.,  $b_{i-1} \gg 1 / \sqrt{n}$, it follows from above result that $a_{i}(t_i + t') / b_{i}(t_i + t') \geq (\alpha_{i-1})^2 (1 - o(1))$.

\paragraph{Concentration during the Propagation Phase}
We consider some fixed generation $i$ and assume that at time $t_i +t'$ the base station starts allowing propagation steps. In the time frame $[t_i +t', t_{i+1}]$, nodes may join generation $i$ via \cref{ln:prop-asy} of \cref{alg:asynchronous} only. One can see this as generation $i$ being spread by pull broadcasting until the base station confirms that at least $n/2$ of all nodes belong to generation $i$ (see \cref{ln:one-leader-ln6} of \cref{alg:leader-increment}).
As discussed in \cref{sec:sl-overview}, the color fractions $c_{j,i}(t)$ for  $t \in [t_i +t', t_{i+1}]$ form a martingale when sequentialized by the points in time at which nodes join generation $i$. However standard techniques, e.g. Azuma-Hoeffding, fail to provide tight enough bounds.

Assuming we start at $t_i +t'$ we are interested in the \emph{absolute} amount of color $j$ nodes at $t_{i+1}$. We can model this value by the following urn process. The urn initially contains $c_{j,i}(t_i + t') \cdot n \cdot g_i(t_i +t')$ many black balls, i.e., as many black balls as there are nodes of generation $i$ and color $j$ at $t_i +t'$. Furthermore, we add a white ball for each remaining node in generation $i$ that is \emph{not} of color $j$ at $t_i + t'$. Now, each step of the process starts with drawing a random ball from the urn. Then, an additional ball is placed inside the selected urn corresponding to the color of the drawn ball. This experiment is then repeated until $n/2 - n\cdot g_i(t_i + t')$ balls have been added, leading to both urns combined containing $n/2$ balls in total. In our original process, each time a node joins generation $i$, a step of the process is triggered. Hence, answering the question of how many black balls throughout the process, gives us the number of nodes that join generation $i$ and take color $j$ until time $t_{i+1}$.

The urn process we just described is called Pólya-Eggenberger process (with $s=1$). The corresponding distribution exactly describes the number of added black balls as desired. A more detailed discussion, including some useful tail-bounds on this distribution can be found in \cref{sec:polya-eggenberger} and allows us to achieve the following result.

\propconc*

\begin{proof} We start by showing the bounds on $b_{i}$.
The \textit{absolute} number of nodes of generation $i$ and color $b$ in the time frame $[ t_i + t', t_{i+1} ] $ follows a Pólya-Eggenberger process. Let $\mathcal{G}$ with $|\mathcal{G}| := n \cdot g_i(t_i + t')$ denote the initial set of generation $i$ nodes at the end of the two-choices phase at $t_i+t'$ . Assuming $\mathcal{G}$ and $b_i(t_i + t')$ to be fixed, we consider the random variable $X$ with
\[
 	X \sim |\mathcal{G}| \cdot b_{i}(t_i + t') +  \text{PE}_1 \Big(~b_{i}(t_i + t') \cdot |\mathcal{G}| ~,~ (1 - b_{i}(t_i + t')) \cdot |\mathcal{G}|~,~ (n/2) - |\mathcal{G}| ~\Big),
\]
 modeling the value $n \cdot g_{i+1}(t_i) \cdot b_{i} = (n/2) \cdot b_{i}$. Here we used the notation $\text{PE}_1(.)$ as defined in \cref{sec:polya-eggenberger} to describe the Pólya-Eggenberger distribution introduced in the paragraph above this lemma.

Applying the result of \cref{pe-thm-main} together with $\delta= c_2^{-1/2} \cdot \sqrt{\log n}$ immediately yields, \whp, that
\begin{align*}
\label{eq:polya-colors}
	X &=  b_{i}(t_i + t') \cdot  (n/2) \pm c_2^{-1/2} \cdot \sqrt{b_{i}(t_i + t')} \cdot  \frac{(n/2)}{\sqrt{|\mathcal{G}|}} \sqrt{\log n}  \\
	&= b_{i}(t_i + t') \cdot (n/2) \left(1 \pm \sqrt{ \frac{\log n}{b_{i}(t_i + t') \cdot |\mathcal{G}| \cdot c_2}}\right) \\
    &= b_{i}(t_i + t') \cdot (n/2) \left( 1  \pm O\left(\frac{1}{b_{i-1}} \cdot \sqrt{\frac{\log n}{n}} \right)  \right).
\end{align*}
 The last line follows by \cref{lem:initial-bias-1} and \cref{lem:1st-step-1} which imply that, \whp,
\[
	b_{i}(t_i +t') = \Omega \left( \frac{(b_{i-1})^2}{p_{i-1}} \right) = \Omega \left( \frac{(b_{i-1})^2}{g_{i-1}} \right) = \Omega \left( \frac{n \cdot (b_{i-1})^2}{|\mathcal{G}|} \right).
\]
As the proof w.r.t. the concentration of $a_{i}(t_i + t')$ is similar, we omit a detailed proof.
\end{proof}
Hence, we established that the color fractions do not deviate much throughout the propagation phase of generation $i$. Moreover, the error terms are of the same order as those in \cref{lem:initial-bias-1}.

\paragraph{Combining Two-Choices and Propagation}
In \cref{lem:initial-bias-1} we established that $a_{i}(t_i + t') / b_{i}(t_i + t') = \alpha_{i-1,t_{i}-1}^2 (1 - o(1))$ as long as $b_{i-1}$ is still of significant size. Furthermore, by \cref{lem:polya-color-bound} we get that this fraction remains close to $\alpha_{i-1}^2$ throughout the propagation phase. That means, the bias between $a$ and $b$ roughly squares throughout the two-choices phase of generation $i$ and remains concentration until generation $i+1$ is allowed by the base station.

The following lemma formalizes above notion of `roughly squaring'. Additionally, we show that the initial additive bias of $\sqrt{n} \log n$ does not diminish over time. This implies that the initial majority color remains dominant in every generation  \whp.

\phaseconc*

\begin{proof}
Starting at time $t_i$, fix the values of $a_{i-1}$ and $b_{i-1}$, and assume they indeed follow the lemmas requirements. Combining the concentration results of both the two-choices and propagation phase -- stated in \cref{lem:polya-color-bound} and \cref{lem:initial-bias-1} respectively -- we immediately get that
\begin{equation}
\label{eq:squaring-error}
	\frac{1}{\alpha_{i,t_{i+1}}} = \frac{b_i}{a_i} < \left( \frac{b_{i-1}}{a_{i-1}} \right)^2 \left(1 + O\left( \frac{1}{b_{i-1}} \cdot \sqrt{\frac{\log n}{n}}\right) \right).
\end{equation}
Now, using $b_{i-1} \gg 1 / \sqrt{n}$, we can initiate the following inequality chain
\begin{align*}
\left(1 + O\left( \frac{1}{b_{i-1}} \cdot \sqrt{\frac{\log n}{n}}\right) \right)^2 &< \left( 1 + \frac{1}{b_{i-1}} \frac{\log n}{\sqrt{n}}\right) \\
&< \left( \frac{b_{i-1}}{b_{i-1}} + \frac{a_{i-1} - b_{i-1}}{b_{i-1}} \right) = \frac{a_i}{b_i},
\end{align*}
where we used in the second step that $a_{i-1} - b_{i-1} > \log n / \sqrt{n}$. Combining this result with (\ref{eq:squaring-error}) immediately yields that
\[
	\frac{b_i}{a_i} < \left( \frac{b_{i-1}}{a_{i-1}} \right)^{1.5}.
\]
Note that it is possible that $(\exists j \neq a,b:~c_{j,i} > b_i)$, i.e., color $b$ is overtaken. However, it is easy to see that for every $x\geq 0$ it holds that $P(c_{j,i} > x) \leq P(b_{i} > x)$ as smaller colors are less likely to be selected in both two-choices and propagation steps of our protocol. Hence, we apply union bounds over $k-1$ colors and deduce that $\alpha_{i} > ( a_{i-1} / b_{i-1})^{1.5}$.

 To show the third statement, we again make use the of concentration statements in  \cref{lem:polya-color-bound} and \cref{lem:initial-bias-1} to derive that \whp
\begin{align*}
	a_i - b_i &> \frac{a_{i-1}^2 - b_{i-1}^2}{p_{i-1}} - O\left(\frac{(a_{i-1}+b_{i-1}) \sqrt{\log n}}{p_{i-1} \sqrt{n}}\right)\\
	& = \frac{a_{i-1} + b_{i-1}}{p_{i-1}} \left( (a_{i-1} - b_{i-1}) - O\left(\frac{\sqrt{\log n}}{\sqrt{n}}\right) \right).
\end{align*}
Note that $p_{i-1} = \sum_{j} c_{j,i-1}^2 \leq \sum_{j} c_{j,i-1} \cdot a_{i-1} = a_{i-1}$ as $a$ is the majority color. In case $a_{i-1} - b_{i-1} > \log^2 n / \sqrt{n}$, the result follows immediately as $(a_{i-1}+ b_{i-1}) / p_{i-1} > 1$ and the difference between $a_{i-1}$ and $b_{i-1}$ dominates the error term.
In case $ \log n / \sqrt{n} \leq a_{i-1} - b_{i-1} \leq \log^2 n / \sqrt{n}$ it holds that $a_{i-1} = b_{i-1} (1 + o(1))$ because of $a_{i-1} \geq p_{i-1} \geq 1/k$. Hence, in this case it holds for $n$ large enough and \whp that
\[
	a_i - b_i > (2 - o(1)) \left(\frac{\log n}{\sqrt{n}} - O \left(\frac{\sqrt{\log n}}{\sqrt{n}} \right) \right) > \frac{\log n}{\sqrt{n}}.
\]
Just as before, we conclude with a union bound argument, yielding that also every other color that had less (or equal) support than $b$ at time $t_i$ adheres to this required absolute bias.
\end{proof}

Next, we consider how the bias evolves over multiple generations. The following is an immediate consequence of a repeated application of above lemma.

\begin{corollary}
\label{lem:async-num-generations}
	Consider an initial bias of $\alpha_{0} > 1 + \frac{1}{b_0} \cdot \frac{\log n}{\sqrt{n}}$. Then, \whp,
	\begin{enumerate}
		\item  after at most $\lceil \log_{1.5} \log_{\alpha} k \rceil $ generations the bias will exceed $k$, and
		\item after at most  $\lceil \log_{1.5} \log_{\alpha} n \rceil$ generations the bias is at least asymptotically similar ($\sim$) to $\sqrt{n}$.
	\end{enumerate}
\end{corollary}

As soon as the bias reaches value roughly $\sqrt{n}$, it follows by \cref{rem:alpha-b} that the second-largest color is no longer of significant size. That is, our previous concentration results, including the squaring in \cref{cor:one-phase-bias-1}, are no longer applicable. However, we can use the fact that at this point at least a $(1-o(1))$ fraction of nodes in the highest generation belong to the same color \whp. This way, we can deduce that after at most $2$ further generations, the first monochromatic generation will be created.

\begin{lemma}
\label{lem:async-alpha>k-steps}
	If in generation $i-1$ it holds that  $\alpha_{i-1} \sim \sqrt{n}$, then ,\whp., $\alpha_{i} \gg \sqrt{n}$. Likewise, if in generation $i-1$ it holds that $\alpha_{i-1} \gg \sqrt{n}$, then generation $i+1$ will be monochromatic.
\end{lemma}
\begin{proof}
	First assume that $\alpha_{i-1} \sim \sqrt{n}$ and let $a$ and $b$ be the largest opinions in generation $i-1$ at time $t_i$. Given the configuration at time $t_i$, consider result of the two-choices phase of generation $i$ which takes place in the time frame $[t_i, t_i + t']$.
Similar as in the proof of \cref{lem:initial-bias-1}, we denote by $\mathcal{G}$ the set of nodes that join generation $i$ by two-choices steps with $|\mathcal{G}|:= n \cdot g_{t_i + t'}(i)$. Just as in the proof of \cref{lem:initial-bias-1} we apply \cref{lem:tc-parent-stable} and deduce that the probability that one of these nodes sets its color to $b$ is exactly $b_{i-1}^2 / p_{i-1}$. This way, we model $|\mathcal{G}| \cdot b_{i}(t_i + t')$ as $\Bin \left( |\mathcal{G}| ~,~ b_{i-1}^2 / p_{i-1} \right)$ 
with expected value $\mu \sim 1$. This expected value is implied by $b_{i-1} \sim 1/\sqrt{n}$ and $p_{i-1} = \Omega(1)$, which follows from $\alpha_{i-1} \sim \sqrt{n}$. Hence, a Chernoff bound application yields that $|\mathcal{G}| \cdot b_i(t_i + t') < n^{\varepsilon}$ \whp for any arbitrary small $\varepsilon > 0$.

Now, let $\mathcal{N} = n/2$ denote the number of nodes of generation $i$ just before the start of generation $i+1$ at $t_{i+1}$. Then, we may model $\mathcal{N} \cdot b_i$ as $|\mathcal{G}| \cdot b_i(t_i + t') + \text{PE}_1 ( |\mathcal{G}| \cdot b_i(t_i + t') ~,~ |\mathcal{G}| -  |\mathcal{G}| \cdot b_i(t_i +t') ~,~ \mathcal{N} - |\mathcal{G}|)$. According to \cref{pe-thm-small-a} we can bound a r.v.\ that follows such a distribution by
\[
	\mathcal{N} \cdot b_i < \max\{1 ~,~ \frac{n}{|\mathcal{G}|}\} \cdot \max \{  3 |\mathcal{G}| \cdot b_i(t_i + t') ~,~ O(\log n)\}
\]
\whp. As $p_{i-1} = \Omega(1)$ it follows by \cref{lem:1st-step-1} that $|\mathcal{G}| = \Omega(n)$ \whp. Therefore, \whp,  $\mathcal{N} \cdot b_i < 3n^{\varepsilon}$. Setting $\varepsilon$ to some constant value less than $1/2$, this implies that $b_i \ll 1/ \sqrt{n}$. We now apply a repetition of this whole argument to every other color $j \neq a,b$. This way, a union bound application yields that $c_{j,i} \ll 1/ \sqrt{n}$ for every color besides $a$, which in turn implies $\alpha_{i} \gg 1 / \sqrt{n}$.

To show the second statement of the lemma we assume that $\alpha_{i-1,t_i} \gg \sqrt{n}$ and note that the proof for this case is similar to the previous one. Following the previous approach it is easy to see that $|\mathcal{G}| \cdot b_i(t_i + t') = O(\log n)$ , \whp, as $E[|\mathcal{G}| \cdot b_i(t_i + t')] \ll 1$. Applying the same Pólya-Eggenberger result as before, we now derive that $\mathcal{N} \cdot b_i = O(\log n)$ \whp. That is, color $b$ only has support of $O(\log n)$ in generation $i$ at the start of generation $i+1$. The probability for color $b$ to survive the following two-choices phase, i.e., $b_{i+1}(t_{i+1} + t') \neq 0$,  is now at most $1 - (1 - O(\frac{\log^2 n}{n^2}))^{n} < \polylog / n$. A final union application yields that \emph{no} color besides $a$ will be present in generation $i+1$.
\end{proof}

When combining all the statements we derived during \cref{sec:Asynchronous-1leader}, the proof of \cref{thm:async-centralized} follows. Most notably \cref{lem:async-num-generations} together with \cref{lem:async-alpha>k-steps} state the number of required generations to reach the first monochromatic one. Additionally \cref{cor:1leader-gossiping-one-generation} indicates that the time between the birth of two consecutive generations is constant as soon as the bias reaches value $k$. The following result finalizes the proof.

\paragraph{Moving on from the monochromatic generation}

From \cref{lem:async-num-generations} and \cref{lem:async-alpha>k-steps} we get that a monochromatic generation emerges among the first $O(\log \log_{\alpha} n)$ generations. At the end of this generation, at least $1/2$ of all nodes will be of the same color. We now show the following

\begin{lemma}
\label{lem:spreading-monochromatic}
	Let $i^*$ denote the first monochromatic generation.
	Then, at time $t_{i^*} + O(\log \log n)$, partial consensus will be reached. After further $O(\log n)$ steps, every node shares the same opinion.
\end{lemma}
\begin{proof}
	Let $a$ denote the dominating color of generation $i^*$. Clearly, if $i$ is monochromatic then so will be every generation $i > i^*$. Also, every node of generation at least $i^*$ must be of color $a$. Fix, now such a generation $i > i^*$ and some node $v$ of generation less than $i^*$. If it finishes an execution of \cref{alg:asynchronous} during the two-choices phase of generation $i$, it will with probability at least $1^2 \cdot g_{i-1}^2 = \Omega(1)$ promote to generation $i$. This follows from \cref{lem:tc-parent-stable} and $a_{i-1} = 1$. On the other hand, if it finishes an execution during the propagation phase in $[t_i + t', t_{i+1}]$, it will with probability at least $g_{i}(t_i + t') = \Omega(1)$ sample a node of generation $i$ and promote to generation $i$ via propagation. Hence, each time $v$ finishes \cref{prop:1leader-gossiping-one-step} it will promote to generation $i$ -- and thereby also take color $a$ -- with at least constant probability. According to the definition of a time unit, $v$ will perform such an execution with probability $0.9$ in each time unit. Hence, $v$ will be of color $a$ after $O(\log \log n)$ time with probability $1/\polylog n$ -- and after $O(\log n)$ time \whp.
\end{proof}

We are now ready to finalize the proof of \cref{thm:single-leader}. According to  \cref{lem:async-num-generations} the bias reaches $k$ after $O(\log \log_{\alpha} k)$ generations. Now, by \cref{prop:1leader-gossiping-one-step} we have that the time between two generations can always be bounded above by $O(\log k)$ \whp. The remaining $O(\log \log_{k} n)$ generations that are required for the bias to hit $n$ (see again \cref{lem:async-num-generations}), each take constant time only (because $\alpha_{i-1} > k$ implies that $p_{i-1} = \Omega(1)$). This time is dominated by the $O(\log \log n)$ time requirement of \cref{lem:spreading-monochromatic}. In total we therefore reach partial consensus after $O(\log \log_{\alpha} k \cdot \log k + \log \log n)$ time units. By \cref{lem:spreading-monochromatic} we have that $O(\log n)$ time later, full consensus is reached.

%% file: 94-analysis-decentralized.tex
\section{Analysis of the Decentralized Algorithm} \label{sec:analysis-decentralized}

\subsection{A Simple Clustering Algorithm} \label{sec:clustering}
In the following we will describe a simple clustering algorithm, which satisfies the desired property of clustering all but $O(1/ \polylog n)$ nodes into clusters of polylogarithmic size. Later in \cref{sec:extended-clustering}, we extend this algorithm and describe how nodes may transition into the consensus protocol after the leader election has been completed.

The simple clustering works as follows. At the beginning, each node flips a coin and with 
probability $1/\log^c n$, the node becomes a leader, where $c$ is a sufficiently 
large constant. The other nodes are followers. Whenever the clock of a node ticks, this node 
establishes communication channels to its 
own leader (if any), and to 
three other nodes chosen uniformly at random\footnote{It would be enough to just contact one randomly selected node. However, in order to select the same number of nodes as in the consensus algorithm, we allow here the selection of three randomly chosen neighbors as well.}. These neighbors send the address of their leaders to 
the node they were contacted by, and then one of these leaders is called by
that node. If a follower (not assigned to a cluster so far) contacts a leader,
then it joins 
the cluster of that leader as long as the cluster 
has size less than $\log^{c-1} n$. The leader nodes keep track of the size of their clusters, 
and if a follower joins 
the cluster of some leader, then this leader notifies the follower that the request to join 
was successful (recall that establishing a communication channel requires time, but 
the exchange of messages is instant). The nodes in a cluster keep sending $0$-signals to 
their leader at each tick of their individual clocks, which enables 
the leader to count the time (similar as in the centralized procedure).
Once the size $\log^{c-1} n$ is reached, the leader starts counting $0$-signals, and 
rejects any further request until its counter reaches value 
$\mathcal{H}(c^2 \log \log n \cdot C_1) \cdot \log^{c-1} n$. Remember, according to \cref{lem:counting-time}, this counting ensures that at least $c^2 \log \log n$ time units pass \whp (note that the constant $c$ needs to be chosen s.t. $c-1 > 3$).
Throughout this phase we say such a leader is in the \emph{waiting} state.
As soon as the $\mathcal{H}(c^2 \log \log n \cdot C_1) \cdot \log^{c-1} n$'th signal is received, the leader starts indefinitely accepting further followers to its cluster. After further $\mathcal{S}(c^2 \log \log n \cdot C_1) / C_1 = O(\log \log n)$ time units most leaders have stopped waiting, and $O(\log \log n)$ time units later, all but a $1/\polylog n$ fraction of nodes belong to clusters. 
In the following we let $L$ denote the set of cluster leaders. It is easy to see, that the initial coin flip guarantees $|L| = (n/\log^{c}) (1 \pm o(1))$ \whp.

\begin{lemma}
\label{lem:simple-clustering}
	Let $t^{(w)}_f$ denote the time when the first leader stopped waiting. Let $B$ be the set of leaders with clusters of size less than $\log^{c-1} n$ at time $t^{(w)}_f$. Then, $|B| < |L| / \log^{C'} n$ and at time $t^{(w)}_f + \mathcal{S}(c^2 \log \log n \cdot C_1)/C_1 = O(\log \log n)$ all clusters in $L \setminus B$ stopped sleeping \whp. Here $C'>0$ is a constant depending on $c$.
\end{lemma} 
\begin{proof}
	As described in the algorithm, each node starts by flipping a coin and becomes a leader with some probability 
$1/\log^c n$. Using simple Chernoff bounds, it follows that there will be 
$n(1\pm o(1))/\log^c n$ leaders \whp. We assume in this proof that all nodes flip their coins at the beginning, and 
flipping a coin is not related to the ticks of the clocks; however, this could also be relaxed by 
assuming that the nodes flip their coins at their first tick, and the result of the theorem would not change.
Let the set of leaders be denoted by $L$.

First, we show that within $c \log \log n$ time there will be at least $|L| (1-1/\log^{2C'} n)$ 
leaders having at least $c' \log \log n$ members in its cluster \whp, where $C'$ and $c'$ are constants 
depending on $c$. As in the centralized case, we call a time unit the period of time $C_1$
in which a node performs a complete execution of one clustering step with probability $9/10$. That is, in this case a time 
unit is the time needed for a node to perform a good tick and to establish connections to a leader and two 
randomly chosen nodes with probability $9/10$. We know that 
a time unit has constant length. We divide now the time frame of length 
$c \log \log n$ into a sequence of non-overlapping time units.
Having in mind that for a time frame of length at least $c (1-o(1)) \log \log n$ no leader will have more than $\log^{c-1} n$ members in its cluster, there will be \whp $9n/10 \cdot (1-o(1))$ nodes 
communicating with another node in a time unit of the sequence of time units defined above. Thus, a leader
is contacted with probability at least
\[ 1-\left(1-\frac{1}{n} \right)^{9n/10 \cdot (1-o(1))} = 1-e^{-9(1-o(1))/10} . \]
Using Chernoff bounds, we obtain that in $\Theta(\log \log n)$ time units, all but 
$|L| (1/\log^{2C'} n)$ leaders have been contacted by at least $c' \log \log n$ other nodes \whp, where 
the constant hidden in $\Theta(\log \log n)$ governs $C'$ and $c'$. Thus, choosing $c$ accordingly we 
obtain our claim.  

We consider now the next $(c^2 -c) \log \log n$ time steps and, again, we divide the time 
into a sequence of time units. As long as no cluster has larger size than $\log^{c-1} n$, 
in each time unit $9n/10 \cdot (1-o(1))$ nodes try to join a cluster. Note that the counting of $0$-signals during the waiting phase guarantees that no leader exceeds size $\log^{c-1} n$ before time $c^2 \log \log n$ . Let $L_v$ be the cluster 
of a leader $v$, and assume that $|L_v| \geq c' \log \log n$ at the beginning of the sequence
of time units defined above. We call a time unit successful, if the size of the cluster grows 
by a factor of $3/2$ in this time unit or the cluster has size $\log^{c-1} n$ at the end of the time 
unit. As before, we know that within a time unit, a node of the cluster is contacted with 
probability at least 
\[ 1-\left(1-\frac{1}{n} \right)^{9n/10 \cdot (1-o(1))} = 1-e^{-9(1-o(1))/10} . \]
Using simple Chernoff bounds, we obtain that a time unit is successful with probability at least 
$1-1/\log^{2C'} n$, where $C'$ depends on the size of that cluster at the beginning of the time unit, i.e., 
in the first time unit of the sequence, $C'$ depends on $c'$. Hence, if there are enough time units 
in the sequence of length $\Theta(\log \log n)$, then there will be $(c-1) \log \log n$  
successful time units for $L_v$, with probability at least $1-1/\log^{2C'-1} n$. Thus, the expected 
number of clusters, for which the number of successful time units is less than $(c-1) \log \log n$, 
is less than $|L|/\log^{2C'-2} n$. Note that these events are not independent between clusters.
However, applying the method of bounded differences, we obtain that at most $|L|/\log^{C'} n$
clusters have size less than $\log^{c-1} n$ at the end of this sequence of time units, \whp, 
provided the constant $c$ is large enough. We denote these cluster leaders by the set $B$.

In the following $\mathcal{S}(c^2 \log \log n \cdot C_1) /C_1 $ time units all cluster leaders in $L \setminus B$ will stop waiting (see \cref{lem:counting-time}). The lemmas results follow.
\end{proof}

At time $t^{(w)}_f + O(\log \log n)$ at least $n/\polylog n$ nodes lie in clusters that passed the waiting phase and accept further followers. In the following $O(\log \log n)$ time units, the set of unclustered nodes follows the behavior of uninformed nodes in pull-broadcasting \cite{DBLP:conf/focs/KarpSSV00}. Therefore, after further $O(\log \log n)$ time at least $n(1 - 1/\log n)$ nodes lie in clusters.

\begin{corollary}
\label{sec:clustering-simple-time}
	In the $O(\log \log n)$ time units following $t^{(w)}_f + \mathcal{S}(c^2 \log \log n \cdot C_1) / C_1$, all but an $O(1/ \log n)$ fraction of nodes lies in some cluster of size at least $\log^{c-1} n$ \whp.  This corresponds to a total time requirement of $O(\log \log n)$.
\end{corollary}

\subsection{Global Sampling Gadget} 
\label{sec:sampling}
Consider some time unit $t$, node $v \in V$ and property $R: V \rightarrow \{\text{true},\text{false}\}$. We say $R(v)$ is true, or holds, in case $R$ is satisfied by $v \in V$.
Now, for the set $R_t := \{v \in V~|~R(v) \text{ holds at time } t\}$ we can define $r_t := |R_t| / n$, which denotes the \emph{ratio} of nodes satisfying property $R$. Imagine that some leader wants an estimation on this global ratio $r_t$. Assume that  every follower $v$ of the leader executes a routine upon a tick (e.g. something similar to \cref{alg:asynchronous} or the clustering routine).
Furthermore assume that throughout this routine, a node $v$ waits until communication channels to at least one randomly chosen node $w$ and $v$'s own leader $l$ are established.
Just before the routine would terminate, leading to $v$ closing the established communication channels, we employ an extension as follows.
The node $v$ collects the state information from $w$ and evaluates $R(w)$. Finally, $v$ informs its leader $l$ whether $R(w)$ holds or not. Both these operations can be performed via the already established communication channels. This way we may interweave the nodes usual execution (for example the nodes routine throughout the leader election) with a sampling gadget without requiring additional time spent.
Note that, when also opening a channel to the leader $l_w$ of $w$, the node $v$ may even evaluate properties of the form $R(w,l_w)$. We will make use of this special case in \cref{sec:multileader_proc}. This still can be seen as a property $R(w)$, as the leader $l_w$ belongs to the state of $w$ and communication via established channels is instant. Finally, observe that $R(w)$ can alternatively also be evaluated one the end of $l$, in case $v$ transfers all the necessary state information to its leader $l$.

On the leaders side two additional counters of $O(\log \log n)$ bits are employed, denoted by $r_1$ and $r_2$ -- both initially set to $0$. Each time the leader is informed by one of his follower w.r.t.\ one such evaluation of the property $R$, it increments $r_1$ by one and, in case $R(w)$ holds, also increments $r_2$. After $r_1$ reaches value $0.8 \cdot \log^{2+ \varepsilon} n$ for some small constant $\varepsilon$, the value $r' = r_2 / r_1$ is evaluated. Hence, $r'$ can be seen as an approximation of the ration $r_t$. The leader can then react depending on $r'$ and/or restart the sampling by setting $r_2=r_1=0$.

In the following we say that some leaders sampling \emph{started} at time $t'$, if at time $t'$ the counters $r_1$ and $r_2$ were set to $0$. Similarly we say that the sampling \emph{ended} at time $t''$, if at this time the counter $r_1$ reached value $0.8 \cdot \log^{2+\varepsilon} n$. If the leaders cluster has size at least $\log^{2+\varepsilon} n$, this estimation $r'$ will be accurate, and completed in at most \emph{one} time unit. More precisely the following holds.

\begin{theorem}
\label{ml-sampling-theorem}
	Consider some fixed leader $l$ with at least $\log^{2+\varepsilon} n$ followers. Assume the leader starts a sampling at $t'$, which ends at time  $t'' > t'$ and results in the ratio $r'$. Let $r_t$  denote the global ratio of nodes satisfying the sampled property $R$ at time $t$, and assume $r_t \in [a,b]$ for $t' \leq t \leq t''$.
Then, with probability $1-n^{-\omega(1)}$ it holds that
	\begin{enumerate}
		\item if $a = \Omega(1/\log n)$, then $r' > a (1 + o(1))$
		\item if $b = \Omega(1/ \log n)$, then  $r' < b (1 + o(1))$
		\item $t'' - t'  < 1$
	\end{enumerate}
\end{theorem}
\begin{proof}
	Let $S$ be a sampling performed by leader $l$, as assumed in the theorems statement. Let $\{s_i ~|~ 1 \leq i \leq 0.8 \cdot \log^{2+\varepsilon} n \}$ be the set of all samples    the leader receives in the time frame $[t',t'']$. Fix some such sample $s_i$ sent by node $v$. It contains the information whether $R$ holds w.r.t.\ some node $w$, sampled u.a.r at some time $t$.  As $v$ had already opened channels to $w$ and its leader $l$ at the time point of  sending $s_i$, the evaluation of $R(w)$ takes place at the same time as $l$ receives $s_i$. Therefore it must hold for $t$ that $t' \leq t \leq t''$ and therefore $P(R(w) \text{ is true}) \in [a,b]$.  The number $X$ of received messages, which contain a property that was evaluated to true, can therefore be majorized by $\Bin(0.8 \log^{2+\varepsilon} n , b)$ and minorized by $\Bin(0.8 \log^{2 + \varepsilon} n, a)$. Applying Chernoff bounds immediately yields the first two statements. Similar, the third claim follows from Chernoff bounds, as Postive Aging (\cref{prop:positive-aging}) guarantees that each node prepares with probability greater 0.9 at least one sample throughout one time unit.
\end{proof}

\subsection{Extended Clustering Algorithm}
\label{sec:extended-clustering}
In the following we describe the clustering algorithm that allows nodes and leaders to properly transition into the consensus algorithm (see \cref{sec:multileader_proc}). It consists mainly of the \emph{simple clustering algorithm}, described in \cref{sec:clustering} extended by a \emph{global sampling gadget} (see \cref{sec:sampling}) as follows.
We consider the property $R(w) \Leftrightarrow (w$ is \emph{not} assigned to a cluster) and assume the above described \emph{global sampling gadget} is employed by followers as soon as they have a leader, and on the leaders end as soon as their clusters reach size $\log^{c-2} n$ (note that $c$ is the clustering constant from \cref{sec:clustering} -- it needs to be set such that $c > 4$). The leader repeats the sampling process, until it witnesses that $r' < 0.9 /\log n$, ensuring \whp that less than a $1 / \log n$ fraction of nodes remains un-clustered.
In this case, the leader sets up a counter, initiated by $0$, and counts incoming follower $0$-signals sent by the first $\log^{c-2} n$ nodes that joined the cluster. Such a leader keeps following the simple clustering protocol as usual, but until its counter reaches $\mathcal{H}(C_1)\log^{c-2} n$ we say that this leader \emph{prepares} for consensus mode.
As soon as the counter reaches value $\mathcal{H}(C_1) \log^{c-2} n$, the leader \emph{decides} whether to switches to \emph{consensus mode} by checking the size of its cluster. If it's size is at least $\log^{c-1} n$, then it participates in the consensus protocol (see \cref{sec:multileader_proc}) and signals its followers do to so as well.
If the cluster's size is less than $\log^{c-1} n$, the leader rejects any requests related to the consensus protocol. In any case, the leaders no longer allow nodes to join its clusters anymore.
This extended leader election algorithm, guarantees the following.

\begin{theorem}
\label{thm:ml-extended-clustering}
	Let $c > 4$ be an arbitary constant. When following the Extended Clustering Algorithm, all but $O(1/\log n)$ many nodes each belong to one of the at least $n/\log^{c} \cdot (1 - o(1))$ clusters of size at least $\log^{c-1} n$ that switch to consensus mode after $O(\log \log n)$ time units \whp. Furthermore, the cluster leaders of such nodes will enter consensus mode with a time difference of at most $C_{\ell}= \mathcal{S}(C_1) + 2 \cdot C_1$ time steps, and the remaining leaders will not participate in the consensus protocol.
\end{theorem}
\begin{proof}
	We consider the property $R$ to be defined as just above the lemma and utilize the notation of \cref{ml-sampling-theorem}. Clearly $r_t$, the fraction of un-clustered nodes,  decreases monotonically for increasing time $t$. Let $t_f$ be the first time that $r_{t_f} \leq \frac{1}{\log n}$. Consider a sampling with starting and ending times $t'$ and $t''$. Then, if $t' \leq t'' \leq t_f$ it follows that $r_t \in [1 / \log n ~,~ 1]$. Together with \cref{ml-sampling-theorem} and a union bound application, this implies \whp that \emph{no} leader will perform a sampling s.t. $r' < (1- o(1)) / \log n$. Hence, no leaders starts to prepare for consensus mode before time $t_f$.
	
	A similar argument can be mode to show that, every sampling started after $t_l$, with $t_l$ being the first time such that $r_{t_l} \leq \frac{0.8}{\log n}$, will succeed. We will now argue that $t_f  - t_l \leq 1$.  Let $L$ with $|L| = \frac{n}{\log^c  n} (1 \pm o(1))$ be the set of all leaders that were initialized after the coin flip of the simple consensus protocol. By a simple counting argument, it follows that at most  $|L| \cdot \log^{c-1} n = O(n/\log n)$ nodes belong to waiting clusters (see \cref{sec:clustering} for the description of the waiting phase) at any point in time. Hence, in the time unit following $t_f$, an unclustered node will remain unclustered with probability at most $O(1/\log n)$. It follows that $t_f - t_l \leq 1$ (note that the counting of $0$-signals prevents leaders from exiting the clustering algorithm before $t_l$ is reached \whp). Any leader that is of size $\log^{c-2} n$ at $t_f$ therefore starts preparing for consensus mode before time $t_l + 3$ \whp. Summarizing, we have:
	
\begin{enumerate}
	\item The first leader enters the preparation phase after $t_f$.
	\item Every leader that is of size $\log^{c-2} n$ at $t_f$ starts to prepare for consensus mode before $t_l \leq t_f + 3$.
\end{enumerate}

We now partition the leaders into $3$ sets depending on their size at $t_f$. $S_1$ is the set of leaders of size larger or equal $\log^{c-1} n$, $S_2$ contains the leaders of size smaller $\log^{c-1} n$ but larger (or equal) $\log^{c-2} n$, and $S_3$ contains the remaining leaders of size less than $\log^{c-2} n$. We will now show that the following holds \whp

\begin{enumerate}
	\item All leaders of $S_1$ enter the consensus mode at most $C_\ell$ time steps after the first leader.
	\item Only some leaders of $S_2$ enter consensus mode. However, all of them decide whether or not to enter consensus mode at most $C_\ell$ time steps after the first leader.
	\item No leader in $S_3$  enters the consensus mode.
\end{enumerate}

We prove the first and second point at the same time. Consider some leader  $l$ in $S_1 \cup S_2$. As established above, such a leader will start to prepare for consensus mode before $t_f + 3$. It then decides whether or not to enter the consensus mode after reaching $\mathcal{H}(C_1) \cdot \log^{c-2} n$ many $0$-signals. If it's cluster is of size at least $\log^{c-1} n$ (which is true for all $l \in S_1$), it will decide to enter the consensus mode. Otherwise it will remain inactive. The leader $l$ finishes this counting of $0$-signals before $t_l + \mathcal{S}(C_1)$ (see \cref{lem:counting-time}) \whp. Following a similar argument, the first leader will \emph{not} enter consensus mode before $t_f +1$ due to the required counting of $0$-signals. Hence, if $l$ enters the consensus mode it does so at most $t_l + \mathcal{S}(C_1) - (t_f + 1)$ time after the first leader. This corresponds to the time difference we denoted by $C_\ell$ in the theorems statement.

Now, consider the last point. We know that a leader $l \in S_3$ is not of size $\log^{c-2} n$ at time $t_f$. Even if it's cluster eventually reaches size $\log^{c-2}$ at some time $\hat{t} > t_f$, 
then it will prepare for consensus before time $\max \{\hat{t} + 2, t_l + 2\} = \hat{t} + O(1)$ \whp, and further $O(1)$ time later decide whether to enter consensus mode or not.
Hence, $l$ only enters consensus mode iff it grows from $\log^{c-2} n$ to $\log^{c-1} n$ in constant time. It is easy too see that this does not happen \whp.

We conclude that the leaders that enter consensus mode do so with a time difference of $C_\ell$ time steps. Also, before the first leader stops waiting at $t_f^{(w)}$ (waiting phase as described in \cref{sec:clustering}), at most $|L| \cdot \log^{c-1} n = O(n/\log n)$ nodes lie in clusters. Therefore, it needs to hold that $t_f > t_f^{(w)}$ as, \whp, no leader performs a successful sampling if only $O(n/\log n)$ nodes lie in clusters. By \cref{lem:simple-clustering} we have that, already at $t_f^{(w)}$, most clusters are of size at least $\log^{c-1} n$. In other words, $|S_1| > n/\log^{c} n \cdot (1 - o(1))$ and all of these leaders enter the consensus mode.
Furthermore, observe that when it comes to the number of unclustered nodes, the extended and simple clustering algorithms behave identically until the first leader entered the preparation phase. As established above, the first leader starts preparing for consensus mode before $t_l$ \whp. Time $t_l$ is reached when the fraction of unclustered nodes hits $0.8 / \log n$. This amount of unclustered nodes can easily be reached by our simple clustering algorithm in $O(\log \log n)$ time. Hence, also the extended clustering algorithm comes with a time requirement of $O(\log \log n)$.
\end{proof}

We finish our discussion of the clustering algorithms with a statement that implies that the congestion of any leader indeed lies in $O(\polylog n)$ \whp.

\begin{lemma}
\label{lem:cluster-size-upper}
	The load is well balanced between all leaders that switch to consensus mode.  That is, none of the clusters created by the Extended Clustering Algorithm will exceed size of $\polylog n$ \whp.
\end{lemma}

\begin{proof}
	We know that the clustering takes time at most $O(\log \log n)$. Using the inequality in Theorem 3.5 of \cite{DBLP:journals/im/ChungL06}, similar as in the proof of \cref{lem:counting-time} we deduce that some fixed node $v$ will tick more than $O(\log \log n)$ times throughout the clustering with probability at most $1/ \log n$.
	The same inequality together with a union bound application shows that \emph{no} node will tick more than $O(\log n)$ times \whp. Since the nodes tick independent from each other, a Chernoff bound application yields that at least an $(1- 1/\log n)$ fraction of nodes tick $O(\log \log n)$ times. The remaining $1/\log n$ fraction of nodes ticks $O(\log n)$ times at most. Therefore in total $O(n \log \log n)$ ticks will occur \whp.
	
	For simplicity assume that a node only contacts a single other random partner per execution during the clustering algorithm.	
Now fix some cluster of size $\log^{c-1} n$ that started accepting followers again and consider the following alternate process: Our system consists of $O(n \log \log n)$ nodes, each sampling one random node upon each tick and if this sample belongs to the cluster, they join the cluster without any additional delay. Observe that in this alternate process the size of the cluster will always be larger than in the original one. In the original process, each node can only join a cluster once, and communication delays need to be accounted for.
We analyze the modified process as follows. Assuming that the cluster has not reached size $2 \log^{c-1} n$ a node will join it part of its next execution with probability $p$ less than $2 \log^{c-1} n / n$. Applying a Chernoff bound with $p$, we deduce that $\frac{n}{2} (1 - o(1))$ many attempts of joining a cluster will not suffice to bring the cluster cardinality to $2 \log^{c-1} n$. We deduce that more than $\frac{n}{2} (1 - o(1))$ attempts are necessary to double the clusters size. 

We repeat this approach for $x= \frac{O(n \log \log n)}{(n/2) (1 - o(1))}$ steps and deduce that no cluster will be of size larger than $2^{x} \log^{c-1}n$ at this point. Clearly this number is some value polylogarithmic in $n$. As the cluster size in this modified process serves as an upper bound, we conclude the proof after applying union bounds over all $n/ \polylog n$ many clusters.
\end{proof}

\subsection{Extended Description of the Decentralized Protocol}
\label{sec:ml-implementing}

\begin{figure}
\centering
\includegraphics[scale=0.8]{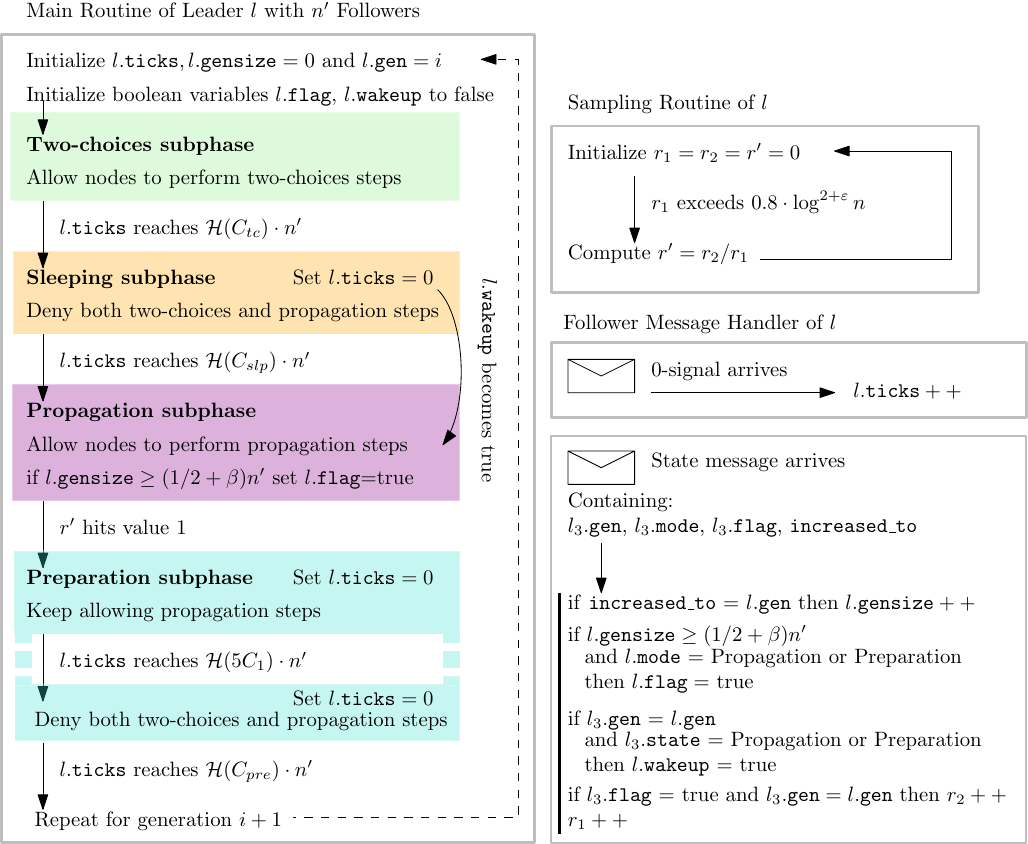}
\caption{Routines executed by leader nodes throughout the consensus mode as well as the handlers for incoming messages of the leaders followers. }
\label{fig:async-leader}
\end{figure}
\begin{figure}
\centering
\includegraphics[scale=0.69]{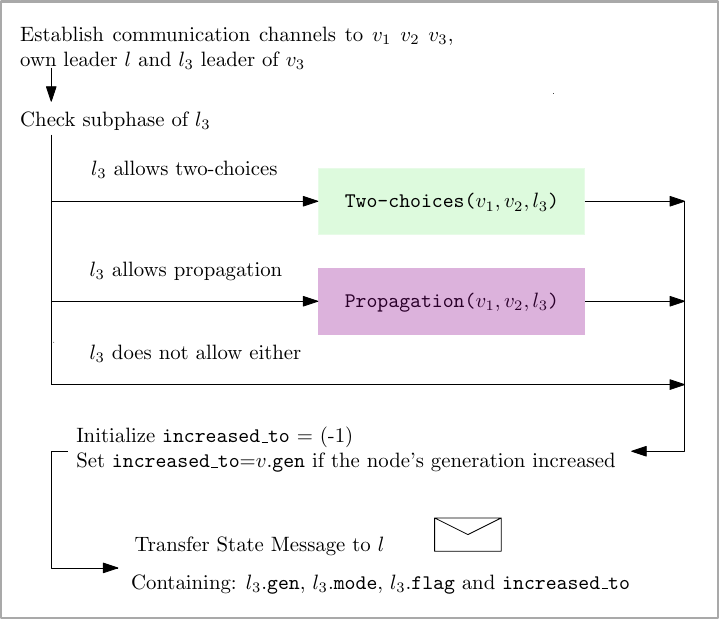}
\caption{The main procedure of follower node $v$. Executed each time $v$ ticks while no previous execution of the procedure is still ongoing.}
\label{fig:async-follower}
\end{figure}

In the following we extend the description of our algorithm in \cref{sec:Decentralized-system}. 
The leaders and followers full procedures are presented \cref{fig:async-leader} and \cref{fig:async-follower}, respectively. We list all required parameters and constants to implement this algorithm in a paragraph that follows later in this section.

\paragraph{The Leaders Routine} We start by discussing the leaders routine in \cref{fig:async-leader} and consider some fixed leader $l$. The main and sampling routines operate passively and only act when information from the followers of $l$ arrives. 
As usual, upon each received $0$-signal, the leader increments its counter $\ticks{l}$,  causing the leader to eventually progress certain phases (e.g. from the two-choices to the sleeping phase) in case a certain threshold is surpassed.
Additionally, another type of message is sent by the followers. It encapsulates the state  information of a randomly sampled leader $l_3$ and is sent by followers each time they establish a communication channel to such a leader $l_3$ (remember, $l_3$ is the leader of a randomly sampled node $v_3$).
Furthermore, this message contains information whether the follower increased its generation due to a two-choices or propagation step. This information is then used by the leader $l$ to (i) track the number of it's followers that are of generation $\gen{l}$, (ii) eventually raise his $\flag{l}$ (set it to true), in case $\gensize{l}$ surpasses half the amount of its total followers, (iii) wake up from the sleeping phase with the help of the $\wakeup{l}$ bit in case a message indicates that another leader $l_3$ already passed the sleeping phase (iv) count the number of leaders $l_3$ that have $\flag{l_3}$ raised, and (v) increase the value of $\gensize{l}$ in case some of the leader's followers promoted to $\gen{v}$.
 
 In the description of the consensus protocol in \cref{sec:Decentralized-system}, we mentioned that a sampling mechanism is employed to control when leaders enter the preparation phase. This is done to guarantee that no leader enters generation $i+1$ before at least $n/2$ of all nodes belong to generation $i$. We may achieve this as follows: Throughout any generation $i$, followers and leaders employ the sampling mechanism described in \cref{sec:sampling} w.r.t.\ the property $R(w) \Leftrightarrow $(the leader $l'$ of $w$ has its $\flag{l'}$ set to \texttt{true} and allows generation $i$).
This samplings are implemented on the leaders end  by incrementing a variable $r_1$ each time a State Message arrives. Only in case this message indicates that $\flag{l_3}$ is raised and $l_3$ allows generation $i$ currently, the variable $r_2$ is incremented as well.
Hence $r_2 / r_1$ contains the ratio of leaders that were recently sampled and lead to $R(w)$ being true. As required in \cref{ml-sampling-theorem} these samplings are performed in batches of size $0.8 \cdot \log^{2 + \varepsilon}$ for some small constant $\varepsilon >0 $, and after each batch is completed the value $r' = r_2 / r_1$ is evaluated.
If $r' = 1$ is observed for the first time , then \cref{sec:sampling} guarantees that, indeed, globally a large fraction of nodes must belong to generation $i$. At this point the leader switches from the propagation into the preparation sub-phase.

\paragraph{The Followers Routine}

\cref{fig:async-follower} depicts the procedure any follower $v$ follows. Most important details were already explained in \cref{sec:Decentralized-system} (e.g. how two-choices and propagation steps are to be performed). The only thing to note is the State Message, which as already described above, contains information about the randomly sampled leader $l_3$ as well as whether the node $v$ itself increased its generation.
It is important to note that, while not reflected in the image, followers still sends $0$-signals upon each tick, and only start an execution of the procedure in \cref{fig:async-follower} in case no previous execution is currently still ongoing (just as ensured by \cref{ln:blocked-tick} of \cref{alg:asynchronous}).

\paragraph{Required Variables, Parameters and Constants}
\label{sec:ml-variables}
In the following we present a list of the most important required variables. We start with the variables needed for the followers routine.
\begin{itemize}
\item the current generation $\gen{v}$ (initially 0) and color values $\col{v}{\cdot}$ just as in the centralized procedure (see \cref{alg:asynchronous}) to be used throughout two-choices and propagation steps.
\item an address of its own leader $l$
\end{itemize}
A leader node $l$ requires the following state information. 
$\mathtt{l}$ set to its own address. However, it also provides followers access to the following public variables 
\begin{itemize}
\item $\gen{l}$, the currently highest allowed generation.
\item $\gensize{l}$, the cardinality of the latest generation in the cluster
\end{itemize}

We note that leaders may also behave as regular follower nodes in addition to following the leaders routine. This allows them to eventually take the initial majority color. However, we want to emphasize that when talking about $\gen{l}$ of a leader $l$, we always talk about the field containing the highest generation he allows. This fields has nothing to do with the (different) field $\gen{l}$ of the same name that is required for $l$ to fulfill his duties as a follower.

Additionally, the following private variables are used by leader nodes throughout the procedure.

\begin{itemize}
	\item $n' \in\mathbb{N}$, the precise cardinality of the cluster, initially set to $1$,
	\item $r_1, r_2, r'\in\mathbb{N}$, the variables used by the sampling gadget as described in \cref{sec:sampling}.
	\item boolean variable $\flag{l}$ used to indicate that at least $(1/2 + \beta) n'$ followers of the cluster are of generation $\gen{l}$.
	\item boolean variable $\wakeup{l}$, indicating that the leader should skip the sleeping phase.
	\end{itemize}

In order to properly execute the leaders routine, the following values, including $\mathcal{H}(\cdot)$ and $\mathcal{S}(\cdot)$ (see \cref{lem:signals-per-time,lem:counting-time} for their definition), need to be known to the leader nodes. Note that all of them can be computed as long as an estimate of $n$ as well as the distributions for the waiting time and channel delays are known to the leader.

\begin{itemize}
\item $C_{1}$ -- the number of time steps in a time unit, see \cref{subsec:Latency-time-statements}.
\item $C_{br} = C_{pre} + \mathcal{S}(C_{pre}) = O(1)$ -- upper bound on the number of time steps between the first and last cluster allowing any fixed generation $i$. 
\item $C_{tc}=C_{br} + C_1 = O(1)$ -- the lower bound for the duration of the two-choices phase in time steps
\item 
$C_{slp}=\mathcal{S}(C_{tc}) + C_{br}= O(1)$ 
-- the lower bound on the number of time steps required to move from start of the sleeping phase to propagation
\item $C_{pre} = \mathcal{S}(5 C_1) = O(1)$ -- time required for any leader to count until $\mathcal{H}(5 \cdot C_1) \cdot n'$.
\item $\beta$ -- an arbitrary constant $0 < \beta <  1/4$, where $(1/2 + \beta) n'$ is the threshold of followers of generation $i$, necessary for the leader to set its flag to \texttt{true}
\item $\varepsilon$ -- an arbitrary constant larger $0$ such that for the clustering constant $c$, it holds that $c \geq 4 + \varepsilon$ (see \cref{thm:ml-extended-clustering}). In other words, $\varepsilon$ is chosen such that each (active) cluster is of size at least $\log^{3 + \varepsilon} n$ \whp.
\end{itemize}

\subsection{Analysis of the Algorithm}
\label{sec:ml-analysis}
As mentioned in \cref{sec:Decentralized-system}, we assume that before the start of the consensus algorithm all but $n/\polylog n$ of all nodes lie in active clusters of size at least $\log^{3+\varepsilon} n$ for some constant $\varepsilon > 0$.
Furthermore, we assume that the leaders of these clusters start the consensus algorithm with a time difference of at most $C_{br}$ time steps, which can for example by achieved by the Extended Clustering Algorithm described in \cref{sec:extended-clustering} (see \cref{thm:ml-extended-clustering}).\\

\emph{Remark:} During the following analysis we will neglect the existence of nodes in faulty clusters, i.e., in clusters that remain inactive after the clustering procedure. 
As established in \cref{thm:ml-extended-clustering} at most a $1/\polylog n$ fraction of them will exists.
If a node $v$ contacts such a node as either $v_1,v_2$ or $v_3$, it will not reply to consensus requests and instead start a new execution upon its next tick.
Remember, the node $v$ will act in some time unit with probability at least $0.9$. Above scenario will prevent $v$ from acting with probability at most $O(1/\polylog n)$. This way $v$ will act during one time unit with probability $0.9 (1- o(1))$.
It is easy to see that this could be accounted for by elongating the length of a time unit $C_1$ slightly.
This illustrates that accounting for inactive nodes does not change the results of the analysis.
For the sake of easier readability we therefore assume that all nodes lie in active clusters. Additionally, in order to allow nodes in faulty cluster to eventually reach consensus, they can for example periodically contact a random neighbor and adapt its color.\\

In the following, we will reuse the notation of the centralized algorithm, defined at the end of \cref{subsec:notions}. In the context of multiple leaders, we will use $t_{i}$ to denote the point in time when generation $i$ is allowed for the first time by \emph{any} leader. The remaining notation remains unchanged.

\begin{figure*}
\centering
\includegraphics[width=\linewidth]{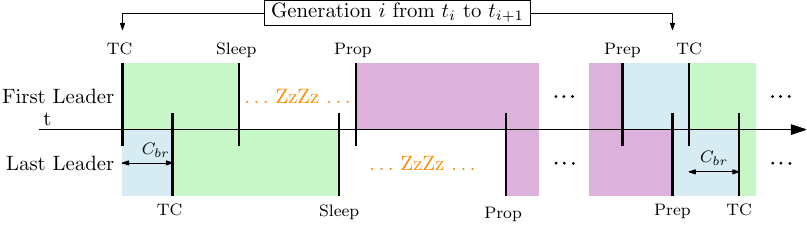}
\caption{Starting times of the subphases a leader allows throughout generation $i$. Color encoding corresponds to \cref{fig:async-leader}. The colored areas indicate whether the leader currently allows two-choice or propagation steps.}\label{fig:asynchronicity-flow}
\end{figure*}

\paragraph{Dealing With Asynchrony}
For any fixed generation, each cluster goes through the following
\emph{(sub)phases} (see \cref{sec:Decentralized-system} for a description):
\begin{enumerate*}
\item the two-choices phase,
\item the sleeping phase,
\item the propagation phase, and
\item the preparation phase.
\end{enumerate*}
While the nodes may be highly dis-synchronized in a given
time-step (a node may wait $\log n$ time units before ticking), this
is not the case for the leaders. Indeed, each leader is contacted
whenever any of its (at least $\log^{3 + \varepsilon} n$) followers ticks, and
therefore we expect the leaders to be much better synchronized.  This behavior is illustrated in \cref{fig:asynchronicity-flow}. The time between the first and last leader allowing specific sub-phases of some generation $i$ might differ by up to $O(1)$. However, among other properties, we will establish that all leaders allow two-choices steps for at least one time unit simultaneously. Additionally, we want that the first leader and last leader enter every generation within a time difference of at most $C_{br}$ time steps, no matter how many generations have already passed.

In what follows, we will fix some arbitrary generation $i \geq 1$
and assume that leaders start allowing generation $i$ within a time difference of at most $C_{br}$ time steps.
An important gadget to achieve some synchronicity among the leaders are the counters $\ticks{l}$, each of which is used by a leader $l$ to switch from the two-choices to the sleeping phase, as well as from the sleeping to the propagation phase. Remember that we stated \cref{lem:signals-per-time} and \cref{lem:counting-time} such that they are applicable for leaders with $|U| > \log^{2+\varepsilon} n$ followers (for some constant $\varepsilon >0$). As in our case clusters are size at least $\log^{3 + \varepsilon} n$, we are able to use these results to derive some statements about the global life-cycle of generation $i$. We start by showing that our algorithm achieves the desired behavior of disjoint two-choices and propagation phases, while allowing $1$ time unit of simultaneous two-choices. A more precise formulation of the statement can be found in the following proposition.

\begin{proposition}
\label{prop:ml-timings}
Fix a generation $i$ and consider the following statements regarding
the flow of the leaders throughout its life cycle. Under assumption that even the slowest leader allows generation $i$ earlier than at time unit $t_{i} + C_{br}/C_1$, it holds \whp that 
\begin{enumerate}
\item \label{enum:same-time-tc}When the fastest leader starts sleeping, \emph{every }cluster leader allowed two-choices to generation $i$ for at least one simultaneous time unit.
\item \label{enum:no-tc-during-prop}The first leader does not wake up before every other leader started sleeping
\item \label{enum:propagation-delay}The slowest leader enters the propagation phase at most $\mathcal{S}(C_{slp}) = O(1)$ time steps after the leader who allowed propagation first.
\end{enumerate}
\label{prop:ml-leader-sync}
\end{proposition}
\begin{proof}
	\cref{lem:counting-time} is the main ingredient of this analysis. As every cluster is of size at least $\log^{2+\varepsilon} n$,  we can apply its results. It states that, if  a leader with $n'$  followers counts to $\mathcal{H}(L) \cdot n'$ for $L = \Omega(1)$, at least $L$ and at most $\mathcal{S}(L) = O(L) $ time steps will pass.
We will now start to proof the statements one after another.
\begin{enumerate}
	\item Remember that $C_{tc} = C_{br} + C_1$, therefore counting until $\mathcal{H}(C_{tc})$ is guaranteed to take $C_{br} + C_1$ time steps \whp. From our assumption we know that all leaders start the two-choices phase within a difference of at most $C_{br}$ time steps.
	\item Slow leaders finish the two-choices phase at most 
	$C_{slp} = C_{br}  + \mathcal{S}(C_{tc})$ time steps after generation $i$ first appeared. At this point in time, the fastest spent at most $C_{slp} - C_1 < C_{slp}$ sleeping \whp. As we require the leaders to count $\mathcal{H}(C_{slp}) \cdot n'$ additional incoming ticks before leaving the sleeping phase, the result follows.
	\item The previous item implies that the last leader enters the  propagation phase at most $\mathcal{S}(C_{slp})$ time steps after the fastest leader. Note that some leaders might even skip parts of the sleeping phase due to being woken up. This only reduces this difference further. \qedhere
\end{enumerate}
\end{proof}

By \cref{thm:ml-extended-clustering} we already know that the leaders allow the \emph{first} generation within a time difference of at most $C_{l} < C_{br} = O(1)$ time steps. In the following we will show that the sampling gadget described in \cref{sec:sampling} allows to establish this property for any later generation as well. Note that this is also depicted in \cref{fig:asynchronicity-flow}: The first leader starts the two-choices phase of some generation $i$ at most $C_{br}/C_1$ time units before the last, which in turn results into leaders allowing the two-choices phase of the next generation within a time difference less than $C_{br}/C_1$ time units.

We start by showing a statement that follows from the fact that a leader may only transition into the preparation phase upon observing that \emph{every} leader out of a sample of size $0.8 \cdot \log^{2+\varepsilon} n$ has its flag set to true. Only clusters with at least a $(1/2 + \beta)$ fraction of followers at generation $i$ set their flag. Therefore, globally, at least half of the nodes belong to generation $i$ when the first leader enters the preparation phase. We formalize this as follows.

\begin{lemma}
\label{lem:first-prep-g}
	Let $t_f$ denote the time unit when the first leader allowed the preparation phase of generation $i$. Then, it holds that
$t_f > t_i(1/2)$, and at $t_f$ at least $(1-\varepsilon') n$ nodes have leaders that set their flag to true \whp. Here $\varepsilon'$ is an arbitrary small constant with $\varepsilon' < \beta $.
\end{lemma}
\begin{proof}
	In the following we employ the notation of \cref{ml-sampling-theorem}. Let $l_f$ be the first leader to enter the preparation phase at $t_f$. To enter the preparation phase, it must have performed a sampling in which all involved followers observed nodes $w$, s.t. $R(w) \Leftrightarrow (w's$ leader has the flag set to true and allows generation $i$) holds. Let $t^*$ denote the time at which $r_{t^*} \geq (1-\varepsilon')$ for any small constant $\beta > \varepsilon' > 0$. Consider some sampling starting at $t'$ and ending at $t''$ with $t' \leq t'' \leq t^*$. Clearly, for any $t$ with $t' \leq t \leq t''$ it holds that $r_t \in [0,(1-\varepsilon')]$. Hence, according to \cref{ml-sampling-theorem}, for any result of such sampling  it holds that $r' < (1-\varepsilon') (1+o(1)) < 1$. This implies that $t_f > t^*$ \whp. 
	In other words, at least $T > (1-\varepsilon') n$ nodes have leaders which have their flag set to true at time $t_f$ (this implies that second statement of the lemma). Let $L$ be the set of clusters leaders that have their flag set to true at $t_f$. Assume $|L_j|$ for $L_j \in L$ denotes the size of the $j$-th such cluster. It follows that $T = \sum_{L_j \in L} |L_j|$. Now, we know that in each cluster $L_j$ at least $|L_j| (1/2 + \beta)$ nodes are of generation $i$, otherwise the leader of $L_j$ would not have set its flag to true. As $T > (1-\varepsilon') n$ we have that  
	\[
		\sum_{L_j \in L} |L_j| (1/2 + \beta) > (1 - \varepsilon') (1/2 + \beta) n > (1/2) n
	\]
many nodes are of generation $i$ at $t_f$.
\end{proof}

Next, we make use of the fact that -- as long as all leaders keep allowing propagation steps -- the ratio of nodes of generation $i$ will quickly approach the global ratio. More specifically we will soon require the following statement in our analysis.

\begin{lemma}
\label{lem:global-g-catchup}
	Let $t \geq t_i(1/2)$ and assume that every leader currently allows propagation steps to nodes of generation $i$ for at least two more time units. Then, before time unit $t+2$ each leader $l$ has $\gensize{l} > (1/2 + 1/4) n'$  and it's $\flag{l}$ set to true \whp.
\end{lemma}
\begin{proof}
	Consider some node $v$ of generation less than $i$ at time $t$. With probability $0.9$, it will perform a full execution throughout the following time unit, and with probability at least $(1-1/4)$ sample at least one node in generation $i$. As every leader currently allows propagation, $v$ will therefore join generation $i$ with probability at least $0.9 \cdot (1-1/4)$. In the worst-case it holds that $g_t(i) = 1/2$. Even in this case a simple Chernoff bound application shows that $g_{t+1} (i) > 1/2 + 0.8 (1-1/4) > 1/2 + 1/4$.
	
	Now, consider the following time unit together with some fixed cluster $C$ of size $n'$. At this point it holds that $g_{t+1}(i) > 1/2 + 1/4$. Hence any node in $C$ that is not of generation $i$, will join generation $i$ with probability at least $0.9 \cdot (1 - 1/16)$. A simply Chernoff bounds application shows that even if $C$ has no generation $i$ nodes yet, in the following time unit at least $(1/2 + 1/4) \cdot n'$ of nodes in $C'$ will be of generation $i$. 
\end{proof}

Now, we again consider the time $t_f$ at which the first leader enters the preparation phase. We make use of the fact that at time $t_f$ most leaders must allow propagation steps already. As stated in \cref{lem:first-prep-g}, most leaders must have their $\mathtt{flag}$ set at $t_f$. These leaders must have passed the sleeping phase already, as they cannot set their $\mathtt{flag}$ to true otherwise. Nodes that encounter such a leader signal their own leaders to wake up in case they are still sleeping. Hence, any remaining sleeping leaders are woken up shortly after $t_f$.
The next statement guarantees that all the nodes enter the preparation phase at roughly the same time, formalized as follows.

\begin{lemma}
\label{lem:ml-prep-prop}
	Let $t_f$ denote the time unit when the first leader entered the preparation phase of generation $i$. Then, the following statements hold \whp.
	\begin{enumerate}
		\item \label{enum:first-prep} Even the fastest leader does not stop allowing propagation steps before time $t_f + 5$.
		\item At time unit $t_f + 3$, every leader has its flag set to true.
		\item  \label{enum:entering-prep} Every leader entered the preparation phase before time unit $t_f + 5$.
	\end{enumerate}
\end{lemma}
\begin{proof}
	The first statement follows from the fact that after entering the preparation phase, each leader with $n'$ followers needs to receive $\mathcal{H}(5 \cdot C_1) \cdot n'$ many $0$-signals in order to stop allowing propagation steps.
	
	Next, the second statement. We know according to \cref{lem:first-prep-g} that after $t_f$, at least $(1-\varepsilon') n$ of all nodes have leaders that have set their flag to true. These leaders are already either in the propagation or preparation phase. Remember that once a follower encounters such a leader, it will inform its own leader, waking it up in case it still is in the sleeping phase. It is easy to see that during one time unit any such sleeping leader is woken up by some follower. Hence, \whp, at time $t_f + 1$ every leader allows propagation steps. Observe that \cref{enum:first-prep} guarantees two more time units of propagation steps following $t_f + 1$.
Hence, \cref{lem:global-g-catchup} guarantees that each leader $l$ has $\gensize{l} > (1/2 + 1/4) n'$ at time $t_f + 3$ and thereby also sets its flag.

Now for the final point. At $t_f + 3$ every leader has the flag set to true. Therefore the next sampling performed by any leader must yield $r' = 1$ and succeed. It takes at most 2 time units to perform such a sampling, as another currently running sampling might need to be concluded first. Note that until this point \emph{no} leader allows generation $i+1$ yet (see \cref{enum:first-prep}). That is, the flags have not been reset for the following generation.
\end{proof}

The following is mostly implied by above statements. We show that we indeed achieve that the leaders allow generation $i+1$ within a time frame of at most $C_{br}$ time steps of each other. Additionally, we state that the during the two-choices phase of the following generation $i+1$, no leader will allow propagation steps anymore.

\begin{proposition}
\label{prop:ml-generation-change}
	Assume that the leaders entered generation $i$ within a time difference of at most $C_{br}$ time steps. Then, the following statements hold \whp.
	\begin{enumerate}
		\item \label{item:prep-time-frame}The first leader enters the preparation phase at time $t_f$ where $t_i(1/2) < t_f < t_i(1/2) + O(1)$.
		\item \label{item:propagation-stop}All leaders entered the second half of the preparation phase (and thereby stopped allowing propagation steps) before time $t_{i+1}$ 
		\item \label{item:next-gen-start}The last leader allows generation $i+1$ at most $C_{br}$ time steps after the first.
	\end{enumerate}
\end{proposition} 
\begin{proof}
	We start with the first statement. The lower bound was already established in \cref{lem:first-prep-g}.
 Assume that at time $t_i (1/2)$ the first leader did not enter the preparation phase yet. We know by \cref{prop:ml-timings} that at most $O(1)$ time later, every leader must allow propagation steps.  By \cref{lem:global-g-catchup} it follows that at $t_i(1/2) + 2$ every leader set its $\mathtt{flag}$ to true \whp. Therefore, after $t_i(1/2) + 2$ every sampling performed by a leader causes it to enter the preparation phase \whp.
 
	Next, the second statement. We know by \cref{enum:entering-prep} of \cref{lem:ml-prep-prop} that every leader entered the propagation phase before time $t_f + 5$. Each such leader counts $\mathcal{H}(5 \cdot C_1) \cdot n'$ many $0$-signals at which point it stops allowing propagation steps, where $n'$ denotes the number of its followers. That is, at time $t_f + 5 + C_{pre}$ for $C_{pre} = \mathcal{S}(5 \cdot C_1)$, no leader allows propagation steps anymore \whp. Observe that every leader -- in particular also the leader that first entered the preparation phase at $t_f$ -- needs \whp at least $5 + C_{pre}$ time steps to count sufficient $0$-signals to pass the second half of the preparation phase. The result follows accordingly.

 Finally, by \cref{enum:first-prep, enum:entering-prep} of \cref{lem:ml-prep-prop} we know that every leader enters the preparation phase before the first leader allows generation $i+1$. Slow leaders require 
$C_{pre} + \mathcal{S}(C_{pre}) = C_{br}$ time to count sufficient $0$-signals for passing on to generation $i+1$. 
\end{proof}

\paragraph{Carrying over the Synchronous Case Analysis}

The results we just established, satisfy some important invariants which allow us to use a similar analysis as in \cref{sec:Asynchronous-1leader}.
Indeed, while the vertices may be far from synchronized, the leaders behave quite synchronized in several aspects. In particular, for any fixed generation $i$:
\begin{itemize}
\item The leaders of all clusters
will be allowing two-choices steps for at least one time unit at the same time.

\item Throughout the time any leader allows two-choices steps to promote to generation $i$, \emph{no} leader allows propagation to generation $i-1$ anymore.

\item No node may be promoted to $i$
as a result of two-choices \emph{after }the first node has joined generation $i$ through a propagation step.

\item Every leader allows promotions via propagation at most $O(1)$ time units after the first leader does so.

\item No leader will allow the next generation to be created before time $t_i(1/2)$.

\item Leaders enters the following generation $i+1$ within a time difference of at most $C_{br}$ time steps from each other.

\end{itemize}

We start by carrying over results considering the growth of some arbitrary generation $i$.
At it's core, this algorithm mimics the behavior of the centralized one analyzed in \cref{sec:analysis-centralized}. It is important to emphasize that a node $v$ determines whether two-choices or propagation steps are allowed by inquiring a leader $l_3$ of a node $v_3$ that is selected uniformly at random -- independently from the other two nodes $v_1$ and $v_2$. Therefore, this  information does not depend on the state of $v_1$ or $v_2$ but rather on the global ratio of nodes that have leaders allowing these steps.

We start with a result, which corresponds to \cref{lem:tc-parent-stable} of the centralized case, implying that the color fraction in generation $i-1$ remain stable throughout the time frame $[t_i, t_i +t']$. As in this case, we consider a set of leaders, we assume $t_i$ to be the time the \emph{first} leader allowed generation $i$, and $t_i + t'$ the time when the \emph{last} leader entered the sleeping phase of generation $i$ (and therefore stopped allowing two-choice steps).

\begin{corollary}
\label{cor:ml-stable-tc}
Consider some fixed generation $i$ throughout $[t_i ,t_i +t']$ and define 
\[ S_{i-1,j}(t) = \{v ~|~ v \text{ has } \col{v}{i-1} = j \text{ at time } t\}.
\]
Assume a node $v$ of generation $i-1$ finished establishing all required communication channels at $t \in [t_i, t_i +t']$. Then,
\begin{enumerate}
	\item \label{item:ml-parent-stable-1}$v$ will promote to generation $i$ and take color $j$ if and only if both sampled nodes $v_1$ and $v_2$ lie in $S_{i-1,j}(t)$,\\
	 and the sampled leader $l_3$ allows two-choices steps;
	\item \label{item:ml-parent-stable-2}$\forall t \in [t_i ,t_i +t']: S_{i-1,j}(t) = S_{i-1,j}(t_i)$ and $|S_{i-1,j}(t_i)| / n = c_{j,i-1}(t_i) \cdot g_{i-1}$;
	\item \label{item:ml-parent-stable-3}$S_{i-1,j} (t) \cap S_{i-1,j'}(t) = \emptyset$ for every pair of colors $j,j'$ with $j \neq j'$. 
\end{enumerate}
\end{corollary}

The above result follows as two-choices steps (on the followers end) are performed almost as in the centralized case with the only difference being that $l_3$ is consulted instead of the own leader. That is, it is still necessary for a nodes to sample two nodes out of the set $S_{i,j}(t)$ to promote to generation $i$ via two-choices. Additionally, \cref{item:ml-parent-stable-1}, needs to account for the fact that not all leaders allow two-choices steps in every time unit of $[t_i, t_i +t']$. Note that, in order for \cref{item:ml-parent-stable-2} to hold, it is required that \emph{no} node may promote to generation $i-1$ via propagation steps anymore. This, however, is guaranteed by \cref{item:propagation-stop} of \cref{prop:ml-generation-change}.

Similar as in the centralized case, we may use above result to deduce that, the next time a node $v$ of generation $i-1$ finishes an execution, it will promote to generation $i$ with probability $p_{i-1} \cdot g_{i-1}^2 \cdot r_{l_3}^2(t)$. Here $r_{l_3}(t)$ denotes the probability that the leader $l_3$ allows two-choices steps at the time $t$ where $t$ denotes the time when $v$ has all its required communication channels established. Additionally, if $v$ promotes to $i$ in $[t_i , t_i + t']$ it will still join color fixed color $j$ with probability exactly $c_{j,i}(t_i) / p_{i-1}$.

Now, observe that $p_{i-1} \cdot g_{i-1}^2 \cdot r_{l_3}^2(t) = p_{i-1} \cdot g_{i-1}^2 = \Omega(p_{i-1})$, in case \emph{all} leaders currently allow two-choices steps. By \cref{enum:same-time-tc} of \cref{prop:ml-timings} there indeed exists $\tilde t, \tilde t' \in [t_i , t_i + t']$ such that $\tilde t = \tilde t' - 1$ and in $[\tilde t, \tilde t']$ every leader allows two-choice steps. Using above notion, this implies that $r_{l_3}(t) = 1$ for $t \in [\tilde t, \tilde t']$. By our definition of a time unit, each node of generation $i-1$ before time $\tilde t$  will join $i$ before time $\tilde t'$ with probability at least $0.9 \cdot \Omega(p_{i-1})$. Hence, the proof of the centralized case (which also considered only 1 time unit of two-choices steps) can easily be adapted to yield.

\begin{corollary}
\label{cor:ml-tc-base}
Fix some generation $i$ and assume that $g_{i-1} \geq 1/2$.
Let $t_i + t'$ denote the time at which the last enters the sleeping phase of generation $i$. Then, $g_{i}(t_i +t') \geq p_{i-1} / 5$ \whp. 
\end{corollary}

Now, consider the time $t_w$ at which the first leader concluded the sleeping phase of generation $i$. As the leaders enter generation $i$ with difference at most $C_{br} = O(1)$, and count signals to approximate constant time frames, it follows that $t_w - (t_i +t') = O(1)$. Furthermore, by \cref{enum:propagation-delay} of \cref{prop:ml-timings} we have that even the slowest leaders will start allowing propagation steps at most $O(1)$ time later. As no leader enters preparation phase before $t_i(1/2)$ (see \cref{item:prep-time-frame} of \cref{prop:ml-generation-change}), it follows that in $[t_w + O(1) ~,~ t_i(1/2)]$ all nodes allow propagation steps and generation $i$ will be spread quickly along the lines of pull gossiping (just as in the centralized case). Therefore, the following result can easily be achieved.

\begin{corollary}
\label{cor:remaining-time}
Fix some generation $i$. Then, $ t_i (1/2) < t_i + O(\log(1 / p_{i-1}))$ \whp.
\end{corollary}

\cref{item:prep-time-frame} of \cref{prop:ml-generation-change} implies that shortly after $t_i(1/2)$, the first leader enters the preparation phase. After $O(1)$ time it will have counted sufficient $0$-signals to switch to generation $i+1$.
It follows that $t_{i+1} < t_{i} + O(\log(1 / p_{i-1}))$, which is a similar result as the one in  \cref{cor:1leader-gossiping-one-generation} w.r.t. the centralized procedure

When it comes to the concentration of color fractions, we start by arguing that \cref{lem:initial-bias-1} of the synchronous case is also applicable in this case.
We already established in \cref{cor:ml-stable-tc} that each time a node in the time frame $[t_i ,t_i + t']$ is promoted to $i$, (i) it does so via a two-choice step, and (ii) it takes color $j$ with probability $c_{j,i-1}^2 / p_{i-1}$, independent from the actions of other nodes throughout this time frame. This, together with \cref{cor:ml-tc-base}, are the main ingredients required in the proof of \cref{lem:initial-bias-1}, which therefore also applies in this setting.

\begin{corollary}[Time $t_{i}\rightarrow t_{i}+t'$.]
\label{cor:tc-conc-ml}
Let $a$ and $b$ be the largest and second largest opinion in generation $i-1$ at time $t_i$ and assume that $b_{i-1} \gg 1/\sqrt{n}$. If $t_{i}+t'$ corresponds to the time
when the first leader enters the propagation phase, then it holds \whp that
\begin{align*}
a_i(t_i + t')  &= \frac{(a_{i-1})^2}{p_{i-1}} \left(1  \pm \frac{1}{a_{i-1}} \sqrt{\frac{\log n}{n}} \right)  \text{, and } \\
	b_i(t_i + t')& = \frac{(b_{i-1})^2}{p_{i-1}}  \left(1 \pm \frac{1}{b_{i-1}} \sqrt{\frac{\log n}{n}} \right).
\end{align*}
\end{corollary}

Throughout the remaining time frame $[t_i +t', t_{i+1}]$ of generation $i$, leaders will no longer allow two-choice steps. Consider the ordered points in time  $t_{(1)}, t_{(2)}, ... \in [t_i + t', t_{i+1}]$ at each of which some node (i) arrived from generation $i-1$, then (ii) sampled a node $v_1$ or $v_2$ of generation $i$, and (iii) sampled a node $v_3$ that has a leader $l_3$ allowing propagation steps.
Each such step leads to an increase in the number of nodes of generation  $i$. As each node samples $v_1$ and $v_2$ u.a.r. and independently from $v_3$ (see \cref{sec:multileader_proc}), it follows that such a node will join color $j$ with probability proportional to its current support in generation $i$. 
In other words, the sequence of color fractions at the points in time at which nodes join generation $i$ still follows a martingale. Just as explained in \cref{sec:sl-overview}, it can be shown -- with the help of a  Pólya-Eggenberger distribution -- that the color fractions remain concentrated throughout the propagation phase of generation $i$.
At this point it is important that the sleeping phase guarantees that no nodes promote via two-choices in this time frame (otherwise they would interfere with the above martingale). This desired property is established in \cref{enum:no-tc-during-prop} of \cref{prop:ml-timings}.  There is one subtle difference: in the centralized case we could guarantee that the length of the corresponding Pólya-Eggenberger process (as well as the length of the martingale) is exactly $n/2 - n \cdot g_{i-1}(t_i)$, however, by \cref{item:prep-time-frame} of \cref{prop:ml-generation-change} we can only say that the length will be \emph{at least} as much as above value \whp. We note that the proof of \cref{lem:polya-color-bound} can be easily adapted and this does not have any effects on our results.

As \cref{lem:polya-color-bound} and \cref{lem:async-alpha>k-steps} only rely on the possibility of modeling $c_{j,i}$ with the help of above Pólya-Eggenberger process and \cref{cor:tc-conc-ml}, we can apply their results also in the decentralized case. 

This, combined with the concentration result \cref{cor:tc-conc-ml} implies that \cref{lem:async-num-generations} and \cref{cor:one-phase-bias-1} hold in this case as well, allowing us to expresses how the bias evolves over multiple generations. Summarizing, we can therefore say the following

\begin{corollary}
\label{cor:squaring-ml}
	Consider an initial bias of $\alpha_{0} > 1 + \frac{1}{b_0} \cdot \frac{\log n}{\sqrt{n}}$. Then, \whp,
	\begin{enumerate}
		\item  after at most $\lceil \log_{1.5} \log_{\alpha} k \rceil $ generations the bias will exceed $k$, and 
		\item generation $\lceil \log_{1.5} \log_{\alpha} n \rceil + 2$ will be monochromatic.
	\end{enumerate}
\end{corollary}

We conclude the analysis of this section with the following statement, which gives us a similar result as \cref{lem:spreading-monochromatic} in the analysis of the centralized protocol.

\begin{lemma}
\label{lem:ml-final-steps}
	At most $O(\log \log n)$ time units after the first monochromatic generation is reached, all but an $1-1/ \polylog n$ fraction of nodes will be of the same color \whp.
	Additionally, after further $O(\log n)$ time units, every node will be of color $i$.
\end{lemma}
\begin{proof}
	Let $i^*$ be the first such monochromatic generation. Consider some fixed generation  $i > i^*$. Clearly this generation $i$ is monochromatic as well. After \cref{cor:remaining-time} we established, that $t_{i+1} - t_{i} = O(\log (1 /p_{i-1})) = O(1)$. 
	If we denote by $t_f$ the time at which the first leader entered the preparation phase of generation $i$, then it follows by \cref{lem:ml-prep-prop} that in the time frame $[t_f + 3, t_f + 5]$, \emph{every} leader allows propagation to generation $i$. Fix now some node $v$ that is of generation less than $i^*$. Such a node will propagate to $i$ with at least constant probability during the time frame $[t_f + 3, t_f + 5]$, as \cref{cor:ml-tc-base} guarantees that with probability $\Omega(1)$ a node of generation $i$ is sampled throughout this time frame.
Hence $\Theta(\log \log n)$ generations following $i^*$, the node $v$ will remain in a generation less than $i^*$ with probability at most $1/\polylog n$. And, after $O(\log n)$ further generations, \emph{no} node will be of generation less than $i^*$, which in turn implies that all nodes share the same color.
\end{proof}

Putting everything together, we have that -- (i) the number of required generations to reach a monochromatic generation, (ii) the duration of each generation, and (iii) the time required to spread the majority color after the first monochromatic generation is reached -- follow (asymptotically) the same bounds as in the centralized case. Therefore, we conclude the proof of \cref{thm:multi-leader}.

%% file: 95-analysis-accelerated.tex
\section{Analysis of the Accelerated Consensus Protocol}
\label{sec:breaking-the-lower-bound}

In the following section we present a modification of the decentralized protocol given in \cref{sec:Decentralized-system}. We call the resulting algorithm the \emph{Accelerated Consensus Protocol} and assume that the waiting time and channel delay follow distributions that are $q$-dense for some constant $q > 0$ (see \cref{prop:tick-in-log}). This allows us to achieve faster partial consensus than \emph{any} plurality consensus protocol operating in the classical synchronous model, for large ranges of $k$ and initial bias $\alpha$, as long as the maximum congestion lies in $O(\polylog n)$. 

\subsection{The Accelerated Consensus Protocol} The Accelerated Consensus Protocol can be described as follows.
In the first step the Extended Clustering algorithm (see \cref{sec:clustering}) is employed as in the decentralized procedure to partition the nodes into clusters of size $\polylog n$ \footnote{The clustering algorithm (\cref{thm:ml-extended-clustering}), needs to be configured to yield clusters of size at least $\log^{c-1} n$, for constant $c-1\geq 8q + 4$ .}. 
Next, a modified version of the decentralized consensus protocol described in \cref{sec:multileader_proc} is executed. We will now list the required modifications.
	After the leader election is complete, follower nodes discard their generation and color values. Only the cluster leaders $l$ keep their initial color value and store it in $\clustercol{l}{0}$. Here $\clustercolarr{l}$ is an array used to store color values (just as $\colarr{u}$ in \cref{alg:asynchronous}). Additionally, $l$ is equipped with a variable $\clustergen{l}$ which is initially set to $0$. Conceptually, this two new fields should be seen as shared memory that is accessible by the followers of $l$.
That is, each time a follower $v$ of $l$ attempts a two-choices or propagation step, it does so based on $\clustergen{l}$ and $\clustercolarr{l}$ instead of consulting its own $\gen{v}$ and $\colarr{u}$ variables.

Similarly, each time a follower $v$ would read the color and generation of the two sampled nodes $v_1$ and $v_2$ as part of the decentralized protocol (see description in \cref{sec:Decentralized-system}), it reads $\clustergen{l_1}$ and $\clustercolarr{l_1}$  as well as $\clustergen{l_2}$ and $\clustercolarr{l_2}$ instead. Here $l_1$ and $l_2$ denote the leaders of $v_1$ and $v_2$ respectively. In order to make this possible, we assume that follower nodes inquire the addresses of $l_1$ and $l_2$ and also opens communication channels to these leader nodes.
	
In some sense, this causes only the leaders to increase in generation and change their colors, with followers acting as relays to facilitate communication between the leaders.
Additionally, all followers in a cluster share the generation and color information stored at their leaders. This way, a successful propagation or two-choices step performed through one single follower suffices to modify the generation and/or color values of a whole cluster.

Note that the leaders still possess their leadership variables and flags as described in \cref{sec:ml-implementing} and progress through the leaders procedure described in \cref{sec:multileader_proc} as usual.
The only exception concerns the variable $\gensize{l}$, which is now set to $n'$ as soon as the cluster of $l$ increases its generation to $\gen{l}$ (remember $n'$ denotes the cluster's size and $\gensize{l}$ denotes the number of nodes of the current generation in the leaders clusters). This reflects the fact that the whole cluster increases its generation at the same time. 

Intuitively, this approach solves the plurality consensus problem among leader nodes, where nodes "help" their leaders to reach said consensus at an accelerated rate.
Remember, a key property used in the previous analysis was that the nodes $v_1$ and $v_2$ are sampled u.a.r. when reading their values of $\mathtt{gen}$ and $\mathtt{col}$.  Therefore, we need to make sure that $l_1$ and $l_2$ appear to be sampled u.a.r as well. However, this would only be the case whenever all the clusters are of equal size.
As this is not guaranteed, we need to implement another modification. Each time some follower $v$ requests information stored in $\clustercolarr{l_i}$ and  $\clustergen{l_i}$ fields of some leader $l_i$ with $i \in \{1,2\}$, then $l_i$ sends with probability $1 - \log^{c-1} n / n'$ the values $\clustergen{l_i} = -1$ and $\clustercol{l_i}{i'}=\NIL$ (for any $i'>0$) instead of its real values. Remember, $\log^{c-1} n$ is a lower bound of the clusters size, where the constant $c$ can be controlled by the clustering algorithm, and $n'$ denotes the cluster size of $l_i$. This way, the probability that some leader is contacted as $l_i$ for $i \in \{1,2\}$ \emph{and} provides some information that does not immediately lead to a failed two-choices or propagation step, is the same for every leader. 

In order to allow for all nodes to eventually reach consensus (not only the leaders), we assume that followers periodically copy the color values that is stored at their leader. This way, followers reach consensus shortly after their leaders.

\paragraph{Enabling Acceleration} 
To achieve an improvement upon the algorithm in \cref{sec:Decentralized-system} we require an additional property. Remember that each time a follower ticks, it appears as if the leaders performed an action according to the follower routine described in \cref{sec:multileader_proc}.
Hence, we want to guarantee that the time between these actions lies in $o(1)$, which would imply that the leader acts multiple time per time unit.
In the following we assume that the distributions $\mathcal{T}_0, \mathcal{T}_\ell$ and $\mathcal{T}_f$ fulfill \cref{prop:tick-in-log} for some constant $q > 0$. In other words, $\mathcal{T}_0, \mathcal{T}_\ell$ and $\mathcal{T}_f$ are $q$-dense.

While this property might seem artificial at first glance, it is indeed fulfilled by most of the distributions, which are used to model waiting times. Most notably the following holds for exponentially distributed waiting times.

\begin{example}
\label{ex:exp-dense}
	Let $X \sim \text{Exp}(1)$. Then, it holds for $x \leq 1$ that 
	\[
		P \left(X < x \right) \geq 0.6 \cdot x.
	\]
	Furthermore,  $\text{Exp}(1)$ is $(1+\varepsilon)$-dense for any arbitrary constant $\varepsilon > 0$.
\end{example}
\begin{proof}
For $0 \leq x \leq 1$ and $q \geq -1$ we have $(1+q)^x \leq 1+qx$ per Bernoulli's inequality.
Setting  $q=- (e-1) /e$, we get that $e^{-x} \leq 1 - \frac{e-1}{e}x < 1 - 0.6 \cdot x$.
Now, consider some $X \sim \text{Exp}(1)$. Then, 
	\[
		P\left(X < x \right) = 1  - e^{-x} \geq 1 -  (1 - 0.6x) = 0.6x. 
	\] 
Note that $0.6 x > x^{(1+\varepsilon)}$ holds for every constant $\varepsilon > 0$ as long as we consider small enough values of $x$. More precisely for $0 \leq x \leq t(\varepsilon)$ where $t(\varepsilon)=0.6^{1/\varepsilon}$ is a constant that depends on $\varepsilon$. This implies that $\text{Exp}(1)$ is $(1+\varepsilon)$-dense for any $\varepsilon > 0$.
\end{proof}

Using the $q$-density property we now deduce that within a time frame of $O(1/\log n)$ every leader will have at least one follower that manages to open all necessary communication channels as long as the Extended Clustering Algorithm (see \cref{sec:extended-clustering}) was configured to partition the nodes into clusters of size at least $8q + 3$.

\begin{lemma}
\label{lem:tick-fast}
	Assume that \cref{prop:tick-in-log} holds for $\mathcal{T}_0, \mathcal{T}_\ell$ and $\mathcal{T}_f$. Fix an arbitrary leader $l$ of size at least $\log^{8q + 3} n$ and at time $t$. Then, independent of events prior to $t$ the following statements hold \whp:
\begin{enumerate}	
		\item \label{enum:tick-fast-1}  in the time frame $[t,t + O(1/\log n)]$ some follower of  $ l$ has its communication channels established and observes that $\clustergen{l_1}, \clustergen{l_2} \geq 0$, and
		\item \label{enum:tick-fast-2}if the event in \cref{enum:tick-fast-1} occurs, then the leaders $l_1$ and $l_2$ appear to be sampled uniformly at random.
	\end{enumerate}
\end{lemma}
\begin{proof}
	Fix some follower node $v$ at time $t$. 
To tick the next time, it needs to pass at most $8$ waiting times. Specifically, it might need to tick and then contact $l,v_1,v_2,v_3,l_1,l_2$ and $l_3$. By the $q$-dense property (set $s = 1/\log n$) and positive aging,
it follows that, with probability $1/\log^{8q} n$, $v$ will have all its channels opened within $8/\log n$ time steps.
Now consider $q_i$ with $i \in \{1,2\}$ denoting the probability that the channel to $l_i$ has been accepted, i.e , the probability that $l_i$ answers with  $\gen{l_i} = -1$.
The node $v$ will hit a fixed cluster of size $n'$ and be accepted by its leader $l_i$ with probability exactly
\[
	\frac{\log^{c-1}n}{n'} \cdot \frac{n'}{n} = \frac{\log^{c-1} n}{ n}.
\]	
As this probability is the same for every cluster, the second statement follows. After the extended clustering algorithm (see \cref{thm:ml-extended-clustering}) at least $(n/ \log^{c} n) (1 - o(1))$ active leaders of size at least $n/\log^{c-1}$ exist \whp.
Therefore, we can easily lower bound $q_i$ by
\[
	\frac{n}{\log^{c} n} (1- o(1)) \cdot \frac{\log^{c-1}}{n} = \frac{1}{\log n} (1 - o(1)).
\] 
As the leaders $l_i$ for $i \in \{1,2\}$ result from independent samplings, it follows that with probability $q > 1/\log^2 n \cdot (1- o(1))$ both of the leaders answer with $\mathtt{gen} \geq 0$
 
Combining our results we have that with probability at least $1/\log^{8q + 2} n \cdot (1- o (1))$, a fixed follower opens channels to all partners without receiving $\mathtt{gen} = -1$ after $O(1/\log n)$ time.
In case a cluster contains at least $\log^{c-1} n$ many followers with constant  
$c-1 > 8q+3$, it follows that such a cluster will have at least one follower throughout every time frame of length $O(1/ \log n)$ \whp.
\end{proof}

\subsection{Analysis of the Accelerated Consensus Protocol}
\label{sec:accelerated-analysis}
The correctness of this algorithm follows largely from the analysis of the decentralized consensus protocol in \cref{sec:ml-analysis}.
In the following, we say that a cluster is of generation $i$ or color $j$, if the leader $l$ of the cluster has $\clustergen{l}=i$ and $\clustercol{l}{\gen{l}} = j$. 

\emph{Generation Lifecycle}
In the following we will examine how the set of leaders progresses a fixed generation $i$ as part of their leaders routine. Luckily, most results can be carried over from the decentralized analysis. Leaders pass most of the sub-phases by counting $0$-signals of its followers until a certain threshold is reached (see \cref{fig:async-leader} on page  \pageref{fig:async-leader}). Note that this mechanism remains completely unchanged in the Accelerated Protocol. 
This allows us to carry over multiple results of the decentralized analysis such as \cref{prop:ml-timings}.

Additionally, the switch from the propagation into the preparation phase is still made by estimating whether $ 0.8 \log^{2 + \varepsilon} n$ sampled leaders have their $\texttt{flag}$ set to true. In the Accelerated Protocol, a leader only sets this $\texttt{flag}$ in case it's cluster reaches generation $i$. 
Hence, \cref{ml-sampling-theorem} of the sampling analysis section, indicates that \emph{no} leader enters the preparation phase before at least a $(1-\varepsilon')$ fraction, for any small constant $\varepsilon' > 0$, of all nodes have clusters of generation $i$.
Using the notation we employed in the analysis of the decentralized case, this means for the time $t_f$ at which the first leader enters the preparation phase, that $t_f > t_i(1-\varepsilon')$. This guarantee is stronger than the one we could make in the decentralized case, where we only stated $t_f > t_i(1/2)$ (see \cref{lem:first-prep-g}). 
Furthermore, it is easy to see that once $t_i(1)$ is reached, every leader will have set its flag only $O(1)$ time later. The main benefit of the Accelerated Consensus Protocol is the speed in which this time $t_i(1)$ can be reached.
\begin{lemma}
	Consider some fixed generation $i$. Assume that all leaders allow generation $i$ before $t_i + C_{br}$. Then, it holds that $t_f > t_i(1/2)$ and $t_{f} - t_i = O(1)$ \whp.
\end{lemma}
\begin{proof}
	In the paragraph above the lemma we already established that the first statement holds.
	Hence, we start with the second statement. 
	As \cref{enum:same-time-tc} of \cref{prop:ml-timings} still holds, each leader will allow one time unit of two-choices simultaneously. Due to the acceleration described in \cref{lem:tick-fast}, every cluster appears if having attempted $\Omega(\log n)$ two-choices steps throughout this time unit. For the purpose of this lemma, it is enough to state that at least \emph{one} cluster will promote to generation $i$ during the two-choices phase, which easily holds \whp. Due to the counting of $0$-signals, every leader will allow propagation steps before time $t'' = t_i + O(1)$ \whp. Assume $t_f > t''$. As each leader allows propagation steps at $t''$, it is easy that at time $t'' + O(1)$, \emph{all} clusters are of generation $i$, and therefore have set their \texttt{flag}. This follows as generation $i$ can be seen as being spread between clusters along the lines of pull broadcasting at an $\Omega(\log n)$ accelerated rate. Further $2$ time units after $t_i(1)$ is reached, every leader has performed a successful sampling and enters the preparation phase \whp.
\end{proof}

Also, note that $t_i(1) -  t_i(1/2) = o(1)$ in case every leader currently allows propagation steps. This is because generation $i$ is spread among clusters via pull broadcasting at an $\Omega(\log n)$ accelerated rate. Soon after $t_i(1)$ is reached, every leader must have entered the propagation phase (every sampling will succeed) and all leaders will enter the preparation phase within a time difference of $O(1)$ \whp.
All above statements allow the results of \cref{lem:ml-prep-prop} and  \cref{prop:ml-generation-change} to be established also in case of this Accelerated Consensus Protocol.

\paragraph{Concentration of Colors}

In the following we consider $c_{j,i}(t)$ , $g_{i}(t), p_{i}(t)$ and $\alpha_{i}(t)$  as well as $t_i(\gamma)$ to be defined w.r.t. the generation and color of \emph{clusters} instead of \emph{individual nodes}. For example, $c_{j,i}(t)$ denotes the fraction of clusters at generation $i$ and time $t$ that are of color $j$.

Throughout the previous paragraph --just as in the decentralized case analysis-- we established the following two crucial properties: (i) all leaders allow two-choices steps for at least $1$ simultaneous time unit, and (ii) after the two-choices phase, propagation steps will be performed until $t_i(1)$, and (iii) the following generation begins shortly after and leaders enter this generation withing a time difference of $O(1)$. Also in this accelerated scenario leaders behave synchronous enough to guarantee that two-choices and propagation steps never overlap \whp.
Note, that for the two-choices and propagation steps in (i) and (ii), \cref{enum:tick-fast-2} of \cref{lem:tick-fast} is important. It guarantees that clusters appear to be performing two-choices and propagation steps based on the color and generation of randomly sampled clusters.

Just as in the decentralized analysis in \cref{sec:ml-analysis}, the above statements allow us to reuse multiple analysis results of the centralized case. In the centralized case, one time unit of two-choices was already enough to create a sufficient foundation of nodes of generation $i$ before the start of propagation steps. However, in case of the Accelerated Consensus Protocol, clusters appear to attempt $\Omega(\log n)$ two-choices attempts throughout this time frame (see \cref{enum:tick-fast-1} of \cref{lem:tick-fast}). Therefore, it is easy to see that \cref{lem:1st-step-1} can be carried over, when denoting by $t_i + t'$ the time at which even the last leader stops allowing promotion via two-choices steps to generation $i$.

Furthermore, the proof of \cref{lem:initial-bias-1} only depends on \cref{lem:1st-step-1} together with the fact that two-choices steps are performed w.r.t. randomly sampled partners. Similar, throughout the time frame $[t_i + t', t_{i+1}]$ when only propagation steps are allowed by any leader, the clusters joining generation $i$ and some fixed color $j$ can again be modeled with the help of a Pólya-Eggenberger distribution. This allows all concentration results to be carried over (most notably \cref{lem:polya-color-bound,cor:one-phase-bias-1}), and thereby guarantees that the bias indeed roughly squares with every further generation and \whp.

There remains one thing to check. Remember that the above mentioned concentration results require an initial absolute bias of $\sqrt{n} \log n$ in favor of the majority opinion. However, this accelerated approach only operates on the set of colors initially assigned to leaders. To guarantee a bias of at least $\sqrt{n} \log n$ among clusters, we need a slightly larger initial bias. We use the fact that the set of elected leaders can be seen as a uniform sample of size $n/\polylog n$ drawn out of all nodes.
Note that the constant $c$ in the following result denotes the clustering constant.
\begin{lemma}
\label{lem:initial-accelerate-bias}
	Let $A$ and $\alpha$ denote the initial absolute and relative bias of colors among all nodes, respectively. Similar, let $A'$ and $\alpha'$ be the initial biases when only considering the colors of active leaders. Then, if $A > 2 \cdot \sqrt{n} \log^{c/2 + 1} n$ and $k \ll \sqrt{n}$ it holds that
	\begin{enumerate}
		\item $\log \log_{\alpha'} m = \max\{O(\log \log n), O(\log \log_{\alpha} n) \}$
		\item $A' > \sqrt{m} \log m$, where $m$ denotes the number of active leaders
		\item $k \ll \sqrt{m}$
	\end{enumerate}
\end{lemma}
\begin{proof}
	Nodes become leaders by successfully flipping a biased coin. Hence, it follows that the color distribution of the $m \geq (n/\log^{c}n) (1 - o(1))$ active leaders (see \cref{thm:ml-extended-clustering} for a bound on the number of leaders) can be modeled by a uniform sampling without replacement out of the global color distribution.
	
	Assume $a$ and $b$ are the initially largest and second largest opinion. Let $B'$ denote the initial absolute number of leaders with color $b$. Observe that $B'$ follows a hypergeometric distribution. That is, to determine $B'$ we draw $m$ balls out of $n$ total balls of which $b_0 \cdot n$ are colored black, and ask the question how many of the drawn balls are black.  The corresponding distribution follows the \emph{negative association} property \cite{JP83}, which according to  Theorem 3.1 of \cite{DBLP:books/daglib/0025902} allows us to bound $B'$ via Chernoff bounds on $\Bin(m,b_0)$. More specifically for $b_0' := B'/m$, it holds that 
	\begin{equation}
	\label{eq:leader-color-b-bound-5}
		b_0' \asymp \frac{\Bin(m,b_0)}{m} \overset{\text{\whp}}{=} b_0 \cdot  \left(1 \pm \frac{C'}{\sqrt{b_0}} \cdot \frac{\log^{c/2} n}{ \sqrt{n}} \right)
	\end{equation} 
	where $C'$ is a large enough constant, and assuming that $b_0 \gg 1/n$.
In case $b_0 \sim 1/n$ or even $b_0 \ll 1/n$ (which implies $a_0 =  1 -o(1)$) it is easy to see that $\log \log_{\alpha'} m = O( \log \log n)$. Now, if $b_0 \gg 1/n$,  
we repeat above approach to derive the color fraction $a'_0$. Then, we apply union bounds, and argue that all colors besides $a'$ also adhere to the upper bound on $b'_0$ in (\ref{eq:leader-color-b-bound-5}). This in turn implies for the bias of colors among leaders $\alpha'$ that 
\begin{align}
\label{eq:new_bias}
	\alpha'  > \alpha \cdot \left(1 - \frac{3C'}{\sqrt{b_0}} \cdot \frac{\log^{c/2} n}{ \sqrt{n}} \right) 
	 > \left( 1 + \frac{2}{b_0} \frac{\log^{c/2 + 1} n }{\sqrt{n}} \right) \cdot \left(1 - \frac{3C'}{\sqrt{b_0}} \cdot \frac{\log^{c/2} n}{ \sqrt{n}} \right), 
\end{align}
where we assumed in the second step that the initial bias $A >2\sqrt{n }\log^{c/2 + 1} n$. Since $b_0 < \sqrt{b_0}$ and $\log^{c/2 + 1} n > \log^{c/2} n$ it is easy to see that the rightmost term is dominated by $\alpha$, even if $\alpha$ is chosen to correspond to the smallest initially allowed bias. Therefore, $\log \log_{\alpha'} m= O( \log \log_{\alpha} m) = O(\log \log_{\alpha} n)$ follows accordingly.

The second statement follows from the fact that $n =  m \log^{c} n (1 \pm o(1))$ \whp. In case of $b_0 \sim 1/n$ or even $b_0 \ll 1/n$ it follows that $\alpha = 1-o(1)$ and the statement easily follows by a Chernoff bound application. If $b_0 \gg 1/n$ we have that $b_0' = b_0 (1 \pm o(1))$ \whp and, using the two rightmost factors in (\ref{eq:new_bias}), we get 
\[
	\alpha'  > \left(1 + \frac{1}{b_0'} \frac{\log m}{ \sqrt{m}}\right).
\]
The term on the right hand side implies that  $A' > \sqrt{m} \log m$, which concludes the proof.
The final statement follows immediately as $n \sim m$.
\end{proof}

By \cref{cor:one-phase-bias-1} we have that after $O(\log \log_{\alpha'} m)$ generations, the first monochromatic generation is reached. The first item of \cref{lem:initial-accelerate-bias} guarantees that this time lies in $O(\log \log_\alpha n + \log \log n) = O(\log \log_\alpha k + \log \log n)$ as desired.
It is easy to see, that in the two-choices phase of the following generation, every cluster will take this majority color value. After further $O(\log \log n)$ time partial consensus among all nodes is reached, as followers periodically copy the color values of their clusters. The result of \cref{thm:accelerated} follows.

%% file: 96-analysis-extensions.tex
\section{Extending our Protocols}
\label{sec:extensions}

\subsection{Extension 1: Termination}
\label{sec:ext-termination}

While our previous algorithms guaranteed fast partial and complete consensus, the nodes themselves are unaware of the fact that consensus has been reached. That is, nodes do not know \emph{when} they are done with the protocol and may consider their current color value as the final result. In the following we present an extension to our algorithm, circumventing this problem.

\paragraph{Centralized Algorithm}

We start by considering the following modification of the centralized algorithm in \cref{sec:Asynchronous-1leader}.
To allow proper termination, we extend each node (including the base station) with two additional state variables \texttt{terminated} and \texttt{final\_color}. The idea is that as soon as \texttt{terminated} is set to true, the nodes may consider the color stored in \texttt{final\_color} as result of the consensus algorithm.

Additionally, we employ a counter $t'$ and variable $c$ on the leaders end, initiated to $0$ and $null$ at the start of each generation. Each time a follower increases its generation, it also notifies the base station with its color (e.g. by appending $\col{v}{\gen{v}}$ to the notification in \cref{ln:sl-ln:notify-1, ln:sl-ln:notify-2} of \cref{alg:The-basic-procedure-asy-1}). 
If the base station $\ell$ receives such a notification while $\gensize{\ell} = 0$, then it sets $c$ to the color value contained in this notification. Throughout the two-choices phase (i.e while $\mode{l} = \TC$ on the leaders end), the leader counts in $t'$ the number of followers that joined the current generation and are of color $c$.

As soon as the condition in \cref{ln:gossip} of \cref{alg:leader-increment} is fulfilled, and the leader stops allowing promotion via two-choices, it checks whether $t' = \gensize{\ell}$. If this is the case, all nodes in the current generation must have taken the color stored in $c$. The leader may now set \texttt{final\_color} to $c$, and \texttt{terminated} to true.

On the followers end, we assume that they read the \texttt{terminated} bit and the \texttt{final\_color} variable of their leader each time they establish communication channels. For example just after \cref{ln:est-channels} in \cref{alg:The-basic-procedure-asy-1}. As soon as a follower $v$  witnesses that the leader set \texttt{terminated} to true, $v$ sets its own \texttt{terminated} variable to true, and copies the leader's value of \texttt{final\_color} into its own respective variable. From this point on $v$ does no longer need to actively execute \cref{alg:The-basic-procedure-asy-1}, and $v$ can consider the color in \texttt{final\_color} as the result of the consensus protocol.  

\begin{proposition}
	The results of \cref{thm:async-centralized} still hold after performing above modifications to the centralized algorithm. Furthermore, after $O\left(\log\log_{\alpha}k\cdot\log k+\log\log n\right)$ time, all but $n/\polylog n$ nodes have \texttt{final\_color} set to $a$, and after further $O(\log n)$ steps every node has set \texttt{final\_color} to $a$ \whp. Here $a$ denotes the initial plurality opinion.
\end{proposition}
\begin{proof}
	Clearly, the leaders \texttt{terminated} flag will be set exactly when the first monochromatic generation $i^*$ is reached. In \cref{subsec:The-Analysis-1leader} we established that this takes at most $O\left(\log\log_{\alpha}k\cdot\log k+\log\log n\right)$ time.
From this point on, every node will pull the $\texttt{terminated}$ flag together with $\texttt{final\_color}$ upon the next time it contacts the leader, and the result follows.
\end{proof}

\paragraph{Decentralized Case}

A termination mechanism employed in the decentralized algorithm follows a similar idea. That is, nodes and leader also employ the \texttt{terminated} and \texttt{final\_color} variables.
However, it is not enough that one cluster leader observes that all his followers belong to the same color after the two-choices phase, as this might not be discovered by all leaders in the same generation. Instead, we employ another instance of the sampling gadget, described in \cref{sec:sampling} into our algorithm.

Throughout the execution of the consensus protocol, the leaders perform consecutive samplings $R(w) \Leftrightarrow ($ w is currently of color $j)$.
These samplings are performed one after another, until a fraction $r' > (1- 1/\log n)$ of received samples confirm that $R(w)$ is indeed true. The idea is that if this sampling succeeds, then $j$ is the majority color \whp. Observe that the color $j$ needs to be specified for such a sampling to be properly defined, as otherwise the leader would require $\Omega(k \cdot \log \log n)$ bits to maintain samplings w.r.t. all colors simultaneously. As explained in \cref{sec:sampling}, the leader may evaluate $R(w)$ on its end. That is, the followers will instead of sending the evaluated $R(w)$, send the color of $w$ to the leader (e.g as part of the State Message in \cref{fig:async-follower}). At the start of each sampling process, the leader sets $j$ to the \emph{first} color it received by some of its followers.

Upon performing a successful sampling, the leader sets \texttt{final\_color} to $j$, and stops evaluating further samples. The leader (of size $n')$ now counts $0$-signals until in total  $\Theta(\log \log n) \cdot n'$ of them have been received \footnote{the exact counting threshold of $\mathcal{H}(C_{term}) \cdot n'$ is given in the proof of \cref{ext:ml-term-proof}}. Then, it sets the \texttt{terminated} flag to true, and stops following the consensus protocol in \cref{sec:multileader_proc} actively. From this point on the leader only needs to let nodes read its values of \texttt{terminated} and \texttt{final\_color}.

Followers encountering a leader (the leader $l_3$ of $v_3$ or their own leader, see \cref{fig:async-follower}) or any other node with \texttt{terminated} set to true, adopt the values of \texttt{terminated} and \texttt{final\_color}. In sequel such  nodes may stop following the consensus protocol actively, and only need to keep letting other nodes read their \texttt{terminated} and \texttt{final\_color} fields.

Additionally we make a modification similar to the mechanism of weaking up leaders from the sleeping phase, described in \cref{sec:multileader_proc}. Each time a follower node observes a \texttt{terminated} flag of some leader to be true, it informs its own leader of this fact together with the observed value of \texttt{final\_color}. This leader then also sets its \texttt{terminated} to true and sets \texttt{final\_color} to the received color, if it has not set \texttt{final\_color} any time earlier.

\begin{proposition}
\label{ext:ml-term-proof}
The results of \cref{thm:async-centralized} still hold after performing above modifications to the decentralized algorithm. Furthermore, after $O\left(\log\log_{\alpha}k\cdot\log k+\log\log n\right)$ time, all but $n/\polylog n$ nodes have \texttt{final\_color} set to $a$, and after further $O(\log n)$ steps every node has set \texttt{final\_color} to $a$ \whp. Here $a$ denotes the initial plurality opinion.
\end{proposition}

\begin{proof}

	Along the lines of \cref{ml-sampling-theorem} it is easy to see that no leader will set its \texttt{terminated} flag to true before a $1 - 2 / \log n$ fraction of nodes belong to the same color globally.
	Let now $i'$ denote the currently allowed generation at the point in time $t_f$ -- the point in time when the first leader performed a successful sampling.
Similar let $t_\ell$ denote the time at which at least $(1- 1/\log n)$ of all leaders managed to perform such a successful sampling.	
	
As established above, it holds \whp that almost every node is of color $a$.  Therefore, globally, color $a$ is $\polylog n$ times more dominant than any other color. Without giving a detailed proof, it is easy to see that this must also hold for the currently highest generation $i'$, i.e., $\alpha_{i'} > \polylog n$. According to \cref{cor:one-phase-bias-1}, which also holds in the decentralized case, the bias is roughly squared with each subsequent generation. Along the lines of \cref{lem:async-num-generations} and \cref{lem:async-alpha>k-steps}, this implies that a  monochromatic generation is reached after $\log \log n +2$ further generations. As $a_{i'} = (1 - o(1))$, each of these generations takes $O(1)$ time at most. A similar argument as in the proof of \cref{lem:ml-final-steps} shows that after $O(\log \log n)$ further steps,  at least a $(1 - 1/(2\log n))$ fraction of nodes will be of a color $a$. At this point, every leader will perform a successful sampling. It can be shown that the time required for this whole process can be bounded by $C_{term} = 3 \log \log n \cdot (C_{br} + \mathcal{S}(C_{tc}) + \mathcal{S}(C_{slp}) + \mathcal{S}(5 C_1) +\mathcal{S}(C_{pre}) + 10 \cdot C_1) = O(\log \log n)$ time steps \whp

As leaders are required to count to $\mathcal{H}(C_{term})$ before setting the \texttt{terminated} flag, every leader is able to perform a successful sampling before any leader stops following the consensus protocol. Hence, $C_{term} + \mathcal{S}(C_{term}) = O(\log \log n)$ time following $t_f$, every leader will have set the $\texttt{terminated}$ flag. This leads to \texttt{final\_color} being set at a $(1- 1/\polylog)$ fraction and all nodes after $O(\log \log n)$ and $O(\log n)$ further time, respectively.
\end{proof}

\subsection{Extension 2: Poisson Clocks and the Accelerated Consensus Protocol}
\label{sec:ext-poisson-accelerated}

Throughout the analysis of the Accelerated Consensus Protocol in \cref{sec:accelerated-analysis} we established that the $q$-dense property together with the fact that followers act as relays to feed information to their leaders allowed us to speed-up the propagation phase by a factor of $\Omega(\log n)$.
 In the following we will expand upon this idea and show that also other parts of the protocol can be improved. For now, we will focus on the consensus part of the protocol. That is, we assume that nodes follow the Accelerated Consensus Protocol and already lie in clusters of sufficient polylogarithmic size. 

For further simplification, assume that communication channels are opened instantly and the ticking time of nodes follows $\Exp(1)$. As illustrated part of an example (see \cref{ex:exp-dense} on page \pageref{ex:exp-dense}) this distribution is $(1+\varepsilon)$-dense for any constant $\varepsilon >0$, and in particular for $X \sim \Exp(1)$ it holds that  $P(X < 1/\log^2 n) > 0.6/\log^2 n$.  This way, a large enough polylogarithmic cluster size implies the following observations.

 \begin{enumerate}
 	\item  As communication takes no time, it follows that throughout any $1/\log^2 n$ time frame each cluster has a follower that ticks and opens all communication channels \whp.
 	\item \label{ext:item-2}In case all leaders currently allow propagation to generation $i$, the spreading of generation $i$ can  be seen as pull gossiping at an $1/\log^2 n$ accelerated rate.
 	\item If two-choices steps are allowed for at least $1/\log^2 n$ time steps simultaneously by all leaders then every cluster has at least one follower that performs a two-choices step for its cluster.
 	\item Leaders can employ the Sampling Gadget which yields a full sampling after at most $1/\log^2 n$ time (i.e. the time $t'' - t'$ in \cref{ml-sampling-theorem} may be bounded by  $O(1/\log^2 n)$)
 \end{enumerate}

We note that all the above can also be achieved even when accounting for channel opening delays under the assumption that all waiting time distributions are $q$-dense and follow the positive aging property. This makes it seem as if the time between two generations $t_{i+1} - t_i$ could be reduced to length $O(1/\log n) = o(1)$ and raises the question why we only sped-up the propagation as part of the Accelerated Consensus Protocol. The reason for this is that the counting of $0$-signals performed by the leaders (see \cref{fig:async-leader} on \pageref{fig:async-leader}) only allows us to accurately approximate time frames that are of at least constant length (see \cref{lem:counting-time}).
This is mostly due to the following two reasons: (i) considering a time interval $[t,t+L]$ of length $L$, there may be many $0$-signals arriving that were sent before time $t$, and (ii) the $q$-dense property alone does not exclude the possibility of multiple nodes ticking at roughly the same time, causing the leader to be flooded with $0$-signals in the aforementioned interval. However, making use of instant communication as well as the \emph{memoryless} property of the exponential distribution, we can overcome these two challenges and show the following.

\begin{lemma}
	Assume that all nodes are equipped with Poisson clocks with rate $\lambda = 1$ and that the establishment of communication channels takes no time. If a leader with $|U| > \log^{3+\varepsilon} n$ followers (for some arbitrary constant $\varepsilon > 0$) starts counting incoming $0$-signals at time step $t$, then the counter will reach value $W := 2 |U| / \log^2 n$ in the time interval $[t + \frac{1}{\log^2 n} ~,~ t + \frac{4}{\log^2 n}]$ \whp.
\end{lemma}
\begin{proof}	
	We start by showing that in the $4/\log^2 n$ time steps following $t$, at least $2|U| / \log^2 n$ many $0$-signals are received by the leader. 
		Let the r.v.\ $X^{(v)}$ denote whether the first tick of some node $v$ following time $t$ lands in the interval $[t, t+4/\log n]$. Due to memorylessness it follows that $X^{(v)} \sim \text{Exp}(1)$ and by \cref{ex:exp-dense} we have that $P(X^{(v)} < 4/\log^2 n) > 2.4 / \log^2 n$. We define the indicator variable $Y^{(v)}$ with $Y^{(v)}=1$ iff $X^{(v)} < 3/\log^2$ and $0$ otherwise. As the variables $Y^{(v)}$ for $v \in V$ are independent, we apply Chernoff bounds w.r.t.\ $X= \sum_{v \in U} Y^{(v)}$ and deduce that \whp $X >  |U| (2/ \log^2 n)$. Hence, the leaders counter will reach $W$ before $t+  4/\log^2 n$.
		
		Next, we consider how many signals the leader will at \emph{most} receive in the interval $[t,t+1/\log n]$. Let the r.v. $Y^{(v)}_i$ now indicate whether the $i$-th tick of $v$ lands in the interval $[t,t + 1/\log n]$. Let $Z \sim \text{Exp}(1)$, then it follows that
		\begin{align*}
			P\left(Y^{(v)}_1 = 1 \right) &= P\left(Z \leq 1/\log^2 n \right) = 1 - \exp(-1/\log^2 n) \\
			&\leq 1 - \left( \left(1 - \frac{1}{\log^2 n} \right)^{\log^2 n} \right)^{1/\log^2 n} = 1/\log^2 n,
		\end{align*}		
		where the first step holds due to memorylessness and we used that $(1-x)^{(1/x)} \leq 1/e$ for $0 < x \leq 1$.
		Let now $Y_i = \sum_{v \in U} Y^{(v)}_i$. It follows that $E(Y_1) < |U|/\log^2 n$ and when applying the Chernoff bound we deduce that $Y_{1} <  |U| (1/\log^2 n) (1+ o(1))$ \whp. Observe that, for $i > 0$ and fixed $v$,  $P(Y^{(v)}_i = 1 | Y^{(v)}_{i-1} = 0) = 0$ as well as $P(Y^{(v)}_i = 1 | Y^{(v)}_{i-1} = 1) \leq 1 /\log^2 n$. That is, node $v$ can only tick $i$ times inside $[t,t+1/\log n]$ if the previous $i-1$ ticks landed in $[t,t+1/\log n]$ as well. Hence, considering the values $Y_i$ for $i> 0$ in sequence, we can majorize $Y_{i}$ by $\Bin(Y_{i-1}, 1 / \log^2 n)$. Until $Y_{i-1} = o(\log^3 n)$ for the first time we thereby get that 
\[
	Y_i  \overset{\text{\whp}}{<}  |U| \left( \frac{1}{\log^2 n} (1 + o(1)) \right)^i.
\]
Hence, it is easy to see that total number of ticks made in the time interval $[t,t+1/\log^2 n]$ -- equaling to $Y = \sum_{i=1}^{\infty} Y_i$ -- can be upper bounded by $2 |U| / \log^2 n$. As we assume communication channels to be established instantly, this upper bounds the number of $0$-signals received by the leader in $[t,t+1/\log^2 n]$.
\end{proof}
A repetition of the above also allows to deduce that by counting until $\mathcal{W}(y) := y  \cdot (2|U|) / \log^2 n$ many $0$-signals are received,  a leader can guarantee that at least $\frac{y}{\log^2 n}$  and at most $\frac{4y}{\log^2 n}$ time passes for any $y \geq 1$. Hence, arbitrary time frames with length in multiples of $1/\log^2 n$ can be approximated.
Throughout the execution of our previous consensus protocols, leaders may count to $\mathcal{H}(T)$ for some $T$ to ensure that at least $T$ time steps passed. These occurrences are now replaced by having the leader count to $\mathcal{W}(T)$ instead. This way, at least $T$ time slots of length $1/\log^2 n$ pass until the counter hits $\mathcal{W}(T)$, while at the same time guaranteeing that at most $T /\log^2 n$ time passes. 

Remember, throughout each such time frame each cluster leader will perform a two-choices or propagation step, using its followers as a relay. Hence, many parts of the protocol that originally required $\Omega(1)$ time, e.g. the consecutive time all leaders allowed two-choices in the decentralized protocol or the sleeping phase, can now be reduced to $\Theta(1/\log^2 n)$. This corresponds to reducing the length of a time unit to $O(1/\log^2 n)$ and  leads to an improved running time of $O(1)$ to reach consensus among leaders. Further $O(\log \log n)$ time later partial consensus is reached, leading to the following statement.

\poissonextension*

The required initial bias (and cluster size) is determined as follows. First, we make a similar argument as in \cref{lem:tick-fast} (and use the fact that communication channels are opened instantly). This yields that configurating the clustering procedure to generate clusters of size at least $\log^{5} n$ is sufficient (i.e., the clustering algorithm needs to be configured with $c=6$ or larger- see \cref{sec:clustering}).   Second, we apply \cref{lem:initial-accelerate-bias} and deduce that this cluster size implies a required initial bias of $2\sqrt{n} \log^{3} n$.

\paragraph{Adapting the Clustering Procedure}

Initially we assumed that all nodes already lie in clusters.
To achieve this, the clustering procedure in \cref{sec:extended-clustering} needs to be employed before the start of the consensus routine, just as in case of the Decentralized and Accelerated Consensus protocols. Note that here we still need to work with the usual notions of a time unit, which lasts for constant time. 

There is one modification that needs to be mode, however. As the consensus protocol described in this section operates on time units of length $O(1/\log^2 n)$ we need to make sure that leaders also transition from the clustering to the start of the consensus routine within time difference at most $O(1/\log^2 n)$. The protocol stated in \cref{sec:extended-clustering} only guarantees a time difference of $C_\ell = \Theta(1)$ (see \cref{thm:ml-extended-clustering}) . However, this can be overcome as follows. Leaders that enter the consensus mode first wait for at least $C_\ell$ time by counting $0$-signals and then trigger the broadcast of a massage. This message is again spread among leaders by using their followers as relays. By \cref{ex:exp-dense} we have for $X \sim \Exp(1)$ that $P(X < 1/\log^3 n) = \Omega(1/\log^3 n)$ and therefore a follower of every large enough leader ticks every $O(1/\log^3 n)$ time steps \whp. This implies that this broadcast requires less than $O(1/\log^2 n)$ time to be completed. Each leader that receives such a message immediately enters consensus mode (without any additional waiting), yielding the desired $O(1/\log^2 n)$ time difference between the first and last leader entering consensus mode.

%% file: 97-polya-eggenberger.tex
\section{The Pólya-Eggenberger Distribution}
\label{sec:polya-eggenberger}

In what follows we describe a simple urn process consisting of a single urn and balls that are colored either black or white. The process consists of a sequence of $n$ steps and in every such step, the total amount of balls inside the urn increases by $s$. The description of such a step $k$, for $1 \leq k \leq n$, is quite simple: first, a random ball is drawn and put back into the urn; then $s$ additional balls that match the color of the drawn ball are added to the urn. Observe that this implies that the probability of drawing a ball of a certain color evolves with each further step. Furthermore, this process is subject to a ``the rich get richer'' effect.

What we just described is the so-called Pólya-Eggenberger process. We define by $PE_s(a,b,n)$ the corresponding distribution, denoting number of black balls \textit{added} throughout this process. Here $s$ describes the batch size of balls added per step (we will only consider $s=1$) and $a,b$ denote the number of initially present black and white balls, respectively. Finally, $n$ denotes the number of steps, which in case of $s=1$ corresponds to the total number of balls added throughout the process.

To the best of our knowledge there do not exist any tight tail bounds on this Pólya-Eggenberger distribution that are simple to work with.
To achieve such a result, we look at the Pólya-Eggenberger process from a different perspective. It can also be seen as the result of the following two step process. Instead of considering a dynamic process where the probability to hit the white urn evolves over time, we employ a static probability $T$ drawn from a beta distribution with parameters $a$ and $b$ at the start of the process. The total number of balls added to the white urn can then be described by $\text{Bin} (n,T)$ -- a binomial distribution consisting of $n$ experiments each succeeding with probability $T$. In other words, for $T \sim \text{Beta}(a,b)$ and $0 \leq w \leq n$, we have that 
\begin{equation*}
	P \Big(  \text{PE}_1 (a,b,n) = w \Big) = P \Big( \text{Bin}(n,T) = w \Big).
\end{equation*}
 A simple proof that this equality indeed holds can be found on page 181 of \cite{JK77}.
In order to derive a concentration result for $A_n \sim \text{PE}_1 (a,b,n)$, we account for (i) the deviation of the value $T$ from its mean, and (ii) the concentration of the binomial distribution conditioned on $T$.  Luckily, among other interesting concentration results, a tight tail bound on the beta-distribution is given in \cite{ZZ19}. We state a slightly modified version of their result as follows.

\begin{theorem}[simplified Theorem 8 of \cite{ZZ19}]
\label{pe-thm-8}
	Let $T \sim \Beta(\alpha,\beta)$ where $\alpha,\beta \geq 1$. Then, it holds for $0 < \delta < \sqrt{\alpha}$  and some universal constant $c_1 >0$ that
	\[
		P \left(T \geq \frac{\alpha}{\alpha +\beta} + \frac{\sqrt{\alpha}}{\alpha + \beta} \cdot \delta \right) < 2 \exp \left( -c_1 \delta^2 \right)
	\] 
and 
	\[
		P \left( T \leq \frac{\alpha}{\alpha +\beta} - \frac{\sqrt{\alpha}}{\alpha + \beta} \cdot \delta \right) < 2 \exp \left( -c_1 \delta^2 \right)
	\] 
\end{theorem} 
\begin{proof}

	The second bound follows immediately from the second inequality in Theorem 8 of \cite{ZZ19}, when setting $x=\frac{\sqrt{\alpha}}{\alpha+\beta} \cdot \delta$ for $0 < \delta < \sqrt{\alpha}$. 
	Now to the bound for the right tail. We set $x$ just as before and this time apply the first inequality of Theorem 8 \cite{ZZ19}. Note that this inequality requires $x < \frac{\beta}{ \beta + \alpha}$ and therefore only yields the desired result for $\delta < \beta / \sqrt{\alpha}$. This might be more restrictive than $\delta < \sqrt{\alpha}$ in case of $\alpha > \beta$. However, for $\delta \geq \beta /\sqrt{\alpha}$ we can use that the Beta distribution has non-zero support in $(0,1)$ only, i.e., 
	\[
	    P \left( T \geq \frac{\alpha}{\alpha + \beta} + \frac{\sqrt{\alpha}}{\alpha + \beta} \delta \right) \leq P \left( T \geq \frac{\alpha}{\alpha + \beta} + \frac{\beta}{\alpha + \beta}  \right) =  P(T \geq 1) = 0. \qedhere
	\]
\end{proof}

Above result allows us to derive the following.

\begin{theorem}
\label{pe-thm-main}
	Let $A_n \sim \PE_1 (a,b,n)$ with $\mu= (a/(a+b)) \cdot n$ as well as $a+b \geq 1$. \\
	 If $n \geq (a+b)$ then it holds for any $0 < \delta < \sqrt{a}$ that 
	\begin{align}
		&P \left(A_n > \mu + \sqrt{a} \cdot \frac{n}{a+b} \cdot \delta \right) < 4 \exp(-c_2 \cdot \delta^2), \text{ and}  \label{eq:pe-main-upper-1}\\
		&P \left(A_n < \mu - \sqrt{a} \cdot \frac{n}{a+b} \cdot \delta \right) < 4 \exp(-c_2 \cdot \delta^2).\label{eq:pe-main-lower-1}
	\end{align}
	Furthermore, if $n < (a+b)$ it holds for any $0 < \delta < \sqrt{a} \cdot \sqrt{n / (a+b)}$ that
	\begin{align}
		&P \left(A_n > \mu + \sqrt{a} \cdot \sqrt{\frac{n}{a+b}} \cdot \delta \right) < 4 \exp(-c_2 \cdot \delta^2), \text{ and} \label{eq:pe-main-upper-2} \\
		&P \left(A_n < \mu - \sqrt{a} \cdot \sqrt{\frac{n}{a+b}} \cdot \delta \right) < 4 \exp(-c_2 \cdot \delta^2). \label{eq:pe-main-lower-2}
	\end{align}
	Finally, if $n < (a+b)$ and $\sqrt{a} \cdot \sqrt{n / (a+b)} \leq \delta <  \sqrt{a}$, we have 
	\begin{align}
		P \left(A_n > \mu + \delta^2 \right) < 4 \exp(-c_2 \cdot \delta^2). \label{eq:pe-main-upper-3}
	\end{align}
	Here $c_2 \geq\min\{1/48, c_1 / 4\}$ is a universal constant with $c_1$ originating from \cref{pe-thm-8}. 
\end{theorem}
\begin{proof}
	\textbf{Lower Tail}. We start with showing (\ref{eq:pe-main-lower-1})  and (\ref{eq:pe-main-lower-2}).  Let $\Delta(\delta) =  \frac{\sqrt{a}}{a+b} \cdot n \cdot \delta$. For $T \sim \text{Beta}(a,b)$ we define the event $\mathcal{E} :\Leftrightarrow \{ T \geq   a/(a+b) - \sqrt{a}/ (a+b) \cdot (\delta/2) \}$ and consider any $\delta$ constrained to $0 < \delta < \sqrt{a}$. Let now $M(\delta) := \max \{ \Delta(\delta), \sqrt{\mu}\cdot \delta\}$ and observe that $\Delta(\delta) / n$ reflects the error term of  \cref{pe-thm-8}. Then, by the law of total probability we have that 
	\begin{align}
	\label{eq:eggen} \nonumber
		P[A_n \leq \mu - M(\delta)] &= P(\text{Bin}(n,T) \leq \mu - M(\delta) ~|~ T )   \\ \nonumber
		&= P ( \text{Bin}(n,T) \leq \mu - M(\delta) ~|~ \mathcal{E}) \cdot P(\mathcal{E})  \\ \nonumber
		&+ P ( \text{Bin}(n,T) \leq \mu - M(\delta) ~|~ \neg \mathcal{E})\cdot (1 - P(\mathcal{E})) \\ 
		&\leq  P ( \text{Bin}(n,T) \leq \mu - M(\delta) ~|~ \mathcal{E}) + 2 \exp \left( {-\frac{c_1}{4} \cdot \delta^2 } \right).
	\end{align}
In the last line we crudely bounded some factors by 1 and applied \cref{pe-thm-8} to bound the term $(1- P(\mathcal{E}))$. 
Observe that $\text{E}(\text{Bin}(n,T) | \mathcal{E}) \geq \mu - \Delta(\delta)/2 =: \mu'$ as the conditioning on $\mathcal{E}$ can be seen as an a priori requirement on the success probability of the binomial distribution. Hence, we may  apply Chernoff bounds and deduce that
\[
    P \left(  \text{Bin}(n,T) \leq  \mu' \cdot  \left( 1 - \frac{\delta}{2 \sqrt{\mu'}} \right) ~\Big|~ \mathcal{E}\right) < \exp({-\delta^2 / 12}).
\]
Note that this Chernoff bound application requires $\delta/ (2\sqrt{\mu'}) < 1$.
As $\mu' > \frac{a}{a+b} \cdot \frac{n}{2}$ is implied by $\delta < \sqrt{a}$, this can be achieved by the additional constraint $\delta < \sqrt{a} \sqrt{\frac{n}{a+b}} \cdot \sqrt{2}$. Initially we considered $\delta < \sqrt{a}$, therefore the combined requirements on $\delta$ can be stated  as 
\[
    0 < \delta < \sqrt{a} \cdot \min\{1, \sqrt { \frac{n}{a+b}} \}.
\]
Next, observe that
\[
    \mu' \cdot \left(1 - \frac{\delta}{2\sqrt{\mu'}}\right) > \left( \mu - \frac{\Delta(\delta)}{2} \right) - \frac{1}{2} \sqrt{\mu} \cdot \delta \geq \mu - \max \{\Delta(\delta), \sqrt{\mu} \delta \} = \mu - M(\delta)
\]
 Hence, we deduce that $P(\text{Bin}(n,T) \leq  \mu - M(\delta)| \mathcal{E}) < \exp{(-\delta^2 / 12)}$. When combining this with (\ref{eq:eggen}), the inequalities (\ref{eq:pe-main-lower-1})  and (\ref{eq:pe-main-lower-2}) follow.\\
\textbf{Upper Tail}. In order to show the bounds  (\ref{eq:pe-main-upper-1}), (\ref{eq:pe-main-upper-2}) and (\ref{eq:pe-main-upper-3}) for the upper tail, we follow a similar approach and consider some arbitrary $\delta < \sqrt{a}$. We again let $\Delta(\delta) = \frac{\sqrt{a}}{a+b} \cdot n \cdot \delta$ and define the event $\mathcal{E} :\Leftrightarrow \{T \leq a / (a+b) + \sqrt{a} / (a+b) \cdot ( \delta / 2) \}$. We slightly extend the definition of $M(\delta)$ to $M(\delta) : = \max \{\Delta(\delta), \delta \sqrt{\mu}, \delta^2\}$. Just as with  (\ref{eq:eggen}), we  employ \cref{pe-thm-8} and the law of total probability to establish the following bound 
\begin{equation}
\label{eq:eggen2}
	P(A_n \geq \mu + M(\delta)) \leq P \left( \Bin(n,T) \geq \mu + M(\delta) ~\Big|~ \mathcal{E} \right) +  2 \exp\left( -\frac{c_1}{4} \cdot \delta^2 \right).
\end{equation}
This time, we observe that $E(\Bin(n,T) | \mathcal{E}) \leq \mu + \Delta(\delta) / 2 := \mu'$ and apply Chernoff bounds to deduce that
\begin{equation}
\label{eq:eggen3}
	P\left( \Bin(n,T) \geq \mu' \cdot (1 +  \max\{ \frac{\delta}{4\sqrt{\mu}}, \frac{\delta^2}{16 \mu}\}) ~|~ \mathcal{E} \right) < \exp (-\delta^2 / 48).
\end{equation}
Next, we make the following observation when using that $\mu' = \mu + \Delta(\delta) / 2 $ and $\Delta(\delta) / 2 \leq \mu/2 < \mu$ in the first step 
\begin{align*}
	&\mu' \cdot  (1 +  \max\{ \frac{\delta}{4\sqrt{\mu}}, \frac{\delta^2}{16 \mu}\}) = \mu + \frac{\Delta(\delta)}{2} +  2\mu \cdot \max\{ \frac{\delta}{4\sqrt{\mu}}, \frac{\delta^2}{16\mu}\} \\
	& < \mu + 2 \max\{ \frac{\Delta(\delta)}{2},  \max\{\frac{\delta}{2}\sqrt{\mu}, \frac{\delta^2}{8}\}\}= \mu +  \max \{\Delta(\delta), \delta \sqrt{\mu}, \frac{\delta^2}{4} \} \leq \mu + M(\delta).
\end{align*}
In the second step we just crudely combined all terms with the help of maximas. When combining (\ref{eq:eggen2}) and (\ref{eq:eggen3}) with this result, we get that  $P(A_n \geq \mu + M(\delta)) \leq \exp(-\Omega(\delta^2))$ as desired. Depending on $\delta$ as well as $n$ and $(a+b)$ the expression $M(\delta)$ might take different values. More specifically, the inequalities (\ref{eq:pe-main-upper-1}), (\ref{eq:pe-main-upper-2}) and (\ref{eq:pe-main-upper-3}) of the theorem follow because 
\begin{align*}
	M(\delta) =
	\begin{cases}
		\Delta(\delta) & \text{if } n \geq a+b  \text{ and } 0 < \delta < \sqrt{a} \\
		 \delta \sqrt{\mu} & \text{if } n < a+b \text{ and } 0 < \delta <  \sqrt{a} \sqrt{\frac{n}{a+b}} = \sqrt{\mu} \\
		\delta^2 & \text{if } n < a+b \text{ and } \sqrt{a} \sqrt{\frac{n}{a+b}} \leq \delta < \sqrt{a}. \qedhere
	\end{cases}
\end{align*}
\end{proof}

Often it is useful to consider the \emph{total} number of black balls that reside in the urn after a certain number of balls have been added to the urns.
In the following we will employ the result of \cref{pe-thm-main} to bound the total amount of black balls after $n-(a+b)$ balls have been added. That is, we bound the number of black balls after filling the urn with $n$ balls in total. This leads to the following convenient but slightly weaker result ($c_2$ is the constant defined in \cref{pe-thm-main}).

\pethmtotal*

\begin{proof}
We need to consider multiple cases. \\
\textbf{Case 1.} $n - (a+b) \geq (a+b)$. We only give a proof for the first inequality as the second is derived in a similar manner. We apply (\ref{eq:pe-main-lower-1}) of  \cref{pe-thm-main} to $A$ which implies for $0 < \delta < \sqrt{a}$ that
 \[ 
    P \left( a + A < \mu - \sqrt{a} \cdot \frac{n-(a+b)}{a+b} \delta \right) < 4 \exp(c_2 \cdot \delta^2).
 \]
    The term on the left-hand side can be simplified and lower bounded as follows
 \[
    \mu - \sqrt{a} \cdot \frac{n-(a+b)}{a+b} \delta > \mu - \sqrt{a} \cdot \frac{n}{a+b} \cdot  \delta
 \]
 and the result follows as $P ( X < x_1) \leq P(X < x_2)$ for $x_1 < x_2$. \\
 \textbf{Case 2.} $n - (a+b) < (a+b)$. Here we need to further distinguish depending on $\delta$. \\
 \textbf{Case 2a.} $0 < \delta < \sqrt{a} \cdot \sqrt{\frac{n-(a+b)}{(a+b)}}$. We again only show the proof for the first inequality as the proof for the second inequality is similar.
First, we deduce by (\ref{eq:pe-main-lower-2}) of  \cref{pe-thm-main} that
 \[
    P \left( a + A < \mu - \sqrt{a} \cdot \sqrt{\frac{n- (a+b)}{ (a+b)} } \delta \right) < 4  \exp(-c_2 \cdot \delta^2).
 \]
 The error term in this expression is smaller than the desired term $\sqrt{a} \frac{n}{a+b} \cdot \delta$. To observe this consider the following, where the second inequality follows from  $n-(a+b)$ and $a+b$ being smaller than $n$ 
 \begin{align*}
    \sqrt{\frac{n - (a+b)}{a+b}} < \frac{n}{a+b} \Leftrightarrow \sqrt{n - (a+b)} \cdot \sqrt{a+b} <  n.
 \end{align*}
 \textbf{Case 2b.} $\sqrt{a} \cdot \sqrt{\frac{n-(a+b)}{(a+b)}} \leq \delta < \sqrt{a}$. 
 We start by showing that the first inequality holds in this setting. Clearly it holds that $P(a + A < a) = 0$, i.e., in the worst case not a single black ball is added to the urn. 
 We show that in this setting $\mu - \sqrt{a} \frac{n}{a+b} \delta \leq a$ holds. This implies that $P(a + A < \mu - \sqrt{a} \frac{n}{a+b} \delta) = 0$ and the desired result follows.
Using that $\sqrt{a} \cdot \sqrt{\frac{n-(a+b)}{(a+b)}} \leq \delta$ and $n \geq \sqrt{n - (a+b)} \cdot \sqrt{a+b}$ in the first and second step, respectively, we observe 
 \begin{align*}
    \sqrt{a} \frac{n}{a+b} \delta \geq a \cdot \frac{n}{a+b} \sqrt{\frac{n - (a+b)}{(a+b)}} \geq \frac{a}{a+b} (n - (a+b)).
 \end{align*}
This intermediate result can then be used to deduce that 
\[
    \mu - \sqrt{a} \frac{n}{a+b} \delta \leq \mu - \frac{a}{a+b} (n- (a+b)) = \frac{a}{a+b}n - \frac{a}{a+b}n + a = a
\]
as desired.

To show the second inequality of the theorem we need to resort to \cref{pe-thm-main}. Using inequality  (\ref{eq:pe-main-upper-3}) we get that 

\[
	P(a + A > \mu + \delta^2) = P \left( A < a \cdot \frac{n-(a+b)}{(a+b)} + \delta^2 \right) < 4e^{-c_2\delta^2}.
\]
Hence, the desired statement follows in case $\mu + \delta^2 \leq \mu + \sqrt{a} \frac{n}{a+b} \delta$. It is easy to see that this indeed holds as $\delta < \sqrt{a} \leq \sqrt{a} \frac{n}{a+b}$.
\end{proof}

All our previous theorems require the $\delta$ factor in the error term to be bounded by $\sqrt{a}$ from above. In case $a$ lies in $o(\sqrt{\log n})$ our bounds cannot be employed to achieve probabilistic guarantees of order $n^{-\Omega(1)}$.
To circumvent this we present the following theorem.

\begin{theorem}
\label{pe-thm-small-a}
Let $A \sim \PE_1 (a,b,n - (a+b))$ with $1 \leq a \leq b$ and $n \geq (a+b)$. Then, it holds that 
\[
    P \Big( a + A  > M \cdot ( 3a + c_4 \log n) \Big) < 2n^{-2}
\]
where $M:= \max \{1 , (n - (a+b)) / (a+b) \}$ and $c_4 > 0$ is a universal constant.
\end{theorem}
\begin{proof}
As the proof of this similar to the one of \cref{pe-thm-main} we keep it short.
We again model $A$ as $A \sim \text{Bin}(n- (a+b),T)$ with $T\sim\text{Beta}(a,b)$. 
We let $c_4 = (\frac{4}{c'} +6)$, where $c'$ is a constant we will specify later, and distinguish two cases depending on the size of $b$. \\
\textbf{Case 1.} $b\leq \frac{2}{c'}\log n$,\\
In this case, observe that
\[
   P \Big( a + A  > M \cdot ( 3a + c_4 \log n) \Big) \leq P \Big( A  > M \cdot ( 2a + c_4 \log n) \Big) \leq P \Big( A  > n - (a+b) \Big) = 0,
\]
where we used that $(2a + c_4 \log n) > (a+b)$  and $M \cdot (a+b) \geq n -(a+b)$ in the second step. \\
\textbf{Case 2.} $b > \frac{2}{c'} \log n$. \\
The first tail bound in Theorem 8 of \cite{ZZ19} can be used to achieve the following bound for any positive $\delta$ subject to $\delta \cdot a < b$ 
\[
    P \left( T > \frac{a}{a+b} + \delta \cdot  \frac{a}{a+b} \right) < 2 \exp \{-c' \cdot \delta \cdot a\}, 
\]
when we use that $a \leq b$ and assume that the constant $c' > 0$ is chosen accordingly. Then, setting  $\delta$ such that $\delta \cdot a = (2/c') \cdot \log n$ implies that $\delta \cdot a < b$ and we can employ above result to derive 
\begin{equation}
\label{pe-eq-2}
    P \left( T \leq \frac{1}{a+b} \left( a + \frac{2}{c'} \log n\right) \right) \geq 1 -  n^{-2}.
\end{equation}
Now, for any arbitrary binomially distributed random variable $B$, Chernoff bounds give us that $P(B > \max \{ 2 \cdot E[B] ~,~ 6 \log n \}) < n^{-2}$.
Using this and abbreviating the probabilistic event in (\ref{pe-eq-2}) with $\mathcal{E}$, we derive that
\[
    P \left( A  > \max \left\{ \frac{n-(a+b)}{a+b} \left(2a + \frac{4}{c'} \log n\right) ~,~ 6 \log n  \right\} ~\Big|~ \mathcal{E} \right) < n^{-2}.
\]
Finally, we set $M := \max \{1, (n - (a+b)) / (a+b)\}$ and translate above result into a bound on $a + A$. We can express the previous bound in the following slightly weaker form when using that $\max(x,y) \leq x+y$ for $x,y \geq 0$.
\[
    P \left(a + A > M \cdot \left(3a +  (4/c' + 6) \cdot \log n \right) ~\Big|~ \mathcal{E} \right) < n^{-2}.
\]
The result follows from the law of total probability as $\neg \mathcal{E}$ occurs with probability at most $n^{-2}$.
\end{proof}